\documentclass[final,1p,times]{elsarticle}
\usepackage{amssymb}
\usepackage{amsfonts}
\usepackage{amsmath}
\usepackage{geometry}
\usepackage{graphicx}
\usepackage{epsfig}
\usepackage{epstopdf}
\usepackage{bm}
\numberwithin{equation}{section}
\newtheorem{thm}{Theorem}

\newdefinition{rmk}{Remark}

\newproof{pf}{Proof}
\newproof{pot}{Proof of Theorem  
}
\journal{}
\begin{document}

\begin{frontmatter}
\title{Power series and integral forms of Lame equation in Weierstrass's form and its asymptotic behaviors}

\author{Yoon Seok Choun
}
\ead{Yoon.Choun@baruch.cuny.edu; ychoun@gradcenter.cuny.edu; ychoun@gmail.com}
\address{Baruch College, The City University of New York, Natural Science Department, A506, 17 Lexington Avenue, New York, NY 10010}
\begin{abstract}

We consider the power series expansion of Lame equation in Weierstrass's form and its integral forms applying three term recurrence formula \cite{chou2012b}. We investigate asymptotic expansions of Lame equation for the cases of infinite series and polynomials. And we show why Poincar\'{e}-Perron theorem is not always applicable to the Lame equation.

We show how the power series expansion of Lame functions in  Weierstrass's form can be converted to closed-form integrals for all cases of infinite series and polynomial. One interesting observation resulting from the calculations is the fact that a $_2F_1$ function recurs in each of sub-integral forms: the first sub-integral form contains zero term of $A_n's$, the second one contains one term of $A_n$'s, the third one contains two terms of $A_n$'s, etc.

This paper is 7th out of 10 in series ``Special functions and three term recurrence formula (3TRF)''. See section 7 for all the papers in the series. Previous paper in series deals with the power series expansion and the integral formalism of Lame equation in the algebraic form and its asymptotic behavior \cite{Chou2012f}. The next paper in the series describes the generating functions of Lame equation in Weierstrass's form\cite{Chou2012h}.

Nine examples of 192 local solutions of the Heun equation (Maier, 2007) are provided in the appendix.  For each example, we show how to convert local solutions of Heun equation by applying 3TRF to analytic solutions of Lame equation in Weierstrass's form.

\end{abstract}

                                      
\end{frontmatter}  
\section{Introduction}
In 1837, Gabriel Lame introduced second ordinary differential equation which has four regular singular points in the method of separation of variables applied to the Laplace equation in elliptic coordinates\cite{Lame1837}. Various authors has called this equation as `Lame equation' or `ellipsoidal harmonic equation'\cite{Erde1955}. 

Lame ordinary differential equation in Weierstrass's form and Heun equation are of Fuchsian types with the four regular singularities. Lame equation in Weierstrass's form is derived from Heun equation by changing all coefficients $\gamma =\delta =\epsilon =\frac{1}{2}$, $a =\rho ^{-2}$, $\alpha = \frac{1}{2}(\alpha +1) $, $\beta = -\frac{1}{2}\alpha  $, $q=-\frac{h}{4}\rho ^{-2}$ and an independent variable $x=sn^2(z,\rho)$ \cite{Heun1889,Ronv1995}.

Due to its mathematical complexity there are no analytic solutions in closed forms of Lame equation \cite{Erde1955,Hobs1931,Whit1952}. Because its solution, in the algebraic form or Weierstrass's form, was a form of a power series that is expressed as a 3-term recurrence relation \cite{Hobs1931,Whit1952}. In contrast, most of well-known special functions consist of 2-term recursion relations (Hypergeometric, Bessel, Legendre, Kummer functions, etc).

In the previous paper\cite{Chou2012f}, applying three term recurrence formula.\cite{chou2012b}, we show the power series expansion in closed forms of Lame equation in the algebraic form (infinite series and polynomial) including all higher terms of $A_n$'s by applying three term recurrence formula \cite{chou2012b}.\footnote{`` higher terms of $A_n$'s'' means at least two terms of $A_n$'s.} We obtain representations in form of contour integrals of Lame equation in the algebraic form and its asymptotic behavior of it and the boundary condition for $x$. 

In this paper we show the analytic solution of Lame equation in Weierstrass's form. Its functions in Weierstrass's form appear as we apply the method of separation of variables to Laplace's equation in an ellipsoidal coordinate system (Gabriel Lame 1837\cite{Lame1837}).

The Lame equation in Weierstrass's form is defined by
\begin{equation}
\frac{d^2{y}}{d{z}^2} = \{ \alpha (\alpha +1)\rho^2\;sn^2(z,\rho )-h\} y(z)\label{eq:1}
\end{equation}
where $\rho$ and $\alpha $ are real parameters such that $0<\rho <1$ and $\alpha \geq -\frac{1}{2}$ in general. The Jacobian elliptic function $sn(z,\rho )$ is defined to be the in the inversion of Legendre's elliptic integral of the first kind:
\begin{equation} 
z= \int_{0}^{\rm{am} (z,\rho )} \frac{d\varphi }{\sqrt{1-\rho^2 \sin^2\varphi }}\nonumber
\end{equation}
which gives $sn(z,\rho ) = \sin \left( \rm{am} (z, \rho )\right)$. $\rm{am} (z, \rho )$ is the Jacobi amplitude and $\rho$ is the modulus of the elliptic function $sn(z,\rho )$. 
If we take $sn^2(z,\rho)=\xi $ as an independent variable, Lame equation becomes
\begin{equation}
\frac{d^2{y}}{d{\xi }^2} + \frac{1}{2}\left(\frac{1}{\xi } +\frac{1}{\xi -1} + \frac{1}{\xi -\rho ^{-2}}\right) \frac{d{y}}{d{\xi }} +  \frac{-\alpha (\alpha +1) \xi +h\rho ^{-2}}{4 \xi (\xi -1)(\xi -\rho ^{-2})} y(\xi ) = 0\label{eq:2}
\end{equation}
This is an equation of Fuchsian type with the four regular singularities: $\xi=0, 1, \rho ^{-2}, \infty $. The first three, namely $0, 1, \rho ^{-2}$, have the property that the corresponding exponents are $0, \frac{1}{2}$ which is the same as the case of Lame equation in the algebraic form.
In Ref.\cite{Chou2012f}, Lame equation of the algebraic form is
\begin{equation}
\frac{d^2{y}}{d{x}^2} + \frac{1}{2}\left(\frac{1}{x-a} +\frac{1}{x-b} + \frac{1}{x-c}\right) \frac{d{y}}{d{x}} +  \frac{-\alpha (\alpha +1) x+q}{4 (x-a)(x-b)(x-c)} y = 0\label{eq:3}
\end{equation}
If we compare (\ref{eq:2}) with (\ref{eq:3}), all coefficients on the above are correspondent to the following way.
\begin{equation}
\begin{split}
& a \longrightarrow   0 \\ & b \longrightarrow  1 \\ & c \longrightarrow  \rho ^{-2} \\
& q \longrightarrow  h \rho ^{-2} \\ & x \longrightarrow \xi = sn^2(z,\rho ) 
\end{split}\label{eq:4}   
\end{equation}
We obtain another expression of Lame function in   Weierstrass's form by using (\ref{eq:4}) in Ref.\cite{Chou2012f}.
\section{Power series}
\subsection{Polynomial in which makes $B_n$ term terminated}
There are three types of polynomials in three-term recurrence relation of a linear ordinary differential equation: (1) polynomial which makes $B_n$ term terminated: $A_n$ term is not terminated, (2) polynomial which makes $A_n$ term terminated: $B_n$ term is not terminated, (3) polynomial which makes $A_n$ and $B_n$ terms terminated at the same time.\footnote{If $A_n$ and $B_n$ terms are not terminated, it turns to be infinite series.} In general Lame polynomial (or Lame spectral polynomial) is defined as type 3 polynomial where $A_n$ and $B_n$ terms terminated. Lame polynomial comes from a Lame equation that has a fixed integer value of $\alpha $, just as it has a fixed value of $h$. In three-term recurrence formula, polynomial of type 3 we categorize as complete polynomial. In future papers we will derive type 3 Lame polynomial. In this paper we construct the power series expansion and integral form for Lame polynomial of type 1:  we treat $h$ as a free variable  and $\alpha $ as a fixed value. In the next papers we will work on the power series expansion and integral form for Lame polynomial of type 2.

The general expression of the power series expansion of Lame equation in algebraic form for the polynomial in which makes $B_n$ term terminated in Ref.\cite{Chou2012f} is given by\footnote{In this paper Pochhammer symbol $(x)_n$ is used to represent the rising factorial: $(x)_n = \frac{\Gamma (x+n)}{\Gamma (x)}$.}
\begin{eqnarray}
 y(z)&=& \sum_{n=0}^{\infty } y_n(z)= y_0(z)+ y_1(z)+ y_2(z)+ y_3(z)+\cdots \nonumber\\
&=& c_0 z^{\lambda } \left\{\sum_{i_0=0}^{\alpha _0} \frac{(-\alpha _0)_{i_0} (\alpha _0+ \frac{1}{4}+\lambda )_{i_0}}{(1+\frac{\lambda }{2})_{i_0}(\frac{3}{4} +\frac{\lambda }{2})_{i_0}} \eta ^{i_0} \right.\nonumber\\
&+& \left\{ \sum_{i_0=0}^{\alpha _0} \frac{ (i_0+\frac{\lambda }{2})^2- \Gamma_0^{(P)}}{(i_0+\frac{1}{2}+\frac{\lambda }{2})(i_0+\frac{1}{4}+\frac{\lambda }{2})}\frac{(-\alpha _0)_{i_0} (\alpha _0+\frac{1}{4}+\lambda )_{i_0}}{(1+\frac{\lambda }{2})_{i_0}(\frac{3}{4}+ \frac{\lambda }{2})_{i_0}} \sum_{i_1=i_0}^{\alpha _1} \frac{(-\alpha _1)_{i_1} (\alpha _1+ \frac{5}{4}+\lambda )_{i_1}(\frac{3}{2}+\frac{\lambda}{2})_{i_0}(\frac{5}{4}+\frac{\lambda}{2})_{i_0}}{(-\alpha _1)_{i_0} (\alpha _1+ \frac{5}{4}+\lambda )_{i_0}(\frac{3}{2}+\frac{\lambda}{2})_{i_1}(\frac{5}{4}+\frac{\lambda}{2})_{i_1}} \eta ^{i_1} \right\}\mu  \nonumber\\ 
&+& \sum_{n=2}^{\infty } \left\{ \sum_{i_0=0}^{\alpha _0} \frac{ (i_0+\frac{\lambda }{2})^2- \Gamma_0^{(P)}}{(i_0+\frac{1}{2}+\frac{\lambda }{2})(i_0+\frac{1}{4}+\frac{\lambda }{2})} \frac{(-\alpha _0)_{i_0} (\alpha _0+\frac{1}{4}+\lambda )_{i_0}}{(1+\frac{\lambda }{2})_{i_0}(\frac{3}{4}+ \frac{\lambda }{2})_{i_0}}\right.\nonumber\\
&\times& \prod _{k=1}^{n-1} \left( \sum_{i_k=i_{k-1}}^{\alpha _k} \frac{ (i_k+\frac{k}{2}+ \frac{\lambda }{2})^2- \Gamma_k^{(P)}}{(i_k+\frac{k}{2}+\frac{1}{2}+\frac{\lambda }{2})(i_k+\frac{k}{2}+\frac{1}{4}+\frac{\lambda }{2})}   \frac{(-\alpha _k)_{i_k} (\alpha _k+ k+\frac{1}{4}+\lambda )_{i_k}(1+\frac{k}{2}+\frac{\lambda}{2})_{i_{k-1}}(\frac{3}{4}+\frac{k}{2}+\frac{\lambda}{2})_{i_{k-1}}}{(-\alpha _k)_{i_{k-1}} (\alpha _k+ k+\frac{1}{4}+\lambda )_{i_{k-1}}(1+\frac{k}{2}+\frac{\lambda}{2})_{i_k}(\frac{3}{4}+\frac{k}{2}+\frac{\lambda}{2})_{i_k}}\right) \nonumber\\
&\times& \left.\left. \sum_{i_n= i_{n-1}}^{\alpha _n}\frac{(-\alpha _n)_{i_n} (\alpha _n+ n+\frac{1}{4}+\lambda )_{i_n}(1+\frac{n}{2}+\frac{\lambda}{2})_{i_{n-1}}(\frac{3}{4}+\frac{n}{2}+\frac{\lambda}{2})_{i_{n-1}}}{(-\alpha _n)_{i_{n-1}} (\alpha _n+n+\frac{1}{4}+\lambda )_{i_{n-1}}(1+\frac{n}{2}+\frac{\lambda}{2})_{i_n}(\frac{3}{4}+\frac{n}{2}+\frac{\lambda}{2})_{i_n}} \eta ^{i_n} \right\} \mu ^n \right\}\label{eq:5}
\end{eqnarray}
where
\begin{equation}
\begin{cases} z= x-a \cr
\eta = \frac{-z^2}{(a-b)(a-c)} \cr
\mu  = \frac{-(2a-b-c)z}{(a-b)(a-c)} \cr
\alpha = 2( 2\alpha _i+ i+\lambda )\;\mbox{or}\; -2 (2\alpha _i+i +\lambda )-1\;\;\mbox{where}\; i,\alpha _i =0,1,2,\cdots \cr
\alpha _i\leq \alpha _j \;\;\mbox{only}\;\mbox{if}\;i\leq j\;\;\mbox{where}\;i,j =0,1,2,\cdots
\end{cases}\label{eq:6}
\end{equation}
and
\begin{equation}
\begin{cases} 
\Gamma_0^{(P)} = \frac{a}{(2a-b-c)}\left( (\alpha _0+\frac{\lambda }{2})(\alpha _0+\frac{1}{4}+\frac{\lambda }{2})- \frac{q}{2^4 a} \right) \cr
\Gamma_k^{(P)} = \frac{a}{(2a-b-c)}\left( (\alpha _k+\frac{k}{2}+\frac{\lambda }{2})(\alpha _k+\frac{k}{2}+\frac{1}{4}+\frac{\lambda }{2})-\frac{q}{2^4 a} \right)  
\end{cases}\label{eq:7}
\end{equation}
Put (\ref{eq:4}) in (\ref{eq:5})-(\ref{eq:7}).\footnote{If we take $\alpha \geq -\frac{1}{2}$, $\alpha =  -2 (2\alpha _i+i +\lambda )-1 $ is not available any more in (\ref{eq:7}). In this paper I consider $\alpha $ as arbitrary.} And take $c_0$= 1 as $\lambda =0$  for the first independent solution of Lame equation and $\lambda =\frac{1}{2}$ for the second one into the new (\ref{eq:5})-(\ref{eq:7}).
\footnote{By definition, `Lame polynomial or Lame spectral polynomial' means polynomial which makes $A_n$ and $B_n$ terms terminated: for any non-negative integer value of $\alpha $ there will be $2\alpha  + 1$ values of $h$ for which the solution $y(\xi )$ reduces to a polynomial. In this paper we construct the power series expansion and integral formalism of Lame polynomial which makes $B_n$ term terminated: we treat the spectral parameter $h$ as a free variable. In the next papers we will work on the power series expansion and integral formalism of Lame polynomial which makes $A_n$ term terminated and Lame spectral polynomial.}
\begin{rmk}
The representation in the form of power series expansion of the first kind of independent solution of Lame equation in Weierstrass's form for the polynomial which makes $B_n$ term terminated about $\xi =0 $ as $\alpha = 2(2\alpha_j +j) $ or $-2(2\alpha_j +j)-1 $ where $j,\alpha _j =0,1,2,\cdots$ is
\begin{eqnarray}
 y(\xi )&=& LF_{\alpha _j}\left( \rho, h, \alpha = 2(2\alpha_j +j)\; \mbox{or} -2(2\alpha_j +j)-1; \xi = sn^2(z,\rho ), \mu =(1+\rho ^2) \xi, \eta = -\rho ^2 \xi^2 \right) \nonumber\\
&=& \sum_{i_0=0}^{\alpha _0} \frac{(-\alpha _0)_{i_0} (\alpha _0+ \frac{1}{4})_{i_0}}{(\frac{3}{4})_{i_0} (1)_{i_0}} \eta ^{i_0} \nonumber\\
&&+\Bigg\{ \sum_{i_0=0}^{\alpha _0}\frac{ i_0^2 -\frac{h}{2^4(1+\rho ^2)}}{(i_0+\frac{1}{2})(i_0+\frac{1}{4})}\frac{(-\alpha _0)_{i_0} (\alpha _0+\frac{1}{4})_{i_0}}{(\frac{3}{4})_{i_0}(1)_{i_0}} \sum_{i_1=i_0}^{\alpha _1} \frac{(-\alpha _1)_{i_1} (\alpha _1+ \frac{5}{4})_{i_1}(\frac{3}{2})_{i_0}(\frac{5}{4})_{i_0}}{(-\alpha _1)_{i_0} (\alpha _1+ \frac{5}{4})_{i_0}(\frac{3}{2})_{i_1}(\frac{5}{4})_{i_1}} \eta ^{i_1} \Bigg\} \mu \nonumber\\ 
&&+ \sum_{n=2}^{\infty } \Bigg\{ \sum_{i_0=0}^{\alpha _0} \frac{ i_0^2 -\frac{h}{2^4(1+\rho ^2)}}{(i_0+\frac{1}{2})(i_0+\frac{1}{4})} \frac{(-\alpha _0)_{i_0} (\alpha _0+\frac{1}{4})_{i_0}}{(\frac{3}{4})_{i_0}(1)_{i_0}}  \nonumber\\
&&\times  \prod _{k=1}^{n-1} \Bigg(\sum_{i_k=i_{k-1}}^{\alpha _k} \frac{ (i_k+\frac{k}{2})^2 -\frac{h}{2^4(1+\rho ^2)}}{(i_k+\frac{k}{2}+\frac{1}{2})(i_k+\frac{k}{2}+\frac{1}{4})} \frac{(-\alpha _k)_{i_k} (\alpha _k+ k+\frac{1}{4})_{i_k}(1+\frac{k}{2})_{i_{k-1}}(\frac{3}{4}+\frac{k}{2})_{i_{k-1}}}{(-\alpha _k)_{i_{k-1}} (\alpha _k+ k+\frac{1}{4})_{i_{k-1}}(1+\frac{k}{2})_{i_k}(\frac{3}{4}+\frac{k}{2})_{i_k}}\Bigg) \nonumber\\
&&\times \sum_{i_n= i_{n-1}}^{\alpha _n}\frac{(-\alpha _n)_{i_n} (\alpha _n+ n+\frac{1}{4})_{i_n}(1+\frac{n}{2})_{i_{n-1}}(\frac{3}{4}+\frac{n}{2})_{i_{n-1}}}{(-\alpha _n)_{i_{n-1}} (\alpha _n+n+\frac{1}{4})_{i_{n-1}}(1+\frac{n}{2})_{i_n}(\frac{3}{4}+\frac{n}{2})_{i_n}} \eta ^{i_n} \Bigg\} \mu ^n \nonumber
\end{eqnarray}
\end{rmk}
For the minimum value of Lame equation of the first kind for a polynomial which makes $B_n$ term terminated about $\xi =0 $, put $\alpha _0=\alpha _1=\alpha _2=\cdots=0$ in Remark 1.
\begin{eqnarray}
y(\xi )&=& LF_{0}\left( \rho, h, \alpha = 2j\; \mbox{or} -2j-1; \xi = sn^2(z,\rho ), \mu = (1+\rho ^2) \xi, \eta = -\rho ^2 \xi^2 \right) \nonumber\\
&=& \; _2F_1\left( \sqrt{\frac{h}{4(1+\rho ^2)}}, -\sqrt{\frac{h}{4(1+\rho ^2)}},\frac{1}{2},(1+\rho ^2)\xi \right) \hspace{1cm}\mbox{where}\;\;|(1+\rho ^2)\xi| < 1 \label{aaa:1}
\end{eqnarray} 
For the special case, if $\xi =\frac{1}{1+\rho ^2}$ in (\ref{aaa:1}),
\begin{eqnarray}
y(\xi )&=& LF_{0}\left( \rho, h, \alpha = 2j\; \mbox{or} -2j-1; \xi = sn^2(z,\rho )=\frac{1}{1+\rho ^2}, \mu = 1, \eta = -\frac{\rho ^2}{(1+\rho ^2)^2} \right) \nonumber\\
&=&  \cos \left( \frac{\pi }{2}\sqrt{\frac{h}{ 1+\rho ^2 }} \right) \nonumber
\end{eqnarray}
\begin{rmk} 
The representation in the form of power series expansion of the second kind of independent solution of Lame equation in Weierstrass's form for the polynomial which makes $B_n$ term terminated about $\xi =0 $ as $\alpha = 2(2\alpha_j +j)+1$ or $-2(2\alpha_j +j+1)$  where $j,\alpha _j =0,1,2,\cdots$  is
\begin{eqnarray}
y(\xi)&=& LS_{\alpha _j}\Bigg( \rho, h, \alpha =  2(2\alpha_j +j)+1\; \mbox{or} -2(2\alpha_j +j+1); \xi = sn^2(z,\rho ), \mu =(1+\rho ^2) \xi, \eta = -\rho ^2 \xi^2 \Bigg)\nonumber\\
&=& \xi ^{\frac{1}{2}}  \Bigg\{\sum_{i_0=0}^{\alpha _0} \frac{(-\alpha _0)_{i_0} (\alpha _0+ \frac{3}{4})_{i_0}}{(\frac{5}{4})_{i_0}(1)_{i_0}} \eta ^{i_0} \nonumber\\
&&+ \Bigg\{ \sum_{i_0=0}^{\alpha _0} \frac{ (i_0+\frac{1}{4})^2 -\frac{h}{2^4(1+\rho ^2)}}{(i_0+\frac{3}{4})(i_0+\frac{1}{2})}\frac{(-\alpha _0)_{i_0} (\alpha _0+\frac{3}{4})_{i_0}}{(\frac{5}{4})_{i_0}(1)_{i_0}} \sum_{i_1=i_0}^{\alpha _1} \frac{(-\alpha _1)_{i_1} (\alpha _1+ \frac{7}{4})_{i_1}(\frac{7}{4})_{i_0}(\frac{3}{2})_{i_0}}{(-\alpha _1)_{i_0} (\alpha _1+ \frac{7}{4})_{i_0}(\frac{7}{4})_{i_1}(\frac{3}{2})_{i_1}} \eta ^{i_1} \Bigg\} \mu \nonumber\\ 
&&+ \sum_{n=2}^{\infty } \Bigg\{ \sum_{i_0=0}^{\alpha _0} \frac{ (i_0+\frac{1}{4})^2-\frac{h}{2^4(1+\rho ^2)}}{(i_0+\frac{3}{4})(i_0+\frac{1}{2})} \frac{(-\alpha _0)_{i_0} (\alpha _0+\frac{3}{4})_{i_0}}{(\frac{5}{4})_{i_0}(1)_{i_0}}\nonumber\\
&&\times \prod _{k=1}^{n-1} \Bigg( \sum_{i_k=i_{k-1}}^{\alpha _k} \frac{ (i_k+\frac{k}{2}+ \frac{1}{4})^2-\frac{h}{2^4(1+\rho ^2)}}{(i_k+\frac{k}{2}+\frac{3}{4})(i_k+\frac{k}{2}+\frac{1}{2})} \frac{(-\alpha _k)_{i_k} (\alpha _k+ k+\frac{3}{4})_{i_k}(\frac{5}{4}+\frac{k}{2})_{i_{k-1}}(1+\frac{k}{2})_{i_{k-1}}}{(-\alpha _k)_{i_{k-1}} (\alpha _k+ k+\frac{3}{4})_{i_{k-1}}(\frac{5}{4}+\frac{k}{2})_{i_k}(1+\frac{k}{2})_{i_k}}\Bigg) \nonumber\\
&&\times \sum_{i_n= i_{n-1}}^{\alpha _n}\frac{(-\alpha _n)_{i_n} (\alpha _n+ n+\frac{3}{4})_{i_n}(\frac{5}{4}+\frac{n}{2})_{i_{n-1}}(1+\frac{n}{2})_{i_{n-1}}}{(-\alpha _n)_{i_{n-1}} (\alpha _n+n+\frac{3}{4})_{i_{n-1}}(\frac{5}{4}+\frac{n}{2})_{i_n}(1+\frac{n}{2})_{i_n}} \eta ^{i_n} \Bigg\} \mu ^n \Bigg\}\nonumber
\end{eqnarray}
\end{rmk}
For the minimum value of Lame equation of the second kind for a polynomial which makes $B_n$ term terminated about $\xi =0 $, put $\alpha _0=\alpha _1=\alpha _2=\cdots=0$ in Remark 2.
\begin{eqnarray}
y(\xi)&=& LS_{0}\Bigg( \rho, h, \alpha =  2j +1\; \mbox{or} -2(j+1); \xi = sn^2(z,\rho ), \mu = (1+\rho ^2)\xi, \eta = -\rho ^2 \xi^2 \Bigg)\nonumber\\
&=& \xi ^{\frac{1}{2}} \; _2F_1\left( \sqrt{\frac{h}{4(1+\rho ^2)}}+\frac{1}{2}, -\sqrt{\frac{h}{4(1+\rho ^2)}}+\frac{1}{2},\frac{3}{2},(1+\rho ^2)\xi \right) \hspace{1cm}\mbox{where}\;\;|(1+\rho ^2)\xi| < 1 \hspace{1cm}\label{aaa:2}
\end{eqnarray} 
For the special case, if $\xi =\frac{1}{1+\rho ^2}$ in (\ref{aaa:2}),
\begin{eqnarray}
y(\xi)&=& LS_{0}\Bigg( \rho, h, \alpha =  2j +1\; \mbox{or} -2(j+1); \xi = sn^2(z,\rho )=\frac{1}{1+\rho ^2}, \mu =1, \eta = -\frac{\rho ^2}{(1+\rho ^2)^2} \Bigg)\nonumber\\ 
&=&  \frac{1}{\sqrt{h}} \sin \left( \frac{\pi }{2}\sqrt{\frac{h}{ 1+\rho ^2 }} \right) \nonumber
\end{eqnarray}
(\ref{aaa:1}) and (\ref{aaa:2}) tell us that Lame polynomials in which makes $B_n$ term terminated, for fixed values of $\alpha $, require $|(1+\rho ^2)\xi| < 1$ for the convergence of the radius.
\subsection{Infinite series}
The general expression of the power series expansion of Lame equation in algebraic form for an infinite series in Ref.\cite{Chou2012f} is given by
\begin{eqnarray}
 y(z)&=& \sum_{n=0}^{\infty } y_n(z)= y_0(z)+ y_1(z)+ y_2(z)+ y_3(z)+\cdots\nonumber\\
&=& c_0 z^{\lambda } \left\{\sum_{i_0=0}^{\infty } \frac{(-\frac{\alpha }{4}+\frac{\lambda }{2})_{i_0} (\frac{\alpha }{4}+\frac{1}{4}+\frac{\lambda }{2})_{i_0}}{(1+\frac{\lambda }{2})_{i_0}(\frac{3}{4}+ \frac{\lambda }{2})_{i_0}} \eta^{i_0} \right.\nonumber\\
&+& \left\{\sum_{i_0=0}^{\infty } \frac{ (i_0 +\frac{\lambda }{2})^2 -\Gamma ^{(I)}}{(i_0+ \frac{1}{2}+ \frac{\lambda }{2})(i_0 + \frac{1}{4}+ \frac{\lambda }{2})}  \frac{(-\frac{\alpha }{4}+\frac{\lambda }{2})_{i_0} (\frac{\alpha }{4}+\frac{1}{4}+\frac{\lambda }{2})_{i_0}}{(1+\frac{\lambda }{2})_{i_0}(\frac{3}{4} +\frac{\lambda }{2})_{i_0}} \sum_{i_1=i_0}^{\infty } \frac{(-\frac{\alpha }{4} + \frac{1}{2} + \frac{\lambda }{2})_{i_1}(\frac{\alpha }{4}+\frac{3}{4}+ \frac{\lambda }{2})_{i_1}(\frac{3}{2}+\frac{\lambda }{2})_{i_0}(\frac{5}{4}+ \frac{\lambda }{2})_{i_0}}{(-\frac{\alpha }{4} + \frac{1}{2} + \frac{\lambda }{2})_{i_0}(\frac{\alpha }{4}+\frac{3}{4}+ \frac{\lambda }{2})_{i_0}(\frac{3}{2}+\frac{\lambda }{2})_{i_1}(\frac{5}{4}+ \frac{\lambda }{2})_{i_1}} \eta ^{i_1} \right\}\mu \nonumber\\
&+& \sum_{n=2}^{\infty } \left\{ \sum_{i_0=0}^{\infty } \frac{ (i_0+\frac{\lambda }{2})^2-\Gamma ^{(I)}}{(i_0+\frac{1}{2}+\frac{\lambda }{2})(i_0+\frac{1}{4}+\frac{\lambda }{2})} \frac{(-\frac{\alpha }{4}+\frac{\lambda }{2})_{i_0} (\frac{\alpha }{4}+\frac{1}{4}+\frac{\lambda }{2})_{i_0}}{(1+\frac{\lambda }{2})_{i_0}(\frac{3}{4} +\frac{\lambda }{2})_{i_0}} \right.\nonumber\\
&\times& \prod _{k=1}^{n-1} \left( \sum_{i_k=i_{k-1}}^{\infty } \frac{ (i_k+\frac{k}{2}+ \frac{\lambda }{2})^2- \Gamma ^{(I)}}{(i_k+\frac{k}{2}+\frac{1}{2}+\frac{\lambda }{2})(i_k+\frac{k}{2}+\frac{1}{4}+\frac{\lambda }{2})}    \frac{(-\frac{\alpha }{4}+\frac{k}{2}+\frac{\lambda }{2})_{i_k} (\frac{\alpha }{4}+\frac{k}{2}+\frac{1}{4}+\frac{\lambda }{2})_{i_k}(1+\frac{k}{2}+\frac{\lambda}{2})_{i_{k-1}}(\frac{k}{2}+\frac{3}{4}+\frac{\lambda}{2})_{i_{k-1}}}{(-\frac{\alpha }{4}+\frac{k}{2}+\frac{\lambda }{2})_{i_{k-1}} (\frac{\alpha }{4}+\frac{k}{2}+\frac{1}{4}+\frac{\lambda }{2})_{i_{k-1}}(1+\frac{k}{2}+\frac{\lambda}{2})_{i_k}(\frac{k}{2}+\frac{3}{4}+\frac{\lambda}{2})_{i_k}}\right) \nonumber\\
&\times& \left.\left. \sum_{i_n= i_{n-1}}^{\infty }\frac{(-\frac{\alpha }{4}+\frac{n}{2}+\frac{\lambda }{2})_{i_n} (\frac{\alpha }{4}+\frac{n}{2}+\frac{1}{4}+\frac{\lambda }{2})_{i_n}(1+\frac{n}{2}+\frac{\lambda}{2})_{i_{n-1}}(\frac{n}{2}+\frac{3}{4}+\frac{\lambda}{2})_{i_{n-1}}}{(-\frac{\alpha }{4}+\frac{n}{2}+\frac{\lambda }{2})_{i_{n-1}} (\frac{\alpha }{4}+\frac{n}{2}+\frac{1}{4}+\frac{\lambda }{2})_{i_{n-1}}(1+\frac{n}{2}+\frac{\lambda}{2})_{i_n}(\frac{n}{2}+\frac{3}{4}+\frac{\lambda}{2})_{i_n}} \eta ^{i_n} \right\} \mu ^n \right\}\label{eq:10}
\end{eqnarray}
where
\begin{equation}
\begin{cases} z= x-a \cr
\eta = \frac{-z^2}{(a-b)(a-c)} \cr
\mu  = \frac{-(2a-b-c)z}{(a-b)(a-c)} \cr
\Gamma ^{(I)} = \frac{a}{2^4(2a-b-c)}\left( \alpha (\alpha +1) -\frac{q}{a}\right) 
\end{cases}\label{eq:10a}
\end{equation}
Put (\ref{eq:4}) in (\ref{eq:10}) and (\ref{eq:10a}). And take $c_0$= 1 as $\lambda =0$  for the first independent solution of Lame equation and $\lambda =\frac{1}{2}$ for the second one into the new (\ref{eq:10}).
\begin{rmk}
The representation in the form of power series expansion of the first kind of independent solution of Lame equation in Weierstrass's form for the infinite series about $\xi=0 $ is
\begin{eqnarray}
 y(\xi )&=& LF \left( \rho, h, \alpha; \xi = sn^2(z,\rho ), \mu = (1+\rho ^2) \xi, \eta = -\rho ^2 \xi^2 \right) \nonumber\\
&=& \sum_{i_0=0}^{\infty } \frac{(-\frac{\alpha }{4})_{i_0} (\frac{\alpha }{4}+\frac{1}{4})_{i_0}}{(\frac{3}{4})_{i_0}(1)_{i_0}} \eta^{i_0}\nonumber\\
&&+ \Bigg\{ \sum_{i_0=0}^{\infty } \frac{ i_0^2 -\frac{h}{2^4(1+\rho ^2)}}{(i_0+ \frac{1}{2})(i_0 + \frac{1}{4})} \frac{(-\frac{\alpha }{4})_{i_0} (\frac{\alpha }{4}+\frac{1}{4})_{i_0}}{(\frac{3}{4})_{i_0}(1)_{i_0}} \sum_{i_1=i_0}^{\infty } \frac{(-\frac{\alpha }{4} + \frac{1}{2})_{i_1}(\frac{\alpha }{4}+\frac{3}{4})_{i_1}(\frac{3}{2})_{i_0}(\frac{5}{4})_{i_0}}{(-\frac{\alpha }{4} + \frac{1}{2})_{i_0}(\frac{\alpha }{4}+\frac{3}{4})_{i_0}(\frac{3}{2})_{i_1}(\frac{5}{4})_{i_1}} \eta ^{i_1} \Bigg\} \mu \nonumber\\
&&+ \sum_{n=2}^{\infty } \Bigg\{ \sum_{i_0=0}^{\infty } \frac{ i_0^2 -\frac{h}{2^4(1+\rho ^2)}}{(i_0+\frac{1}{2})(i_0+\frac{1}{4})} \frac{(-\frac{\alpha }{4})_{i_0} (\frac{\alpha }{4}+\frac{1}{4})_{i_0}}{(\frac{3}{4})_{i_0}(1)_{i_0}} \nonumber\\
&&\times \prod _{k=1}^{n-1} \Bigg( \sum_{i_k=i_{k-1}}^{\infty } \frac{ (i_k+\frac{k}{2})^2-\frac{h}{2^4(1+\rho ^2)}}{(i_k+\frac{k}{2}+\frac{1}{2})(i_k+\frac{k}{2}+\frac{1}{4})}  \frac{(-\frac{\alpha }{4}+\frac{k}{2})_{i_k} (\frac{\alpha }{4}+\frac{1}{4}+\frac{k}{2})_{i_k}(1+\frac{k}{2})_{i_{k-1}}(\frac{3}{4}+\frac{k}{2})_{i_{k-1}}}{(-\frac{\alpha }{4}+\frac{k}{2})_{i_{k-1}} (\frac{\alpha }{4}+\frac{1}{4}+\frac{k}{2})_{i_{k-1}}(1+\frac{k}{2})_{i_k}(\frac{3}{4}+\frac{k}{2})_{i_k}}\Bigg) \nonumber\\
&&\times \sum_{i_n= i_{n-1}}^{\infty }\frac{(-\frac{\alpha }{4}+\frac{n}{2})_{i_n} (\frac{\alpha }{4}+\frac{1}{4}+\frac{n}{2})_{i_n}(1+\frac{n}{2})_{i_{n-1}}(\frac{3}{4}+\frac{n}{2})_{i_{n-1}}}{(-\frac{\alpha }{4}+\frac{n}{2})_{i_{n-1}} (\frac{\alpha }{4}+\frac{1}{4}+\frac{n}{2})_{i_{n-1}}(1+\frac{n}{2})_{i_n}(\frac{3}{4}+\frac{n}{2})_{i_n}} \eta ^{i_n} \Bigg\} \mu ^n \nonumber
\end{eqnarray}
\end{rmk}
\begin{rmk}
The representation in the form of power series expansion of the second kind of independent solution of Lame equation in Weierstrass's form for the infinite series about $\xi=0 $ is 
\begin{eqnarray}
y(\xi )&=&  LS \left( \rho, h, \alpha; \xi = sn^2(z,\rho ), \mu = (1+\rho ^2)\xi, \eta = -\rho ^2 \xi^2 \right) \nonumber\\
&=& \xi^{\frac{1}{2}} \Bigg\{\sum_{i_0=0}^{\infty } \frac{(-\frac{\alpha }{4}+\frac{1}{4})_{i_0} (\frac{\alpha }{4}+\frac{1}{2})_{i_0}}{(\frac{5 }{4})_{i_0}(1)_{i_0}} \eta^{i_0}\nonumber\\
&&+ \Bigg\{ \sum_{i_0=0}^{\infty } \frac{ (i_0 +\frac{1}{4})^2 -\frac{h}{2^4(1+\rho ^2)}}{(i_0+ \frac{3}{4})(i_0 + \frac{1}{2})} \frac{(-\frac{\alpha }{4}+\frac{1}{4})_{i_0} (\frac{\alpha }{4}+\frac{1}{2})_{i_0}}{(\frac{5}{4})_{i_0}(1)_{i_0}} \sum_{i_1=i_0}^{\infty }  \frac{(-\frac{\alpha }{4} + \frac{3}{4})_{i_1}(\frac{\alpha }{4}+1)_{i_1}(\frac{7}{4})_{i_0}(\frac{3}{2})_{i_0}}{(-\frac{\alpha }{4} + \frac{3}{4})_{i_0}(\frac{\alpha }{4}+1)_{i_0}(\frac{7}{4})_{i_1}(\frac{3}{2})_{i_1}} \eta ^{i_1} \Bigg\} \mu \nonumber\\
&&+ \sum_{n=2}^{\infty } \Bigg\{ \sum_{i_0=0}^{\infty } \frac{ (i_0+\frac{1}{4})^2-\frac{h}{2^4(1+\rho ^2)}}{(i_0+\frac{3}{4})(i_0+\frac{1}{2})} \frac{(-\frac{\alpha }{4}+\frac{1}{4})_{i_0} (\frac{\alpha }{4}+\frac{1}{2})_{i_0}}{(\frac{5}{4})_{i_0}(1)_{i_0}} \nonumber\\
&&\times \prod _{k=1}^{n-1} \Bigg( \sum_{i_k=i_{k-1}}^{\infty } \frac{ (i_k+\frac{k}{2}+ \frac{1}{4})^2-\frac{h}{2^4(1+\rho ^2)}}{(i_k+\frac{k}{2}+\frac{3}{4})(i_k+\frac{k}{2}+\frac{1}{2})}   \frac{(-\frac{\alpha }{4}+\frac{k}{2}+\frac{1}{4})_{i_k} (\frac{\alpha }{4}+\frac{k}{2}+\frac{1}{2})_{i_k}(\frac{5}{4}+\frac{k}{2})_{i_{k-1}}(1+\frac{k}{2})_{i_{k-1}}}{(-\frac{\alpha }{4}+\frac{k}{2}+\frac{1}{4})_{i_{k-1}} (\frac{\alpha }{4}+\frac{k}{2}+\frac{1}{2})_{i_{k-1}}(\frac{5}{4}+\frac{k}{2})_{i_k}(1+\frac{k}{2})_{i_k}}\Bigg) \nonumber\\
&&\times \sum_{i_n= i_{n-1}}^{\infty }\frac{(-\frac{\alpha }{4}+\frac{n}{2}+\frac{1}{4})_{i_n} (\frac{\alpha }{4}+\frac{n}{2}+\frac{1}{2})_{i_n}(\frac{5}{4}+\frac{n}{2})_{i_{n-1}}(1+\frac{n}{2})_{i_{n-1}}}{(-\frac{\alpha }{4}+\frac{n}{2}+\frac{1}{4})_{i_{n-1}} (\frac{\alpha }{4}+\frac{n}{2}+\frac{1}{2})_{i_{n-1}}(\frac{5}{4}+\frac{n}{2})_{i_n}(1+\frac{n}{2})_{i_n}} \eta ^{i_n} \Bigg\} \mu ^n \Bigg\}\nonumber
\end{eqnarray}
\end{rmk}
\section{Integral Formalism}
\subsection{Polynomial in which makes $B_n$ term terminated}
The general expression of the representation in the form of integral of Lame equation in algebraic form for the polynomial in which makes $B_n$ term terminated in Ref.\cite{Chou2012f} is given by
\begin{eqnarray}
 y(z)&=& \sum_{n=0}^{\infty } y_{n}(z) = y_0(z)+ y_1(z)+ y_2(z)+y_3(z)+\cdots\nonumber\\
&=& c_0 z^{\lambda } \left\{ \sum_{i_0=0}^{\alpha _0}\frac{(-\alpha _0)_{i_0}(\alpha _0+\frac{1}{4}+\lambda )_{i_0}}{(1+\frac{\lambda }{2})_{i_0}(\frac{3}{4}+ \frac{\lambda }{2})_{i_0}}  \eta ^{i_0} \right.\nonumber\\
&&+ \sum_{n=1}^{\infty } \left\{\prod _{k=0}^{n-1} \left\{ \int_{0}^{1} dt_{n-k}\;t_{n-k}^{\frac{1}{2}(n-k-\frac{5}{2}+\lambda )} \int_{0}^{1} du_{n-k}\;u_{n-k}^{\frac{1}{2}(n-k-2+\lambda )} \right.\right. \nonumber\\
&&\times  \frac{1}{2\pi i}  \oint dv_{n-k} \frac{1}{v_{n-k}} \left( 1-\overleftrightarrow {w}_{n-k+1,n} v_{n-k}(1-t_{n-k})(1-u_{n-k})\right)^{-(n-k+\frac{1}{4}+\lambda )}\nonumber\\
&&\times \left(\frac{(v_{n-k}-1)}{v_{n-k}} \frac{1}{1-\overleftrightarrow {w}_{n-k+1,n}v_{n-k}(1-t_{n-k})(1-u_{n-k})}\right)^{\alpha _{n-k}}\nonumber\\
&&\times \left. \left( \overleftrightarrow {w}_{n-k,n}^{-\frac{1}{2}(n-k-1+\lambda )}\left(  \overleftrightarrow {w}_{n-k,n} \partial _{ \overleftrightarrow {w}_{n-k,n}}\right)^2 \overleftrightarrow {w}_{n-k,n}^{\frac{1}{2}(n-k-1+\lambda )} -\Omega _{n-k-1}^{(P)}\right) \right\}\nonumber\\
&&\times \left.\left. \sum_{i_0=0}^{\alpha _0}\frac{(-\alpha _0)_{i_0}(\alpha _0+\frac{1}{4}+\lambda )_{i_0}}{(1+\frac{\lambda }{2})_{i_0}(\frac{3}{4}+ \frac{\lambda }{2})_{i_0}}  \overleftrightarrow {w}_{1,n}^{i_0}\right\} \mu ^n \right\} \label{eq:31}
\end{eqnarray}
where
\begin{equation}\overleftrightarrow {w}_{i,j}=
\begin{cases} \displaystyle {\frac{1}{(v_i-1)}\; \frac{\overleftrightarrow w_{i+1,j}v_i t_i u_i}{1- \overleftrightarrow w_{i+1,j} v_i (1-t_i)(1-u_i)}}\;\;\mbox{where}\; i\leq j\cr
\eta \;\;\mbox{only}\;\mbox{if}\; i>j
\end{cases}\label{eq:32}
\end{equation}
and
\begin{equation}
\Omega _{n-k-1}^{(P)} =  \frac{a}{(2a-b-c)}\left( \left(\alpha _{n-k-1}+\frac{ n-k-1+\lambda }{2} \right) \left(\alpha _{n-k-1}+\frac{ n-k-\frac{1}{2}+\lambda }{2} \right) -\frac{q}{2^4 a} \right) \label{eq:32a}
\end{equation}
Put (\ref{eq:4}) in (\ref{eq:31})--(\ref{eq:32a}).
\begin{eqnarray}
 y(\xi )&=& \sum_{n=0}^{\infty } y_{n}(\xi ) = y_0(\xi )+ y_1(\xi )+ y_2(\xi )+ y_3(\xi )+\cdots \nonumber\\
&=& c_0 \xi ^{\lambda } \Bigg\{ \sum_{i_0=0}^{\alpha _0}\frac{(-\alpha _0)_{i_0}(\alpha _0+\frac{1}{4}+\lambda )_{i_0}}{(1+\frac{\lambda }{2})_{i_0}(\frac{3}{4}+ \frac{\lambda }{2})_{i_0}}  \eta ^{i_0}\nonumber\\
&&+ \sum_{n=1}^{\infty } \Bigg\{\prod _{k=0}^{n-1} \Bigg\{ \int_{0}^{1} dt_{n-k}\;t_{n-k}^{\frac{1}{2}(n-k-\frac{5}{2}+\lambda )} \int_{0}^{1} du_{n-k}\;u_{n-k}^{\frac{1}{2}(n-k-2+\lambda )} \nonumber\\
&&\times  \frac{1}{2\pi i}  \oint dv_{n-k} \frac{1}{v_{n-k}} \left( 1-\overleftrightarrow {w}_{n-k+1,n} v_{n-k}(1-t_{n-k})(1-u_{n-k})\right)^{-(n-k+\frac{1}{4}+\lambda )}\nonumber\\
&&\times \left(\frac{(v_{n-k}-1)}{v_{n-k}} \frac{1}{1-\overleftrightarrow {w}_{n-k+1,n}v_{n-k}(1-t_{n-k})(1-u_{n-k})}\right)^{\alpha _{n-k}}\nonumber\\
&&\times \Bigg(  \overleftrightarrow {w}_{n-k,n}^{-\frac{1}{2}(n-k-1+\lambda )}\left(  \overleftrightarrow {w}_{n-k,n} \partial _{ \overleftrightarrow {w}_{n-k,n}}\right)^2 \overleftrightarrow {w}_{n-k,n}^{\frac{1}{2}(n-k-1+\lambda )}-\frac{h}{2^4(1+\rho ^2)}\Bigg) \Bigg\}\nonumber\\
&&\times \sum_{i_0=0}^{\alpha _0}\frac{(-\alpha _0)_{i_0}(\alpha _0+\frac{1}{4}+\lambda )_{i_0}}{(1+\frac{\lambda }{2})_{i_0}(\frac{3}{4}+ \frac{\lambda }{2})_{i_0}}  \overleftrightarrow {w}_{1,n}^{i_0}\Bigg\} \mu ^n \Bigg\} \label{eq:33}
\end{eqnarray}
Put $c_0$= 1 as $\lambda =0$  for the first independent solution of Lame equation and $\lambda =\frac{1}{2}$ for the second one into (\ref{eq:33}).
\begin{rmk}
The representation in the form of integral of the first kind of independent solution of Lame equation in Weierstrass's form for the polynomial which makes $B_n$ term terminated about $\xi=0 $ as $\alpha = 2(2\alpha_j +j) $ or $ -2(2\alpha_j +j)-1$ where $j, \alpha _j =0,1,2,\cdots$ is
\begin{eqnarray}
 y(\xi )&=& LF_{\alpha _j}\left( \rho, h, \alpha = 2(2\alpha_j +j)\; \mbox{or} -2(2\alpha_j +j)-1; \xi = sn^2(z,\rho ), \mu = (1+\rho ^2) \xi, \eta = -\rho ^2 \xi^2 \right) \nonumber\\
&=& _2F_1 \left(-\alpha _0, \alpha _0+\frac{1}{4};\frac{3}{4}; \eta \right) + \sum_{n=1}^{\infty } \Bigg\{\prod _{k=0}^{n-1} \Bigg\{ \int_{0}^{1} dt_{n-k}\;t_{n-k}^{\frac{1}{2}(n-k-\frac{5}{2})} \int_{0}^{1} du_{n-k}\;u_{n-k}^{\frac{1}{2}(n-k-2)} \nonumber\\
&&\times  \frac{1}{2\pi i}  \oint dv_{n-k} \frac{1}{v_{n-k}} \left( 1-\overleftrightarrow {w}_{n-k+1,n} v_{n-k}(1-t_{n-k})(1-u_{n-k})\right)^{-(n-k+\frac{1}{4})}\nonumber\\
&&\times \left(\frac{(v_{n-k}-1)}{v_{n-k}} \frac{1}{1-\overleftrightarrow {w}_{n-k+1,n}v_{n-k}(1-t_{n-k})(1-u_{n-k})}\right)^{\alpha _{n-k}}\nonumber\\
&&\times \Bigg( \overleftrightarrow {w}_{n-k,n}^{-\frac{1}{2}(n-k-1)}\left(  \overleftrightarrow {w}_{n-k,n} \partial _{ \overleftrightarrow {w}_{n-k,n}}\right)^2 \overleftrightarrow {w}_{n-k,n}^{\frac{1}{2}(n-k-1)} -\frac{h}{2^4(1+\rho ^2)}\Bigg) \Bigg\} \nonumber\\
&&\times _2F_1 \left(-\alpha _0, \alpha _0+\frac{1}{4};\frac{3}{4};  \overleftrightarrow {w}_{1,n}\right) \Bigg\} \mu ^n \nonumber
\end{eqnarray}
\end{rmk}
\begin{rmk}
The representation in the form of integral of the second kind of independent solution of Lame equation in Weierstrass's form for the polynomial  which makes $B_n$ term terminated about $\xi=0 $ as $\alpha = 2(2\alpha_j +j )+1$ or $-2(2\alpha _j+j+1)$ where $j, \alpha _j=0,1,2,\cdots$ is
\begin{eqnarray}
y(\xi )&=& LS_{\alpha _j}\Bigg( \rho, h, \alpha =  2(2\alpha_j +j)+1\; \mbox{or} -2(2\alpha_j +j+1); \xi = sn^2(z,\rho ), \mu =(1+\rho ^2)\xi, \eta = -\rho ^2 \xi^2 \Bigg) \nonumber\\
&=& \xi ^{\frac{1}{2}} \Bigg\{\; _2F_1 \left(-\alpha _0, \alpha _0+\frac{3}{4};\frac{5}{4}; \eta \right)
+  \sum_{n=1}^{\infty } \Bigg\{\prod _{k=0}^{n-1} \Bigg\{ \int_{0}^{1} dt_{n-k}\;t_{n-k}^{\frac{1}{2}(n-k-2)} \int_{0}^{1} du_{n-k}\;u_{n-k}^{\frac{1}{2}(n-k-\frac{3}{2})} \nonumber\\
&&\times  \frac{1}{2\pi i}  \oint dv_{n-k} \frac{1}{v_{n-k}} \left( 1-\overleftrightarrow {w}_{n-k+1,n} v_{n-k}(1-t_{n-k})(1-u_{n-k})\right)^{-(n-k+\frac{3}{4})}\nonumber\\
&&\times \left(\frac{(v_{n-k}-1)}{v_{n-k}} \frac{1}{1-\overleftrightarrow {w}_{n-k+1,n}v_{n-k}(1-t_{n-k})(1-u_{n-k})}\right)^{\alpha _{n-k}}\nonumber\\
&&\times \Bigg( \overleftrightarrow {w}_{n-k,n}^{-\frac{1}{2}(n-k-\frac{1}{2})}\left(  \overleftrightarrow {w}_{n-k,n} \partial _{ \overleftrightarrow {w}_{n-k,n}}\right)^2 \overleftrightarrow {w}_{n-k,n}^{\frac{1}{2}(n-k-\frac{1}{2})} -\frac{h}{2^4(1+\rho ^2)}\Bigg) \Bigg\}\nonumber\\
&&\times _2F_1 \left(-\alpha _0,  \alpha _0+\frac{3}{4};\frac{5}{4}; \overleftrightarrow {w}_{1,n}\right)  \Bigg\} \mu ^n \Bigg\}\nonumber
\end{eqnarray}
\end{rmk}
\subsection{Infinite series}
The general expression of the representation in the form of integral of Lame equation in algebraic form for the infinite series in Ref.\cite{Chou2012f} is given by
\begin{eqnarray}
 y(z)&=& \sum_{n=0}^{\infty } y_{n}(z) = y_0(z)+ y_1(z)+ y_2(z)+y_3(z)+\cdots \nonumber\\
&=& c_0 z^{\lambda } \left\{ \sum_{i_0=0}^{\infty }\frac{(-\frac{\alpha }{4}+\frac{\lambda }{2})_{i_0}(\frac{\alpha }{4}+\frac{1}{4}+\frac{\lambda }{2})_{i_0}}{(1+\frac{\lambda }{2})_{i_0}(\frac{3}{4}+ \frac{\lambda }{2})_{i_0}}  \eta ^{i_0}\right. \nonumber\\
&&+ \sum_{n=1}^{\infty } \left\{\prod _{k=0}^{n-1} \Bigg\{ \int_{0}^{1} dt_{n-k}\;t_{n-k}^{\frac{1}{2}(n-k-\frac{5}{2}+\lambda )} \int_{0}^{1} du_{n-k}\;u_{n-k}^{\frac{1}{2}(n-k-2+\lambda )} \right.\nonumber\\
&&\times  \frac{1}{2\pi i}  \oint dv_{n-k} \frac{1}{v_{n-k}} \left(\frac{ v_{n-k}-1 }{v_{n-k}} \right)^{\frac{1}{2}(\frac{\alpha }{2}-n+k-\lambda )}   \left( 1-\overleftrightarrow {w}_{n-k+1,n} v_{n-k}(1-t_{n-k})(1-u_{n-k})\right)^{-\frac{1}{2}(\frac{\alpha }{2}+\frac{1}{2}+n-k+\lambda )} \nonumber\\
&&\times \Bigg( \overleftrightarrow {w}_{n-k,n}^{-\frac{1}{2}(n-k-1+\lambda )}\left(  \overleftrightarrow {w}_{n-k,n} \partial _{ \overleftrightarrow {w}_{n-k,n}}\right)^2 \overleftrightarrow {w}_{n-k,n}^{\frac{1}{2}(n-k-1+\lambda )}-\Gamma ^{(I)}\Bigg) \Bigg\}\nonumber\\
&&\times \left.\left. \sum_{i_0=0}^{\infty }\frac{(-\frac{\alpha }{4}+\frac{\lambda }{2})_{i_0}(\frac{\alpha }{4}+\frac{1}{4}+\frac{\lambda }{2})_{i_0}}{(1+\frac{\lambda }{2})_{i_0}(\frac{3}{4}+ \frac{\lambda }{2})_{i_0}} \overleftrightarrow {w}_{1,n}^{i_0}\right\} \mu ^n \right\} \label{eq:36}
\end{eqnarray}
where
\begin{equation}
\Gamma ^{(I)} = \frac{a}{2^4(2a-b-c)}\left( \alpha (\alpha +1) -\frac{q}{a}\right) \label{eq:36a}
\end{equation}
Put (\ref{eq:4}) in (\ref{eq:36}) and (\ref{eq:36a}). And put $c_0$= 1 as $\lambda =0$  for the first independent solution of Lame equation and $\lambda =\frac{1}{2}$ for the second one into the new (\ref{eq:36}).
\begin{rmk}
The representation in the form of integral of the first kind of independent solution of Lame equation in Weierstrass's form for the infinite series about $\xi=0 $ is
\begin{eqnarray}
 y(\xi )&=& LF \left( \rho, h, \alpha; \xi = sn^2(z,\rho ), \mu = (1+\rho ^2) \xi, \eta = -\rho ^2 \xi^2 \right) \nonumber\\
&=& _2F_1 \left( -\frac{\alpha }{4}, \frac{\alpha }{4}+\frac{1}{4};\frac{3}{4}; \eta \right) + \sum_{n=1}^{\infty } \Bigg\{\prod _{k=0}^{n-1} \Bigg\{ \int_{0}^{1} dt_{n-k}\;t_{n-k}^{\frac{1}{2}(n-k-\frac{5}{2})} \int_{0}^{1} du_{n-k}\;u_{n-k}^{\frac{1}{2}(n-k-2)} \nonumber\\
&&\times  \frac{1}{2\pi i}  \oint dv_{n-k} \frac{1}{v_{n-k}} \left(\frac{ v_{n-k}-1 }{v_{n-k}} \right)^{\frac{1}{2}(\frac{\alpha }{2}-n+k )}   \left( 1-\overleftrightarrow {w}_{n-k+1,n} v_{n-k}(1-t_{n-k})(1-u_{n-k})\right)^{-\frac{1}{2}(\frac{\alpha }{2}+\frac{1}{2}+n-k )} \nonumber\\
&&\times \Bigg(  \overleftrightarrow {w}_{n-k,n}^{-\frac{1}{2}(n-k-1)}\left(  \overleftrightarrow {w}_{n-k,n} \partial _{ \overleftrightarrow {w}_{n-k,n}}\right)^2 \overleftrightarrow {w}_{n-k,n}^{\frac{1}{2}(n-k-1)}-\frac{h}{2^4(1+\rho ^2)}\Bigg) \Bigg\} \nonumber\\
&&\times _2F_1 \left( -\frac{\alpha }{4}, \frac{\alpha }{4}+\frac{1}{4};\frac{3}{4}; \overleftrightarrow {w}_{1,n}\right) \Bigg\} \mu ^n \nonumber
\end{eqnarray}
\end{rmk}
\begin{rmk}
The representation in the form of integral of the second kind of independent solution of Lame equation in Weierstrass's form for the infinite series about $\xi=0 $ is
\begin{eqnarray}
y(\xi )&=& LS \left( \rho, h, \alpha; \xi = sn^2(z,\rho ), \mu = (1+\rho ^2) \xi, \eta = -\rho ^2 \xi^2 \right) \nonumber\\
&=& \xi ^{\frac{1}{2}} \Bigg\{\; _2F_1 \left(-\frac{\alpha }{4}+\frac{1}{4}, \frac{\alpha }{4}+\frac{1}{2};\frac{5}{4}; \eta \right) \nonumber\\
&&+  \sum_{n=1}^{\infty } \Bigg\{\prod _{k=0}^{n-1} \Bigg\{ \int_{0}^{1} dt_{n-k}\;t_{n-k}^{\frac{1}{2}(n-k-2)} \int_{0}^{1} du_{n-k}\;u_{n-k}^{\frac{1}{2}(n-k-\frac{3}{2} )} \nonumber\\
&&\times  \frac{1}{2\pi i}  \oint dv_{n-k} \frac{1}{v_{n-k}} \left(\frac{ v_{n-k}-1 }{v_{n-k}} \right)^{\frac{1}{2}(\frac{\alpha }{2}-\frac{1}{2}-n+k )}   \left( 1-\overleftrightarrow {w}_{n-k+1,n} v_{n-k}(1-t_{n-k})(1-u_{n-k})\right)^{-\frac{1}{2}(\frac{\alpha }{2}+1+n-k )} \nonumber\\
&&\times \Bigg( \overleftrightarrow {w}_{n-k,n}^{-\frac{1}{2}(n-k-\frac{1}{2})}\left(  \overleftrightarrow {w}_{n-k,n} \partial _{ \overleftrightarrow {w}_{n-k,n}}\right)^2 \overleftrightarrow {w}_{n-k,n}^{\frac{1}{2}(n-k-\frac{1}{2})} -\frac{h}{2^4(1+\rho ^2)}\Bigg) \Bigg\} \nonumber\\
&&\times \; _2F_1 \left(-\frac{\alpha }{4}+\frac{1}{4}, \frac{\alpha }{4}+\frac{1}{2};\frac{5}{4}; \overleftrightarrow {w}_{1,n}\right) \Bigg\} \mu ^n \Bigg\}\nonumber
\end{eqnarray}
\end{rmk}
\section{Asymptotic behavior of the function $y(\xi )$ and the boundary condition for $\xi =sn^2(z,\rho )$}
The recurrence system of a Lame equation in the algebraic form for the infinite series in Ref.\cite{Chou2012f} is given by
\begin{subequations}
\begin{equation}
c_{n+1}=A\;c_n +B\;c_{n-1} \hspace{1cm};n\geq 1 \label{cc:1}
\end{equation}
where    
\begin{equation}
\lim_{n\gg 1} A_n = A= \frac{-(2a-b-c)}{(a-b)(a-c)} \hspace{2cm} \lim_{n\gg 1} B_n = B= \frac{-1}{(a-b)(a-c)}\label{cc:2}
\end{equation}
with $c_1= A $ and $c_0 =1$ for simplicity.
\end{subequations}
Its condition of convergence and the asymptotic function in Ref.\cite{Chou2012f} are\footnote{for sufficiently large, like an index $n$ is close to infinity, or treat as $n\rightarrow \infty $}
\begin{eqnarray}
\lim_{n\gg 1}y(z) &=& \sum_{n=0}^{\infty }\sum_{m=0}^{\infty } \frac{(n+m)!}{n!\;m!}\left( \frac{(x-a)^2}{(a-b)(a-c)}\right)^n \left( \frac{(2a-b-c)(x-a)}{(a-b)(a-c)}\right)^m \nonumber\\
&=& \frac{1}{1+\left(\frac{(x-a)^2}{(a-b)(a-c)}+ \frac{(2a-b-c)(x-a)}{(a-b)(a-c)}\right)}  \label{eq:53}
\end{eqnarray}
where
\begin{equation}
\left|\frac{(x-a)^2}{(a-b)(a-c)}\right| +\left| \frac{(2a-b-c)(x-a)}{(a-b)(a-c)}\right| <1 \label{eq:54}
\end{equation}
Put (\ref{eq:4}) in (\ref{eq:53}) and (\ref{eq:54}). And its asymptotic function and the boundary condition of $\xi =sn^2(z,\rho ) $ for the infinite series of Lame function is
\begin{eqnarray}
\lim_{n\gg 1}y(\xi )&=& \sum_{n=0}^{\infty }\sum_{m=0}^{\infty } \frac{(n+m)!}{n!\;m!}\left( -\rho ^2 sn^4(z,\rho )\right)^n \left( (1+\rho ^2) sn^2(z,\rho )\right)^m \nonumber\\
&=& \frac{1}{1+\rho ^2 sn^4(z,\rho )-(1+\rho ^2) sn^2(z,\rho )}\label{eq:55}\\
&&\nonumber\\
&&\mbox{where}\; \left| \rho ^2 sn^4(z,\rho )\right| +\left| (1+\rho ^2) sn^2(z,\rho )\right| <1 \label{eq:100}
\end{eqnarray}
For the case of $ z, sn(z,\rho )\in \mathbb{R}$ where $ 0< \rho < 1 $, the boundary condition of $sn^2(z,\rho )$ in (\ref{eq:55}) is given by
\begin{equation}
0 \leq sn^2(z,\rho )< \frac{-(1+\rho ^2)+\sqrt{\rho ^4+6\rho ^2+1}}{2 \rho ^2}\label{eq:56}
\end{equation}
\begin{figure}[htbp]
\centering
\includegraphics[scale=.7]{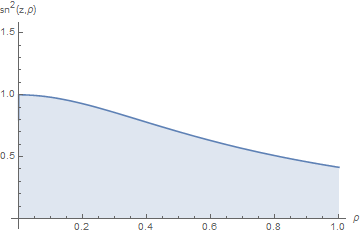}
\caption{Domain of convergence of the series (\ref{eq:55})}
\label{img1}
\end{figure}

Fig.~\ref{img1} represents a graph of (\ref{eq:56}) in the $\rho $-$sn^2(z,\rho )$ plane; the shaded area represents the domain of convergence of the series for a Lame equation around $sn^2(z,\rho )=0$; it does not include solid lines.

In the case of $\rho \approx 0$ assuming $\rho $ is approximately close to $0$, (\ref{eq:55}) turns to be
\begin{eqnarray}
&&\lim_{n\gg 1}y(\xi )\approx  \frac{1}{1 - \sin^2 z}\nonumber
\end{eqnarray}
where  $\left| \sin^2 z \right|<1$. If $ z \in \mathbb{R}$, its radius of convergence is $ 0\leq \sin^2 z <1$.

In this paper we derive asymptotic expansions in closed forms of Lame equation for an infinite series and their radii of convergence are constructed analytically.  As long as the independent variable is existed in the domain of convergence of the series, a solution in series is absolute convergent. From these boundary conditions, we are able to obtain numerical approximations of Lame equation by computer simulations without any serious errors even if we rearrange of its terms for the series solution. 
\section{Poincar\'{e}-Perron theorem and its applications to Frobenius solutions}

\begin{thm}Poincar\'{e}-Perron theorem \cite{Miln1933}: If the coefficients $\alpha _{i,n}$ where $i=1,2,\cdots,k$ of a linear homogeneous difference equation
\begin{equation}
u(n+1)+ \alpha _{1,n} u(n)+ \alpha _{2,n} u(n-1)+ \alpha _{3,n} u(n-2)+\cdots + \alpha _{k,n} u(n-k+1)=0\nonumber
\end{equation}
have limits $\lim_{n\rightarrow \infty }\alpha _{i,n}=\alpha _i$ with $\alpha _{k,n} \ne 0$ and if the roots $\lambda _1, . . . ,\lambda _k $ of the characteristic equation $t^k + \alpha _1 t^{k-1} +  \alpha _2 t^{k-2}+ \cdots + \alpha _k =0$ have distinct absolute values.

H. Poincar\'{e}'s suggested that
\begin{equation}
\lim_{n\rightarrow \infty } \frac{u(n+1)}{u(n)}\nonumber
\end{equation}
is equal to one of the roots of the characteristic equation in 1885 \cite{Poin1885}. And a more general theorem has been extended by O. Perron in 1921 \cite{Perr1921} such that
\begin{equation}
\lim_{n\rightarrow \infty } \frac{u_{i}(n+1)}{u_{i}(n)}= \lambda _i \nonumber
\end{equation}
where $i=1,2,\cdots, k$ and $\lambda _i$ is a root of the characteristic equation, and $n\rightarrow \infty $ by positive integral increments.\label{thm.1}
\end{thm}
The asymptotic recurrence relation for an infinite series of Lame equation in Weierstrass's form is obtained by putting (\ref{eq:4}) in (\ref{cc:1}) and (\ref{cc:2}).
\begin{subequations}
\begin{equation}
c_{n+1}=\alpha _1\;c_n +\alpha _2\;c_{n-1} \hspace{1cm};n\geq 1 \label{gg:1}
\end{equation}
where    
\begin{equation}
\lim_{n\gg 1} \alpha _{1,n} = \alpha _1= 1+\rho ^2 \hspace{2cm} \lim_{n\gg 1} \alpha _{2,n} = \alpha _2=-\rho ^2\label{gg:2}
\end{equation}
\end{subequations}
with seed values $c_1= \alpha _1 c_0$ and $c_0 =0$ for simplicity.
 
The characteristic equation of (\ref{gg:1}) is given by 
\begin{equation}
r^2 -\alpha _{1} r -\alpha _{2} =0
\label{gg:3}
\end{equation}
The roots of a polynomial (\ref{gg:3}) have two different moduli such as
\begin{equation}
r_1 = \frac{\alpha _{1} -\sqrt{\alpha _{1}^2 +4\alpha _{2}}}{2}\hspace{1cm} r_2 = \frac{\alpha _{1} +\sqrt{\alpha _{1}^2 +4\alpha _{2}}}{2} \label{gg:4}
\end{equation}
If $|r_1|< |r_2|$, then lim $ |c_{n+1}/c_n| \rightarrow  |r_2|$ as $n\rightarrow \infty $ in general, so that the radius of convergence for a 3-term recursion relation (\ref{gg:1}) is $|r_2|^{-1}$: as if $|r_2|< |r_1|$, then lim $ |c_{n+1}/c_n| \rightarrow  |r_1|$ as $n\rightarrow \infty $, and its radius of convergence is increased to $|r_1|^{-1}$; and if $|r_1|=|r_2|$ and $r_1 \ne r_2$, $\lim_{n\rightarrow \infty } |c_{n+1}/c_n| $ does not exist; if $r_1 = r_2$, $\lim_{n\rightarrow \infty } |c_{n+1}/c_n| $ is convergent. More explicit details are explained in Appendix B of part A \cite{Ronv1995}, Wimp (1984) \cite{Wimp1984}, Kristensson (2010) \cite{Kris2010} or Erd\'{e}lyi (1955) \cite{Erde1955}.

We obtain two different moduli by putting (\ref{gg:2}) in (\ref{gg:4}) such as 
\begin{equation}
r_1 = \frac{1+\rho ^2 -|1-\rho ^2|}{2}\hspace{1cm} r_2 = \frac{1+\rho ^2 +|1-\rho ^2|}{2} \label{gg:5}
\end{equation}
 $ \rho $ is mostly between 0 and 1, then we have $|r_1 =\rho ^2|<|r_2 =1|$ in (\ref{gg:5}). And Poincar\'{e}-Perron theorem tells us that the radius of convergence for a independent variable $sn^2(z,\rho )$ is $|sn^2(z,\rho )|<1$. If $ z, sn(z,\rho )\in \mathbb{R}$, then the boundary condition of $sn^2(z,\rho )$ is
\begin{equation}
 0\leq sn^2(z,\rho ) <1  \label{gg:6}
\end{equation} 

\begin{figure}[htbp]
\centering
\includegraphics[scale=.7]{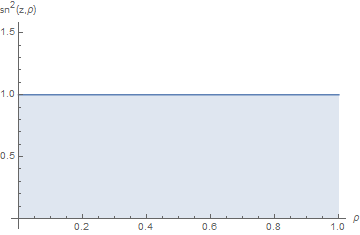}
\caption{Domain of convergence of the series by applying Poincar\'{e}-Perron theorem}
\label{img2}
\end{figure}
The corresponding domain of convergence in the real axis, given by (\ref{gg:6}), is shown shaded in Fig.~\ref{img2}; it does not include solid lines, and maximum modulus of $sn^2(z,\rho )$ is the unity.

As we compare (\ref{eq:56}) with (\ref{gg:6}), both boundary conditions for radius of convergence are not equivalent to each other:
the boundary condition of $sn^2(z,\rho )$ in (\ref{gg:6}) is derived by the ratio of sequence $c_{n+1}$ to $c_n$ at the limit $n\rightarrow \infty $. And radius of convergence of $sn^2(z,\rho )$ in (\ref{eq:56}) is constructed by rearranging coefficients $\alpha _1$ and $\alpha _2$ in each sequence $c_n$ in (\ref{gg:1}).

Let assume that Poincar\'{e}-Perron theorem provides us the radius of convergence for a solution in series. Then we know that a  solution of its power series is absolutely convergent and we can rearrange of its terms for the series solution. 
Consider the following summation series such as
\begin{eqnarray}
y(\xi) = \sum_{n=0}^{N} \sum_{m=0}^{N} \frac{(n+m)!}{n!\;m!} \tilde{x}^n \tilde{y}^m \hspace{1cm} \mbox{where}\; \tilde{x}=-\rho ^2 sn^4(z,\rho ) \;\mbox{and} \; \tilde{y}= (1+\rho ^2) sn^2(z,\rho ) \label{gg:7}
\end{eqnarray}
This equation is equivalent to (\ref{eq:55}) as $N\rightarrow \infty $. 
For instance, put $\rho =0.8$ in (\ref{eq:56}) 
\begin{equation}
0 \leq sn^2(z,0.8)< \frac{-(1+0.8^2)+\sqrt{0.8^4+6\times 0.8^2+1}}{2 \times 0.8^2}\approx 0.50875 \label{gg:8}
\end{equation}  
And if Perron's rule verifies that an infinite series of Lame equation is absolute convergent and gives the corrected radius of convergence, (\ref{gg:6}) must to be satisfied. Then, we also have a solution in series at $0.50875 \leq  sn^2(z,0.8)<1$.

Let consider $sn^2(z,0.8)=0.7$ in (\ref{gg:7}) with various positive integer values $N$ such as $N= 10,50,100,200,300,\cdots,1000$ where $\rho =0.8$ in Mathematica program.

\begin{table}[htbp]
\begin{center}
\tabcolsep 5.8pt
\begin{tabular}{l*{6}{c}|r}
 $N$ &  $y(\xi)$ \\
\hline      
$10$ &  $8.97174$  \\ 
$50$ &  $4.44473\times 10^6$  \\ 
$100$ & $2.62952\times 10^{14}$  \\ 
$200$ & $1.28525\times 10^{30}$ \\
$300$ & $7.23351\times 10^{45}$  \\
$400$ & $4.31499\times 10^{61}$  \\ 
$500$ & $2.65768\times 10^{77}$ \\
$600$ & $1.67043\times 10^{93}$  \\
$700$ & $1.06472\times 10^{109} $ \\
$800$ & $6.85643\times 10^{124} $  \\ 
$900$ & $4.45007\times 10^{140} $ \\
$1000$ & $2.90618\times 10^{156} $  \\ 
\end{tabular}
\end{center}
\caption{ $y(\xi)$ with $\rho =0.8$ and $\xi = sn^2(z,0.8)=0.7$}\label{cb.1}
\end{table}  

Table~\ref{cb.1} informs that $y(\xi)$ is divergent as $N\rightarrow \infty $.  
And we notice that the radius of convergence obtained by Poincar\'{e}-Perron theorem is not available for a solution in series of Lame equation.
\begin{thm}
We can not use Poincar\'{e}-Perron theorem to obtain the radius of convergence for a power series solution.
And a series solution for an infinite series, obtained by applying Poincar\'{e}-Perron theorem, is not absolute convergent but only conditionally convergent.\label{Thm.1} 
\end{thm}
Thm.\ref{Thm.1} is proved by rearranging the order of the terms in series in Sec.2 \cite{Chou2012d}. In order to answer a reason why we take errors of the radius of convergence obtained by Perron's rule, first of all, let us think about a series expansion of (\ref{gg:1}) such as 
\begin{eqnarray}
 \sum_{n=0}^{\infty }c_n \xi^n &=& 1+ \alpha _1 \xi +\left( \alpha _1^2+ \alpha _2 \right) \xi^2 +\left( \alpha _1^3 + 2\alpha _1 \alpha _2 \right) \xi^3 +\left( \alpha _1^4 + 3 \alpha _1^2 \alpha _2 + \alpha _1^2 \right) \xi^4 \nonumber\\
&&+\left( \alpha _1^5 + 4\alpha _1^3 \alpha _2 + 3\alpha _1 \alpha _2^2 \right) \xi^5 + \cdots 
\label{gg:9}
\end{eqnarray} 
where $c_0=1$ for simplicity. In general, a series $\sum_{n=0}^{\infty }c_n \xi^n$  is called absolutely convergent if $\sum_{n=0}^{\infty }|c_n| |\xi|^n$ is convergent. Take moduli of each sequence $c_n$ and $\xi$ in (\ref{gg:9})
\begin{eqnarray}
 \sum_{n=0}^{\infty }|c_n| |\xi|^n &=& 1+ |\alpha _1| |\xi| +\left| \alpha _1^2+ \alpha _2 \right| |\xi|^2 +\left| \alpha _1^3 + 2\alpha _1 \alpha _2 \right| |\xi|^3 +\left|  \alpha _1^4 + 3 \alpha _1^2 \alpha _2 + \alpha _2^2 \right| |\xi|^4 \nonumber\\
&&+\left| \alpha _1^5 + 4\alpha _1^3 \alpha _2 + 3\alpha _1 \alpha _2^2 \right| |\xi|^5 + \cdots 
\label{gg:10}
\end{eqnarray} 
The Cauchy ratio test tells us that a series is absolute convergent if lim $\left| \frac{c_{n+1}}{c_n}\right| \left| \xi\right| $ as $n\rightarrow \infty $ is less than the unit. And Poincar\'{e}-Perron theorem give us the value of $\left| \lim_{n\rightarrow \infty }\frac{c_{n+1}}{c_n}\right| $ in (\ref{gg:10}). However, we can not obtain the radius of convergence using this basic principle. We must take all absolute values inside parentheses of (\ref{gg:9}) such as
\begin{eqnarray}
 \sum_{n=0}^{\infty }|c_n| |\xi|^n &=& 1+ \big| \alpha _1\big| |\xi| +\left( \left| \alpha _1^2\right| + \big| \alpha _2 \big| \right) |\xi|^2 +\left( \left| \alpha _1^3 \right| + \big| 2\alpha _1 \alpha _2 \big| \right) |\xi|^3 +\left( \left|  \alpha _1^4\right| + \left| 3 \alpha _1^2 \alpha _2\right| + \left| \alpha _2^2 \right| \right) |\xi|^4 \nonumber\\
&&+\left( \left| \alpha _1^5 \right| + \left| 4\alpha _1^3 \alpha _2 \right| + \left| 3\alpha _1 \alpha _2^2 \right| \right) |\xi|^5 + \cdots 
\label{gg:11}
\end{eqnarray} 
The more explicit explanation about this mathematical phenomenon is available in Sec.2 of Ref.\cite{Chou2012d}.

Take all absolute values of constant coefficients $\alpha _i$ of the characteristic equation in Thm.\ref{thm.1}
\begin{equation}
t^k + |\alpha _1| t^{k-1} + |\alpha _2| t^{k-2}+ \cdots + |\alpha _k| =0\label{gg:12}
\end{equation}
suggesting the roots of its characteristic equation as $\lambda _1^{\star }, . . . ,\lambda _k^{\star } $.
And lim $\left| \frac{u_{i}(n+1)}{u_{i}(n)} \right|$ as $n\rightarrow \infty $ in Thm.\ref{thm.1} is equivalent to $\left| \lambda _i^{\star }\right|$. With this reconsideration, the corrected radius of convergence for a Lame function is equivalent to (\ref{eq:56}) since $0<\rho <1$ where $ z, sn(z,\rho )\in \mathbb{R}$.

\begin{figure}[htbp]
\centering
\includegraphics[scale=.7]{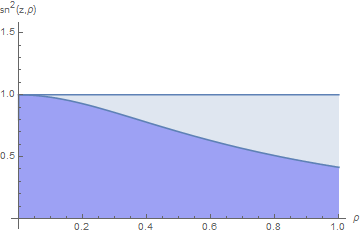}
\caption{Two different domains of absolute and conditional convergence of the Lame equation}
\label{img3}
\end{figure}

Fig.\ref{img3} represents two different shaded areas of convergence in Figs.\ref{img1} and \ref{img2}: In the bright shaded area, the domain of absolute convergence of the series for the Lame equation around $sn^2(z,\rho ) =0$ is not available; it only provides the domain of conditional convergence for it, and the dark shaded region shows the one of its absolute convergence.
\section{Application}
Lame equation appears elsewhere in mathematical physics. For example, Recently, in ``Droplet nucleation and domain wall motion in a bounded interval''\cite{Maie2001}, the authors investigate an extended model (a classical Ginzburg–Landau model) of noise-induced magnetization reversal. Lame equation arises in some specific boundary conditions.(see (8), (9) in Ref.\cite{Maie2001}. In (9) its solution consists of the Jacobi eta, theta, and zeta functions according to Hermite's solution of the Lame equation.)    
In ``Group Theoretical Properties and Band Structure of the Lame Hamiltonian''\cite{Iach2000}, the authors represent a group theoretical analysis of the Lame equation, which is an example of a SGA band structure problem for $su(2)$ and $su(1,1)$. (see (1), (10), (13), (14), (28), (29), (33), (38) in Ref.\cite{Iach2000}) Applying three term recurrence formula\cite{chou2012b}, we can obtain the power series expansion in closed forms and asymptotic behaviors of Lame function analytically. And it might be possible to obtain specific eigenvalues for the Lame Hamiltonian.\footnote{The authors treat the analytic solution of Lame equation as Lame polynomial of type 3: for any non-negative integer value of $\alpha $ there will be $2\alpha  + 1$ values of $h$ (Energy) for which the solution $y(\xi )$ reduces to a polynomial. In this paper we construct the power series expansion and integral formalism of Lame polynomial of type 1: we treat the spectral parameter $h$ as a free variable. Mathematically, it can be one of possible analytic solutions of Lame equation. In future papers we will derive types 2 and 3 Lame polynomial.} Again Lame equation is applicable to diverse areas such as theory of the stability analysis of static configurations in Josephson junctions \cite{Capu2000}, the computation of the distance–redshift relation in inhomogeneous cosmologies\cite{Kant2001}, magnetostatic problems in triaxial ellipsoids\cite{Dobn1998} and etc.
\section{Conclusion}
From the above all, applying three term recurrence formula \cite{chou2012b}, we show the power series expansion in closed forms of Lame equation in Weierstrass's form (for an infinite series and a polynomial which makes $B_n$ term terminated) and its integral forms. 
We show that a $_2F_1$ function recurs in each of sub-integral forms of Lame function in   Weierstrass's form: the first sub-integral form contains zero term of $A_n's$, the second one contains one term of $A_n$'s, the third one contains two terms of $A_n$'s, etc. And we show asymptotic expansions of Lame equation for an infinite series and the special case as $\rho \approx 0$. 
Since we obtain the closed integral forms of Lame equation in Weierstrass's form, Lame function is able to be transformed to other well-known special functions analytically; hypergeometric function, Mathieu function, Lame function, confluent forms of Heun function and etc. 

For type 3 Lame polynomial, various authors argue that the value of $h\rho ^{-2}$ can be chosen properly such that the Lame function is not an infinite series but a polynomial as $\alpha $ parameters of Lame functions is a positive integer.
For type 1 Lame polynomial, since $\alpha $ is $2( 2\alpha _i+ i+\lambda )$ or $-2 (2\alpha _i+i +\lambda )-1$ as $i,\alpha _i\in \mathbb{N}_{0}$ in the analysis of the three term recurrence formula \cite{chou2012b}, Lame functions will be polynomial which makes $B_n$ term terminated; $\lambda $ is an indicial root which is 0 or $\frac{1}{2}$, then all possible $\alpha $ is $\cdots, -3, -2, -1, 0, 1, 2, 3, \cdots$.
 
According to Erdelyi (1940\cite{Erde1940}), ``there is no corresponding representation of simple integral formalisms of the solutions in ordinary linear differential equations with four regular singularities; Heun equation, Lame equation and Mathieu equation. It appears that the theory of integral equations connected with periodic solutions of Lame equation is not as complete as the corresponding theory of integral representations of, say, Legendre functions.''
The reason, why the analytic integral forms of Lame functions can not be obtained, is that the coefficients in a power series expansions do not have two term recursion relations. We have a recursive relation between a 3-term. By using the three term recurrence formula\cite{chou2012b}, we are able to obtain analytic integral solution of any linear ordinary differential equation in which has three term recursion relations.
\section{Series ``Special functions and three term recurrence formula (3TRF)''} 

This paper is 7th out of 10.
\vspace{3mm}

1. ``Approximative solution of the spin free Hamiltonian involving only scalar potential for the $q-\bar{q}$ system'' \cite{chou2012a} - In order to solve the spin-free Hamiltonian with light quark masses we are led to develop a totally new kind of special function theory in mathematics that generalize all existing theories of confluent hypergeometric types. We call it the Grand Confluent Hypergeometric Function. Our new solution produces previously unknown extra hidden quantum numbers relevant for description of supersymmetry and for generating new mass formulas.
\vspace{3mm}

2. ``Generalization of the three-term recurrence formula and its applications'' \cite{chou2012b} - Generalize three term recurrence formula in linear differential equation.  Obtain the exact solution of the three term recurrence for polynomials and infinite series.
\vspace{3mm}

3. ``The analytic solution for the power series expansion of Heun function'' \cite{chou2012c} -  Apply three term recurrence formula to the power series expansion in closed forms of Heun function (infinite series and polynomials) including all higher terms of $A_n$'s.
\vspace{3mm}

4. ``Asymptotic behavior of Heun function and its integral formalism'', \cite{Chou2012d} - Apply three term recurrence formula, derive the integral formalism, and analyze the asymptotic behavior of Heun function (including all higher terms of $A_n$'s). 
\vspace{3mm}

5. ``The power series expansion of Mathieu function and its integral formalism'', \cite{Chou2012e} - Apply three term recurrence formula, analyze the power series expansion of Mathieu function and its integral forms.  
\vspace{3mm}

6. ``Lame equation in the algebraic form'' \cite{Chou2012f} - Applying three term recurrence formula, analyze the power series expansion of Lame function in the algebraic form and its integral forms.
\vspace{3mm}

7. ``Power series and integral forms of Lame equation in   Weierstrass's form and its asymptotic behaviors'' \cite{Chou2012g} - Applying three term recurrence formula, derive the power series expansion of Lame function in   Weierstrass's form and its integral forms. 
\vspace{3mm}

8. ``The generating functions of Lame equation in   Weierstrass's form'' \cite{Chou2012h} - Derive the generating functions of Lame function in   Weierstrass's form (including all higher terms of $A_n$'s).  Apply integral forms of Lame functions in   Weierstrass's form.
\vspace{3mm}

9. ``Analytic solution for grand confluent hypergeometric function'' \cite{Chou2012i} - Apply three term recurrence formula, and formulate the exact analytic solution of grand confluent hypergeometric function (including all higher terms of $A_n$'s). Replacing $\mu $ and $\varepsilon \omega $ by 1 and $-q$, transforms the grand confluent hypergeometric function into Biconfluent Heun function.
\vspace{3mm}

10. ``The integral formalism and the generating function of grand confluent hypergeometric function'' \cite{Chou2012j} - Apply three term recurrence formula, and construct an integral formalism and a generating function of grand confluent hypergeometric function (including all higher terms of $A_n$'s). 

\appendix
\section*{Appendix. Conversion from 9 out of 192 local solutions of Heun equation to 9 local solutions of Lam\'{e} equation in Weierstrass's form for an infinite series and a polynomial of type 1}
\addcontentsline{toc}{section}{Appendix}  
\renewcommand*{\thesection}{\Alph{section}}
A machine-generated list of 192 (isomorphic to the Coxeter group of the Coxeter diagram $D_4$) local solutions of the Heun equation was obtained by Robert S. Maier(2007) \cite{Maie2007}. In appendix of Ref.\cite{Chou2012d}, by applying 3TRF, we obtain power series expansions and integrals of Heun equation (for an infinite series and a polynomial of type 1) of nine out of the 192 local solution of Heun equation in Table 2 \cite{Maie2007}. 

In this appendix, by changing all coefficients and independent variables of the previous nine examples of 192 local solutions of Heun equation into the first kind of independent solutions of Heun equation by applying 3TRF \cite{chou2012c,Chou2012d}, we construct 9 local solutions of Lame equation in Weierstrass's form for Frobenius solutions in closed form (for infinite series and polynomial of type 1) and its integral representations.\footnote{In this appendix, we treat $h$ as a free variable and a fixed value of $\alpha $ to construct polynomials of type 1 for all 9 local solutions of Lame equation. An independent variable $sn^2(z,\rho )$ is denoted by $\xi$. And we consider $\alpha $ as arbitrary. The condition $\alpha \geq -\frac{1}{2}$ is not necessary any more.}

Lame equation in Weierstrass's form is a special case of Heun's equation. Heun equation is a second-order linear ordinary differential equation of the form \cite{Heun1889,Ronv1995}.
\begin{equation}
\frac{d^2{y}}{d{x}^2} + \left(\frac{\gamma }{x} +\frac{\delta }{x-1} + \frac{\epsilon }{x-a}\right) \frac{d{y}}{d{x}} +  \frac{\alpha \beta x-q}{x(x-1)(x-a)} y = 0 \label{eq:1005}
\end{equation}
With the condition $\epsilon = \alpha +\beta -\gamma -\delta +1$. The parameters play different roles: $a \ne 0 $ is the singularity parameter, $\alpha $, $\beta $, $\gamma $, $\delta $, $\epsilon $ are exponent parameters, $q$ is the accessory parameter which in many physical applications appears as a spectral parameter. Also, $\alpha $ and $\beta $ are identical to each other. The total number of free parameters is six. It has four regular singular points which are 0, 1, $a$ and $\infty $ with exponents $\{ 0, 1-\gamma \}$, $\{ 0, 1-\delta \}$, $\{ 0, 1-\epsilon \}$ and $\{ \alpha, \beta \}$.

As we compare (\ref{eq:2}) with (\ref{eq:1005}), all coefficients on the above are correspondent to the following way.
\begin{equation}
\begin{split}
& \gamma ,\delta ,\epsilon  \longleftrightarrow   \frac{1}{2} \\ & a\longleftrightarrow  \rho ^{-2} \\ & \alpha  \longleftrightarrow \frac{1}{2}(\alpha +1) \\
& \beta   \longleftrightarrow -\frac{1}{2} \alpha \\
& q \longleftrightarrow  -\frac{1}{4}h \rho ^{-2} \\ & x \longleftrightarrow \xi = sn^2(z,\rho ) 
\end{split}\label{eq:1006}   
\end{equation}
\section{Power series}
In Ref.\cite{chou2012c}, the representation in the form of power series expansion of the first kind of independent solution of Heun equation for polynomial of type 1 about $x=0$ as $\alpha = -2 \alpha _j-j $ where $j,\alpha _j  \in \mathbb{N}_{0} $ is given by
\begin{eqnarray}
y(x)&=& HF_{\alpha _j, \beta }\left( \alpha _j =-\frac{1}{2}(\alpha +j)\big|_{j\in \mathbb{N}_{0}}; \eta = \frac{(1+a)}{a} x ; z= -\frac{1}{a} x^2 \right) \nonumber\\
&=& \sum_{i_0=0}^{\alpha _0} \frac{(-\alpha _0)_{i_0} \left(\frac{\beta }{2} \right)_{i_0}}{(1 )_{i_0}\left(\frac{1}{2}+ \frac{\gamma}{2}\right)_{i_0}} z^{i_0}  + \left\{\sum_{i_0=0}^{\alpha _0}\frac{ i_0 \left( i_0+ \Gamma_0^{(S)} \right)+ Q}{\left(i_0+ \frac{1}{2} \right)\left( i_0 + \frac{\gamma }{2}\right)} \frac{(-\alpha _0)_{i_0} \left(\frac{\beta }{2} \right)_{i_0}}{(1 )_{i_0}\left(\frac{1}{2}+ \frac{\gamma}{2} \right)_{i_0}} \right. \left.\sum_{i_1=i_0}^{\alpha _1} \frac{(-\alpha _1)_{i_1}\left(\frac{1}{2}+\frac{\beta }{2} \right)_{i_1}\left(\frac{3}{2} \right)_{i_0}\left( 1+\frac{\gamma }{2} \right)_{i_0}}{(-\alpha _1)_{i_0}\left(\frac{1}{2}+\frac{\beta }{2} \right)_{i_0}\left(\frac{3}{2} \right)_{i_1}\left(1+ \frac{\gamma}{2} \right)_{i_1}} z^{i_1} \right\} \eta \nonumber\\
&&+ \sum_{n=2}^{\infty } \left\{ \sum_{i_0=0}^{\alpha _0} \frac{ i_0 \left( i_0+ \Gamma_0^{(S)} \right)+ Q}{\left(i_0+ \frac{1}{2} \right)\left( i_0 + \frac{\gamma }{2} \right)}  \frac{(-\alpha _0)_{i_0} \left(\frac{\beta }{2} \right)_{i_0}}{(1 )_{i_0}\left(\frac{1}{2}+ \frac{\gamma}{2} \right)_{i_0}}\right.\nonumber\\
&&\times \prod _{k=1}^{n-1} \left\{ \sum_{i_k=i_{k-1}}^{\alpha _k} \frac{\left(i_k+\frac{k}{2} \right) \left( i_k+\Gamma_k^{(S)} \right)+ Q}{\left(i_k+ \frac{k}{2}+\frac{1}{2} \right)\left(i_k +\frac{k}{2}+\frac{\gamma }{2} \right)}  \frac{(-\alpha _k)_{i_k}\left(\frac{k}{2}+\frac{\beta }{2} \right)_{i_k}\left(1+ \frac{k}{2} \right)_{i_{k-1}}\left(\frac{1}{2}+\frac{k}{2}+\frac{\gamma }{2} \right)_{i_{k-1}}}{(-\alpha _k)_{i_{k-1}}\left(\frac{k}{2}+\frac{\beta }{2} \right)_{i_{k-1}}\left(1+\frac{k}{2} \right)_{i_k}\left(\frac{1}{2}+ \frac{k}{2}+ \frac{\gamma}{2} \right)_{i_k}}\right\} \nonumber\\
&&\times \left. \sum_{i_n= i_{n-1}}^{\alpha _n} \frac{(-\alpha _n)_{i_n}\left(\frac{n}{2}+\frac{\beta }{2} \right)_{i_n}\left(1+ \frac{n}{2} \right)_{i_{n-1}}\left(\frac{1}{2}+\frac{n}{2}+\frac{\gamma }{2} \right)_{i_{n-1}}}{(-\alpha _n)_{i_{n-1}}\left(\frac{n}{2}+\frac{\beta }{2} \right)_{i_{n-1}}\left(1+\frac{n}{2} \right)_{i_n}\left(\frac{1}{2}+ \frac{n}{2}+ \frac{\gamma}{2} \right)_{i_n}} z^{i_n} \right\} \eta ^n \label{eq:10023}
\end{eqnarray}
where
\begin{equation}
\begin{cases} z = -\frac{1}{a}x^2 \cr
\eta = \frac{(1+a)}{a} x \cr
\alpha _i\leq \alpha _j \;\;\mbox{only}\;\mbox{if}\;i\leq j\;\;\mbox{where}\;i,j= 0,1,2,\cdots
\end{cases}\nonumber 
\end{equation}
and
\begin{equation}
\begin{cases} 
\Gamma_0^{(S)} = \frac{1}{2(1+a)}(-2\alpha _0+ \beta -\delta +a(\delta +\gamma -1 )) \cr
\Gamma_k^{(S)} = \frac{1}{2(1+a)}(-2\alpha _k+ \beta -\delta +a(\delta +\gamma +k-1)) \cr
Q= \frac{q}{4(1+a)}
\end{cases}\nonumber 
\end{equation}
In Ref.\cite{chou2012c}, the representation in the form of power series expansion of the first kind of independent solution of Heun equation for infinite series about $x=0$ is given by
\begin{eqnarray}
y(x)&=& HF_{\alpha , \beta }\left( \eta = \frac{(1+a)}{a} x ; z= -\frac{1}{a} x^2 \right) \nonumber\\
&=& \sum_{i_0=0}^{\infty } \frac{\left(\frac{\alpha }{2} \right)_{i_0} \left(\frac{\beta }{2} \right)_{i_0}}{(1 )_{i_0}\left(\frac{1}{2}+ \frac{\gamma}{2} \right)_{i_0}} z^{i_0} \nonumber\\
&&+ \left\{\sum_{i_0=0}^{\infty }\frac{ i_0 \left( i_0+ \Gamma_0^{(I)}\right)+ Q}{\left(i_0+ \frac{1}{2} \right)\left(i_0 + \frac{\gamma }{2} \right)}\right. \left. \frac{\left(\frac{\alpha }{2} \right)_{i_0} \left(\frac{\beta }{2} \right)_{i_0}}{(1 )_{i_0}\left(\frac{1}{2}+ \frac{\gamma}{2} \right)_{i_0}} \sum_{i_1=i_0}^{\infty } \frac{\left(\frac{1}{2}+\frac{\alpha }{2} \right)_{i_1}\left(\frac{1}{2}+\frac{\beta }{2} \right)_{i_1}\left(\frac{3}{2} \right)_{i_0}\left(1+\frac{\gamma }{2} \right)_{i_0}}{\left(\frac{1}{2}+\frac{\alpha }{2} \right)_{i_0}\left(\frac{1}{2}+\frac{\beta }{2} \right)_{i_0}\left(\frac{3}{2} \right)_{i_1}\left(1+ \frac{\gamma}{2} \right)_{i_1}} z^{i_1} \right\} \eta \nonumber\\
&&+ \sum_{n=2}^{\infty } \left\{ \sum_{i_0=0}^{\infty } \frac{ i_0 \left( i_0+ \Gamma_0^{(I)}\right)+ Q}{\left(i_0+ \frac{1}{2} \right)\left(i_0 + \frac{\gamma }{2} \right)}
 \frac{\left(\frac{\alpha }{2} \right)_{i_0} \left(\frac{\beta }{2} \right)_{i_0}}{(1 )_{i_0}\left(\frac{1}{2}+ \frac{\gamma}{2} \right)_{i_0}}\right.\nonumber\\
&&\times \prod _{k=1}^{n-1} \left\{ \sum_{i_k=i_{k-1}}^{\infty } \frac{\left(i_k+\frac{k}{2} \right) \left( i_k+ \Gamma_k^{(I)}\right)+ Q}{\left(i_k+ \frac{k}{2}+\frac{1}{2} \right)\left(i_k +\frac{k}{2}+\frac{\gamma }{2} \right)} \right.  \left.\frac{\left(\frac{k}{2}+\frac{\alpha }{2} \right)_{i_k}\left(\frac{k}{2}+\frac{\beta }{2} \right)_{i_k}\left(1+ \frac{k}{2} \right)_{i_{k-1}}\left(\frac{1}{2}+\frac{k}{2}+\frac{\gamma }{2} \right)_{i_{k-1}}}{\left(\frac{k}{2}+\frac{\alpha }{2} \right)_{i_{k-1}}\left(\frac{k}{2}+\frac{\beta }{2} \right)_{i_{k-1}}\left(1+\frac{k}{2}\right)_{i_k}\left(\frac{1}{2}+ \frac{k}{2}+ \frac{\gamma}{2} \right)_{i_k}}\right\} \nonumber\\
&&\times \left. \sum_{i_n= i_{n-1}}^{\infty } \frac{\left(\frac{n}{2}+\frac{\alpha }{2} \right)_{i_n}\left(\frac{n}{2}+\frac{\beta }{2} \right)_{i_n}\left(1+ \frac{n}{2} \right)_{i_{n-1}}\left(\frac{1}{2}+\frac{n}{2}+\frac{\gamma }{2} \right)_{i_{n-1}}}{\left(\frac{n}{2}+\frac{\alpha }{2} \right)_{i_{n-1}}\left(\frac{n}{2}+\frac{\beta }{2} \right)_{i_{n-1}}\left(1+\frac{n}{2} \right)_{i_n}\left(\frac{1}{2}+ \frac{n}{2}+ \frac{\gamma}{2} \right)_{i_n}} z^{i_n} \right\} \eta ^n \label{eq:10024}
\end{eqnarray}
where
\begin{equation}
\begin{cases} 
\Gamma_0^{(I)} =  \frac{1}{2(1+a)}(\alpha +\beta -\delta +a(\delta +\gamma -1 ))\cr
\Gamma_k^{(I)} =  \frac{1}{2(1+a)}(\alpha +\beta -\delta +k +a(\delta +\gamma -1+k )) \cr
Q= \frac{q}{4(1+a)}
\end{cases}\nonumber 
\end{equation}
\subsection{ ${\displaystyle (1-x)^{1-\delta } Hl(a, q - (\delta  - 1)\gamma a; \alpha - \delta  + 1, \beta - \delta + 1, \gamma ,2 - \delta ; x)}$ }
\subsubsection{Polynomial of type 1}
Replace coefficients $q$, $\alpha$, $\beta$ and $\delta$ by $q - (\delta - 1)\gamma a $, $\alpha - \delta  + 1 $, $\beta - \delta + 1$ and $2 - \delta$ into (\ref{eq:10023}). Multiply $(1-x)^{1-\delta }$ and (\ref{eq:10023}) together. Put (\ref{eq:1006}) into the new (\ref{eq:10023}) with replacing $\alpha $ by $-2(2\alpha _j+j+1)$ where $j,\alpha _j \in \mathbb{N}_{0}$; apply $\alpha =-2(2\alpha _0+1)$ into sub-power series $y_0(\xi)$, apply $\alpha =-2(2\alpha _0+1)$ into the first summation and $\alpha =-2(2\alpha _1+2)$ into second summation of sub-power series $y_1(\xi)$, apply $\alpha =-2(2\alpha _0+1)$ into the first summation, $\alpha =-2(2\alpha _1+2)$ into the second summation and $\alpha =-2(2\alpha _2+3)$ into the third summation of sub-power series $y_2(\xi)$, etc in the new (\ref{eq:10023}).\footnote{For all 9 local solutions of Lame equation for polynomial of type 1 in this appendix, $\alpha _i\leq \alpha _j$  only if $i\leq j$ where $i,j,\alpha _i,\alpha _j \in \mathbb{N}_{0}$.}
\begin{eqnarray}
&& (1-\xi )^{\frac{1}{2}} y(\xi )\nonumber\\
&=& (1-\xi )^{\frac{1}{2}} Hl\left(\rho ^{-2}, -\frac{1}{4}(h-1)\rho ^{-2}; -2\alpha _j-j, -2\alpha _j-j, \frac{1}{2},\frac{3}{2}; \xi \right)\nonumber\\
&=& (1-\xi )^{\frac{1}{2}} \left\{ \sum_{i_0=0}^{\alpha _0} \frac{(-\alpha _0)_{i_0} \left(\alpha _0+\frac{3}{4} \right)_{i_0}}{(1 )_{i_0}\left(\frac{3}{4} \right)_{i_0}} z^{i_0} \right. \nonumber\\
&&+ \left\{\sum_{i_0=0}^{\alpha _0}\frac{ i_0 \left( i_0+ \Gamma_0 \right)+ Q}{\left(i_0+ \frac{1}{2} \right)\left( i_0 + \frac{1}{4}\right)} \frac{(-\alpha _0)_{i_0} \left(\alpha _0+\frac{3}{4}\right)_{i_0}}{(1 )_{i_0}\left(\frac{3}{4} \right)_{i_0}} \right. \left.\sum_{i_1=i_0}^{\alpha _1} \frac{(-\alpha _1)_{i_1}\left(\alpha _1+\frac{7}{4} \right)_{i_1}\left(\frac{3}{2} \right)_{i_0}\left( \frac{5}{4} \right)_{i_0}}{(-\alpha _1)_{i_0}\left(\alpha _1+\frac{7}{4} \right)_{i_0}\left(\frac{3}{2} \right)_{i_1}\left( \frac{5}{4} \right)_{i_1}} z^{i_1} \right\} \eta \nonumber\\
&&+ \sum_{n=2}^{\infty } \left\{ \sum_{i_0=0}^{\alpha _0} \frac{ i_0 \left( i_0+ \Gamma_0 \right)+ Q}{\left(i_0+ \frac{1}{2} \right)\left( i_0 + \frac{1}{4} \right)}  \frac{(-\alpha _0)_{i_0} \left(\alpha _0+ \frac{3}{4} \right)_{i_0}}{(1 )_{i_0}\left( \frac{3}{4} \right)_{i_0}}\right.\nonumber\\
&&\times \prod _{k=1}^{n-1} \left\{ \sum_{i_k=i_{k-1}}^{\alpha _k} \frac{\left( i_k+\frac{k}{2} \right) \left( i_k+\Gamma_k \right)+ Q}{\left(i_k+ \frac{k}{2}+\frac{1}{2} \right)\left(i_k +\frac{k}{2}+\frac{1}{4} \right)}  \frac{(-\alpha _k)_{i_k}\left(\alpha _k+k+\frac{3}{4} \right)_{i_k}\left(1+ \frac{k}{2} \right)_{i_{k-1}}\left(\frac{3}{4}+\frac{k}{2} \right)_{i_{k-1}}}{(-\alpha _k)_{i_{k-1}}\left(\alpha _k+ k +\frac{3}{4} \right)_{i_{k-1}}\left(1+\frac{k}{2} \right)_{i_k}\left(\frac{3}{4}+ \frac{k}{2} \right)_{i_k}}\right\} \nonumber\\
&&\times \left. \left.\sum_{i_n= i_{n-1}}^{\alpha _n} \frac{(-\alpha _n)_{i_n}\left(\alpha _n+n+\frac{3}{4} \right)_{i_n}\left(1+ \frac{n}{2} \right)_{i_{n-1}}\left( \frac{3}{4} +\frac{n}{2}\right)_{i_{n-1}}}{(-\alpha _n)_{i_{n-1}}\left(\alpha _n+n+\frac{3}{4} \right)_{i_{n-1}}\left( 1+\frac{n}{2} \right)_{i_n}\left( \frac{3}{4} +\frac{n}{2} \right)_{i_n}} z^{i_n} \right\} \eta ^n\right\} \label{eq:10025}
\end{eqnarray}
where
\begin{equation}
\alpha = 2\left( 2\alpha _j +j+\frac{1}{2}\right) \;\mbox{or}\; -2\left( 2\alpha _j +j+1\right) \nonumber
\end{equation}
For the minimum value of Lame equation for a polynomial which makes $B_n$ term terminated about $\xi =0 $, put $\alpha _0=\alpha _1=\alpha _2=\cdots=0$ in (\ref{eq:10025}).
\begin{eqnarray}
 y(\xi ) &=&   Hl\left(\rho ^{-2}, -\frac{1}{4}(h-1)\rho ^{-2}; -j, -j, \frac{1}{2},\frac{3}{2}; \xi \right) \nonumber\\
&=&  \; _2F_1\left(  \frac{1-\sqrt{(1+\rho ^2)h-\rho ^2}}{2(1+\rho ^2)}, \frac{1+\sqrt{(1+\rho ^2)h-\rho ^2}}{2(1+\rho ^2)} , \frac{1}{2}; \eta \right)  \nonumber
\end{eqnarray}  
It tells us that Lame polynomials in which makes $B_n$ term terminated, for fixed values of $\alpha $, require $|\eta|=|(1+\rho ^2)sn^2(z,\rho ) | < 1$ for the convergence of the radius.
\subsubsection{Infinite series}
Replace coefficients $q$, $\alpha$, $\beta$ and $\delta$ by $q - (\delta - 1)\gamma a $, $\alpha - \delta  + 1 $, $\beta - \delta + 1$ and $2 - \delta$ into (\ref{eq:10024}). Multiply $(1-x)^{1-\delta }$ and (\ref{eq:10024}) together. Put (\ref{eq:1006}) into the new (\ref{eq:10024}).
\begin{eqnarray}
&& (1-\xi )^{\frac{1}{2}} y(\xi )\nonumber\\
&=& (1-\xi )^{\frac{1}{2}} Hl\left(\rho ^{-2}, -\frac{1}{4}(h-1)\rho ^{-2}; \frac{\alpha }{2}+1, -\frac{\alpha }{2}+\frac{1}{2}, \frac{1}{2},\frac{3}{2}; \xi \right)\nonumber\\
&=& (1-\xi )^{\frac{1}{2}} \left\{ \sum_{i_0=0}^{\infty } \frac{\left(\frac{\alpha }{4} +\frac{1}{2}\right)_{i_0} \left(-\frac{\alpha }{4} +\frac{1}{4}\right)_{i_0}}{(1 )_{i_0}\left(\frac{3}{4} \right)_{i_0}} z^{i_0}\right. \nonumber\\
&&+ \left\{\sum_{i_0=0}^{\infty }\frac{ i_0 \left( i_0+ \Gamma_0 \right)+ Q}{\left(i_0+ \frac{1}{2} \right)\left(i_0 + \frac{1 }{4} \right)}\right. \left. \frac{\left(\frac{\alpha }{4}+\frac{1}{2} \right)_{i_0} \left(-\frac{\alpha }{4}+\frac{1}{4} \right)_{i_0}}{(1 )_{i_0}\left(\frac{3}{4} \right)_{i_0}} \sum_{i_1=i_0}^{\infty } \frac{\left(\frac{\alpha }{4}+1\right)_{i_1}\left(-\frac{\alpha }{4}+\frac{3}{4} \right)_{i_1}\left(\frac{3}{2} \right)_{i_0}\left( \frac{5}{4} \right)_{i_0}}{\left(\frac{\alpha }{4}+1\right)_{i_0}\left(-\frac{\alpha }{4}+\frac{3}{4} \right)_{i_0}\left(\frac{3}{2} \right)_{i_1}\left( \frac{5}{4} \right)_{i_1}} z^{i_1} \right\} \eta \nonumber\\
&&+ \sum_{n=2}^{\infty } \left\{ \sum_{i_0=0}^{\infty } \frac{ i_0 \left( i_0+ \Gamma_0 \right)+ Q}{\left(i_0+ \frac{1}{2} \right)\left(i_0 + \frac{1}{4} \right)}
 \frac{\left(\frac{\alpha }{4}+\frac{1}{2} \right)_{i_0} \left(-\frac{\alpha }{4}+\frac{1}{4} \right)_{i_0}}{(1 )_{i_0}\left(\frac{3}{4} \right)_{i_0}}\right.\nonumber\\
&&\times \prod _{k=1}^{n-1} \left\{ \sum_{i_k=i_{k-1}}^{\infty } \frac{\left( i_k+\frac{k}{2} \right) \left( i_k+ \Gamma_k \right)+ Q}{\left( i_k+ \frac{k}{2}+\frac{1}{2} \right)\left(i_k +\frac{k}{2}+\frac{1}{4} \right)} \right.  \left.\frac{\left(\frac{k}{2}+\frac{\alpha }{4}+\frac{1}{2} \right)_{i_k}\left(\frac{k}{2}-\frac{\alpha }{4}+\frac{1}{4}  \right)_{i_k}\left( 1+ \frac{k}{2} \right)_{i_{k-1}}\left(\frac{3}{4}+\frac{k}{2} \right)_{i_{k-1}}}{\left(\frac{k}{2}+\frac{\alpha }{4}+\frac{1}{2} \right)_{i_{k-1}}\left(\frac{k}{2}-\frac{\alpha }{4}+\frac{1}{4} \right)_{i_{k-1}}\left(1+\frac{k}{2}\right)_{i_k}\left(\frac{3}{4}+ \frac{k}{2} \right)_{i_k}}\right\} \nonumber\\
&&\times \left. \left. \sum_{i_n= i_{n-1}}^{\infty } \frac{\left(\frac{n}{2}+\frac{\alpha }{4}+\frac{1}{2} \right)_{i_n}\left(\frac{n}{2}-\frac{\alpha }{4}+\frac{1}{4}\right)_{i_n}\left( 1+ \frac{n}{2} \right)_{i_{n-1}}\left(\frac{3}{4}+\frac{n}{2} \right)_{i_{n-1}}}{\left(\frac{n}{2}+\frac{\alpha }{4}+\frac{1}{2} \right)_{i_{n-1}}\left(\frac{n}{2}-\frac{\alpha }{4}+\frac{1}{4} \right)_{i_{n-1}}\left( 1+\frac{n}{2} \right)_{i_n}\left(\frac{3}{4}+ \frac{n}{2} \right)_{i_n}} z^{i_n} \right\} \eta ^n \right\} \label{eq:10026}
\end{eqnarray}
On (\ref{eq:10025}) and (\ref{eq:10026}),
\begin{equation}
\begin{cases} 
\eta =(1+\rho ^2)\xi \cr
z=-\rho ^2\xi^2 \cr
\xi = sn^2(z,\rho ) \cr
\Gamma_0 = \frac{1}{2(1+\rho ^2)}  \cr
\Gamma_k = \frac{k}{2}+\frac{1}{2(1+\rho ^2)} \cr
Q= \frac{1-h}{16(1+\rho ^2)}
\end{cases}\nonumber 
\end{equation}
\subsection{ ${\displaystyle x^{1-\gamma } (1-x)^{1-\delta } Hl(a, q-(\gamma +\delta -2)a-(\gamma -1)(\alpha +\beta -\gamma -\delta +1); \alpha - \gamma -\delta +2}$ \\${\displaystyle, \beta - \gamma -\delta +2, 2-\gamma, 2 - \delta ; x)}$}
\subsubsection{Polynomial of type 1}
Replace coefficients $q$, $\alpha$, $\beta$, $\gamma $ and $\delta$ by $q-(\gamma +\delta -2)a-(\gamma -1)(\alpha +\beta -\gamma -\delta +1)$, $\alpha - \gamma -\delta +2$, $\beta - \gamma -\delta +2, 2-\gamma$ and $2 - \delta$ into (\ref{eq:10023}). Multiply $x^{1-\gamma } (1-x)^{1-\delta }$ and (\ref{eq:10023}) together. Put (\ref{eq:6}) into the new (\ref{eq:10023}) with replacing $\alpha $ by $-2(2\alpha _j+j+3/2)$ where $j,\alpha _j \in \mathbb{N}_{0}$; apply $\alpha =-2(2\alpha _0+3/2)$ into sub-power series $y_0(\xi)$, apply $-2(2\alpha _0+3/2)$ into the first summation and $-2(2\alpha _1 +5/2)$ into second summation of sub-power series $y_1(\xi)$, apply $-2(2\alpha _0+3/2)$ into the first summation, $-2(2\alpha _1+ 5/2)$ into the second summation and $-2(2\alpha _2 +7/2)$ into the third summation of sub-power series $y_2(\xi)$, etc in the new (\ref{eq:10023}).
\begin{eqnarray}
&&\xi ^{\frac{1}{2}} (1-\xi )^{\frac{1}{2}} y(\xi )\nonumber\\
&=&\xi ^{\frac{1}{2}} (1-\xi )^{\frac{1}{2}} Hl\left(\rho ^{-2}, -\frac{1}{4}\left( (h-4)\rho ^{-2}-1\right); -2\alpha _j-j, -2\alpha _j-j, \frac{3}{2},\frac{3}{2}; \xi \right)\nonumber\\
&=& \xi ^{\frac{1}{2}} (1-\xi )^{\frac{1}{2}} \left\{ \sum_{i_0=0}^{\alpha _0} \frac{(-\alpha _0)_{i_0} \left(\alpha _0+\frac{5}{4} \right)_{i_0}}{(1 )_{i_0}\left(\frac{5}{4} \right)_{i_0}} z^{i_0} \right. \nonumber\\
&&+ \left\{\sum_{i_0=0}^{\alpha _0}\frac{ i_0 \left( i_0+ \Gamma_0 \right)+ Q}{\left(i_0+ \frac{1}{2} \right)\left( i_0 + \frac{3}{4}\right)} \frac{(-\alpha _0)_{i_0} \left(\alpha _0+\frac{5}{4}\right)_{i_0}}{(1 )_{i_0}\left(\frac{5}{4} \right)_{i_0}} \right. \left.\sum_{i_1=i_0}^{\alpha _1} \frac{(-\alpha _1)_{i_1}\left(\alpha _1+\frac{9}{4} \right)_{i_1}\left(\frac{3}{2} \right)_{i_0}\left( \frac{7}{4} \right)_{i_0}}{(-\alpha _1)_{i_0}\left(\alpha _1+\frac{9}{4} \right)_{i_0}\left(\frac{3}{2} \right)_{i_1}\left( \frac{7}{4} \right)_{i_1}} z^{i_1} \right\} \eta \nonumber\\
&&+ \sum_{n=2}^{\infty } \left\{ \sum_{i_0=0}^{\alpha _0} \frac{ i_0 \left( i_0+ \Gamma_0 \right)+ Q}{\left(i_0+ \frac{1}{2} \right)\left( i_0 + \frac{3}{4} \right)}  \frac{(-\alpha _0)_{i_0} \left(\alpha _0+ \frac{5}{4} \right)_{i_0}}{(1 )_{i_0}\left( \frac{5}{4} \right)_{i_0}}\right.\nonumber\\
&&\times \prod _{k=1}^{n-1} \left\{ \sum_{i_k=i_{k-1}}^{\alpha _k} \frac{\left( i_k+\frac{k}{2} \right) \left( i_k+\Gamma_k \right)+ Q}{\left(i_k+ \frac{k}{2}+\frac{1}{2} \right)\left(i_k +\frac{k}{2}+\frac{3}{4} \right)}  \frac{(-\alpha _k)_{i_k}\left(\alpha _k+k+\frac{5}{4} \right)_{i_k}\left(1+ \frac{k}{2} \right)_{i_{k-1}}\left(\frac{5}{4}+\frac{k}{2} \right)_{i_{k-1}}}{(-\alpha _k)_{i_{k-1}}\left(\alpha _k+ k +\frac{5}{4} \right)_{i_{k-1}}\left(1+\frac{k}{2} \right)_{i_k}\left(\frac{5}{4}+ \frac{k}{2} \right)_{i_k}}\right\} \nonumber\\
&&\times \left. \left.\sum_{i_n= i_{n-1}}^{\alpha _n} \frac{(-\alpha _n)_{i_n}\left(\alpha _n+n+\frac{5}{4} \right)_{i_n}\left(1+ \frac{n}{2} \right)_{i_{n-1}}\left( \frac{5}{4} +\frac{n}{2}\right)_{i_{n-1}}}{(-\alpha _n)_{i_{n-1}}\left(\alpha _n+n+\frac{5}{4} \right)_{i_{n-1}}\left( 1+\frac{n}{2} \right)_{i_n}\left( \frac{5}{4} +\frac{n}{2} \right)_{i_n}} z^{i_n} \right\} \eta ^n\right\} \label{eq:10027}
\end{eqnarray}
where
\begin{equation}
\alpha = 2\left( 2\alpha _j +j+1\right) \;\mbox{or}\; -2\left( 2\alpha _j +j+\frac{3}{2}\right) \nonumber
\end{equation}
For the minimum value of Lame equation for a polynomial which makes $B_n$ term terminated about $\xi =0 $, put $\alpha _0=\alpha _1=\alpha _2=\cdots=0$ in (\ref{eq:10027}).
\begin{eqnarray}
 y(\xi ) &=&   Hl\left(\rho ^{-2}, -\frac{1}{4}\left( (h-4)\rho ^{-2}-1\right); -j, -j, \frac{3}{2},\frac{3}{2}; \xi \right) \nonumber\\
&=&  \; _2F_1\left(  \frac{2+\rho ^2-\sqrt{(1+\rho ^2)h-\rho ^2}}{2(1+\rho ^2)}, \frac{2+\rho ^2+\sqrt{(1+\rho ^2)h-\rho ^2}}{2(1+\rho ^2)}, \frac{3}{2}; \eta \right)  \nonumber
\end{eqnarray}  
It tells us that Lame polynomials in which makes $B_n$ term terminated, for fixed values of $\alpha $, require $|\eta|=|(1+\rho ^2)sn^2(z,\rho ) | < 1$ for the convergence of the radius.
\subsubsection{Infinite series}
Replace coefficients $q$, $\alpha$, $\beta$, $\gamma $ and $\delta$ by $q-(\gamma +\delta -2)a-(\gamma -1)(\alpha +\beta -\gamma -\delta +1)$, $\alpha - \gamma -\delta +2$, $\beta - \gamma -\delta +2, 2-\gamma$ and $2 - \delta$ into (\ref{eq:10024}). Multiply $x^{1-\gamma } (1-x)^{1-\delta }$ and (\ref{eq:10024}) together. Put (\ref{eq:1006}) into the new (\ref{eq:10024}).
\begin{eqnarray}
&&\xi ^{\frac{1}{2}} (1-\xi )^{\frac{1}{2}} y(\xi )\nonumber\\
&=&\xi ^{\frac{1}{2}} (1-\xi )^{\frac{1}{2}} Hl\left(\rho ^{-2}, -\frac{1}{4}\left( (h-4)\rho ^{-2}-1\right); \frac{\alpha }{2}+\frac{3}{2}, -\frac{\alpha }{2} +1, \frac{3}{2},\frac{3}{2}; \xi \right)\nonumber\\
&=& \xi ^{\frac{1}{2}} (1-\xi )^{\frac{1}{2}} \left\{ \sum_{i_0=0}^{\infty } \frac{\left(\frac{\alpha }{4} +\frac{3}{4}\right)_{i_0} \left(-\frac{\alpha }{4} +\frac{1}{2}\right)_{i_0}}{(1 )_{i_0}\left(\frac{5}{4} \right)_{i_0}} z^{i_0}\right. \nonumber\\
&&+ \left\{\sum_{i_0=0}^{\infty }\frac{ i_0 \left( i_0+ \Gamma_0 \right)+ Q}{\left(i_0+ \frac{1}{2} \right)\left(i_0 + \frac{3 }{4} \right)}\right. \left. \frac{\left(\frac{\alpha }{4}+\frac{3}{4} \right)_{i_0} \left(-\frac{\alpha }{4}+\frac{1}{2} \right)_{i_0}}{(1 )_{i_0}\left(\frac{5}{4} \right)_{i_0}} \sum_{i_1=i_0}^{\infty } \frac{\left(\frac{\alpha }{4}+\frac{5}{4}\right)_{i_1}\left(-\frac{\alpha }{4}+1 \right)_{i_1}\left(\frac{3}{2} \right)_{i_0}\left( \frac{7}{4} \right)_{i_0}}{\left(\frac{\alpha }{4}+\frac{5}{4}\right)_{i_0}\left(-\frac{\alpha }{4}+1 \right)_{i_0}\left(\frac{3}{2} \right)_{i_1}\left( \frac{7}{4} \right)_{i_1}} z^{i_1} \right\} \eta \nonumber\\
&&+ \sum_{n=2}^{\infty } \left\{ \sum_{i_0=0}^{\infty } \frac{ i_0 \left( i_0+ \Gamma_0 \right)+ Q}{\left( i_0+\frac{1}{2} \right)\left(i_0 + \frac{3}{4} \right)}
 \frac{\left(\frac{\alpha }{4}+\frac{3}{4} \right)_{i_0} \left(-\frac{\alpha }{4}+\frac{1}{2} \right)_{i_0}}{(1 )_{i_0}\left(\frac{5}{4} \right)_{i_0}}\right.\nonumber\\
&&\times \prod _{k=1}^{n-1} \left\{ \sum_{i_k=i_{k-1}}^{\infty } \frac{\left( i_k+\frac{k}{2} \right) \left( i_k+ \Gamma_k \right)+ Q}{\left( i_k+ \frac{k}{2}+\frac{1}{2} \right)\left(i_k +\frac{k}{2}+\frac{3}{4} \right)} \right.  \left.\frac{\left(\frac{k}{2}+\frac{\alpha }{4}+\frac{3}{4} \right)_{i_k}\left(\frac{k}{2}-\frac{\alpha }{4}+\frac{1}{2}  \right)_{i_k}\left( 1+ \frac{k}{2} \right)_{i_{k-1}}\left(\frac{5}{4}+\frac{k}{2} \right)_{i_{k-1}}}{\left(\frac{k}{2}+\frac{\alpha }{4}+\frac{3}{4} \right)_{i_{k-1}}\left(\frac{k}{2}-\frac{\alpha }{4}+\frac{1}{2} \right)_{i_{k-1}}\left(1+\frac{k}{2}\right)_{i_k}\left(\frac{5}{4}+ \frac{k}{2} \right)_{i_k}}\right\} \nonumber\\
&&\times \left. \left. \sum_{i_n= i_{n-1}}^{\infty } \frac{\left(\frac{n}{2}+\frac{\alpha }{4}+\frac{3}{4} \right)_{i_n}\left(\frac{n}{2}-\frac{\alpha }{4}+\frac{1}{2}\right)_{i_n}\left( 1+ \frac{n}{2} \right)_{i_{n-1}}\left(\frac{5}{4}+\frac{n}{2} \right)_{i_{n-1}}}{\left(\frac{n}{2}+\frac{\alpha }{4}+\frac{3}{4} \right)_{i_{n-1}}\left(\frac{n}{2}-\frac{\alpha }{4}+\frac{1}{2} \right)_{i_{n-1}}\left( 1+\frac{n}{2} \right)_{i_n}\left(\frac{5}{4}+ \frac{n}{2} \right)_{i_n}} z^{i_n} \right\} \eta ^n \right\} \label{eq:10028}
\end{eqnarray}
On (\ref{eq:10027}) and (\ref{eq:10028}), 
\begin{equation}
\begin{cases} 
\eta =(1+\rho ^2)\xi \cr
z=-\rho ^2\xi^2 \cr
\xi = sn^2(z,\rho ) \cr
\Gamma_0 = \frac{2 +\rho ^2}{2(1+\rho ^2)}  \cr
\Gamma_k = \frac{k}{2}+\frac{2 +\rho ^2}{2(1+\rho ^2)} \cr
Q= \frac{4+\rho ^2-h}{16(1+\rho ^2)}
\end{cases}\nonumber 
\end{equation}
\subsection{ ${\displaystyle  Hl(1-a,-q+\alpha \beta; \alpha,\beta, \delta, \gamma; 1-x)}$} 
\subsubsection{Polynomial of type 1}
Replace coefficients $a$, $q$, $\gamma $, $\delta$ and $x$ by $1-a$, $-q +\alpha \beta $, $\delta $, $\gamma $ and $1-x$ into (\ref{eq:10023}). Put (\ref{eq:1006}) into the new (\ref{eq:10023}) with replacing $\alpha $ by $-2(2\alpha _j+j+1/2)$ where $j,\alpha _j \in \mathbb{N}_{0}$; apply $\alpha =-2(2\alpha _0+1/2)$ into sub-power series $y_0(\varsigma)$, apply $-2(2\alpha _0+1/2)$ into the first summation and $-2(2\alpha _1 +3/2)$ into second summation of sub-power series $y_1(\varsigma)$, apply $-2(2\alpha _0+1/2)$ into the first summation, $-2(2\alpha _1+ 3/2)$ into the second summation and $-2(2\alpha _2 +5/2)$ into the third summation of sub-power series $y_2(\varsigma)$, etc in the new (\ref{eq:10023}).
\begin{eqnarray}
y(\varsigma )&=&  Hl\left( 1-\rho ^{-2}, \frac{1}{4}h\rho ^{-2} -4\left( \alpha _j+\frac{j}{2}\right)\left( \alpha _j+\frac{j}{2}+\frac{1}{4}\right) ; -2\alpha _j-j, -2\alpha _j-j, \frac{1}{2}, \frac{1}{2}; \varsigma \right)\nonumber\\
&=&  \sum_{i_0=0}^{\alpha _0} \frac{(-\alpha _0)_{i_0} \left(\alpha _0+\frac{1}{4} \right)_{i_0}}{(1 )_{i_0}\left(\frac{3}{4} \right)_{i_0}} z^{i_0}  \nonumber\\
&&+ \left\{\sum_{i_0=0}^{\alpha _0}\frac{ i_0^2+ Q_0}{\left(i_0+ \frac{1}{2} \right)\left( i_0 + \frac{1}{4}\right)} \frac{(-\alpha _0)_{i_0} \left(\alpha _0+\frac{1}{4}\right)_{i_0}}{(1 )_{i_0}\left(\frac{3}{4} \right)_{i_0}} \right. \left.\sum_{i_1=i_0}^{\alpha _1} \frac{(-\alpha _1)_{i_1}\left(\alpha _1+\frac{5}{4} \right)_{i_1}\left(\frac{3}{2} \right)_{i_0}\left( \frac{3}{4} \right)_{i_0}}{(-\alpha _1)_{i_0}\left(\alpha _1+\frac{5}{4} \right)_{i_0}\left(\frac{3}{2} \right)_{i_1}\left( \frac{3}{4} \right)_{i_1}} z^{i_1} \right\} \eta \nonumber\\
&&+ \sum_{n=2}^{\infty } \left\{ \sum_{i_0=0}^{\alpha _0} \frac{ i_0^2+ Q_0}{\left(i_0+ \frac{1}{2} \right)\left( i_0 + \frac{1}{4} \right)}  \frac{(-\alpha _0)_{i_0} \left(\alpha _0+ \frac{1}{4} \right)_{i_0}}{(1 )_{i_0}\left( \frac{3}{4} \right)_{i_0}}\right.\nonumber\\
&&\times \prod _{k=1}^{n-1} \left\{ \sum_{i_k=i_{k-1}}^{\alpha _k} \frac{\left( i_k+\frac{k}{2} \right)^2+ Q_k}{\left( i_k+ \frac{k}{2}+\frac{1}{2} \right)\left(i_k +\frac{k}{2}+\frac{1}{4} \right)}  \frac{(-\alpha _k)_{i_k}\left(\alpha _k+k+\frac{1}{4} \right)_{i_k}\left(1+ \frac{k}{2} \right)_{i_{k-1}}\left(\frac{3}{4}+\frac{k}{2} \right)_{i_{k-1}}}{(-\alpha _k)_{i_{k-1}}\left(\alpha _k+ k +\frac{1}{4} \right)_{i_{k-1}}\left( 1+\frac{k}{2} \right)_{i_k}\left(\frac{3}{4}+ \frac{k}{2} \right)_{i_k}}\right\} \nonumber\\
&&\times  \left.\sum_{i_n= i_{n-1}}^{\alpha _n} \frac{(-\alpha _n)_{i_n}\left(\alpha _n+n+\frac{1}{4} \right)_{i_n}\left( 1+ \frac{n}{2} \right)_{i_{n-1}}\left( \frac{3}{4} +\frac{n}{2}\right)_{i_{n-1}}}{(-\alpha _n)_{i_{n-1}}\left(\alpha _n+n+\frac{1}{4} \right)_{i_{n-1}}\left( 1+\frac{n}{2} \right)_{i_n}\left( \frac{3}{4} +\frac{n}{2} \right)_{i_n}} z^{i_n} \right\} \eta ^n \label{eq:10029}
\end{eqnarray}
where
\begin{equation}
\begin{cases}
\alpha = 2\left( 2\alpha _j +j \right) \;\mbox{or}\; -2\left( 2\alpha _j +j+\frac{1}{2}\right) \cr
Q_0 = \frac{1}{4(2-\rho ^{-2})}\left(\frac{1}{4}h\rho ^{-2} +4 \alpha _0 ^2\right)  \cr
Q_k = \frac{1}{4(2-\rho ^{-2})}\left(\frac{1}{4}h\rho ^{-2} +4\left( \alpha _k+\frac{k}{2}\right)^2\right)
\end{cases}\nonumber  
\end{equation}
For the minimum value of Lame equation for a polynomial which makes $B_n$ term terminated about $\varsigma =0 $, put $\alpha _0=\alpha _1=\alpha _2=\cdots=0$ in (\ref{eq:10029}).
\begin{eqnarray}
y(\varsigma ) &=&  Hl\left( 1-\rho ^{-2}, \frac{1}{4}h\rho ^{-2} - j\left( j+\frac{1}{2}\right) ; -j, -j, \frac{1}{2}, \frac{1}{2}; \varsigma \right) \nonumber\\
&=& \sum_{n=0}^{\infty } \frac{\prod_{k=0}^{n-1}\left( \frac{1}{4}h\rho ^{-2}+(3-\rho ^{-2})k^2\right)}{ n! \left( \frac{1}{2}\right)_n} \left( \frac{\eta }{2-\rho ^{-2}} \right)^n  = \cosh \left( \sqrt{\frac{h}{1-3\rho ^2}}\sinh ^{-1}\left( \sqrt{\frac{(1-3\rho ^2)\eta }{2\rho ^2-1}}\right) \right) \nonumber\\
&=& \cosh \left( \sqrt{\frac{h}{1-3\rho ^2}} \ln \left( \sqrt{\frac{(1-3\rho ^2)\eta }{2\rho ^2-1}}+\sqrt{\frac{(1-3\rho ^2)\eta }{2\rho ^2-1}+1}\right) \right) \nonumber
\end{eqnarray}  
It tells us that Lame polynomials in which makes $B_n$ term terminated, for fixed values of $\alpha $, require $\left| \frac{\eta }{2-\rho ^{-2}}\right|=\left|\frac{1-sn^2(z,\rho )}{1-\rho ^{-2}} \right| < 1$ for the convergence of the radius.

\subsubsection{Infinite series}
Replace coefficients $a$, $q$, $\gamma $, $\delta$ and $x$ by $1-a$, $-q +\alpha \beta $, $\delta $, $\gamma $ and $1-x$ into (\ref{eq:10024}). Put (\ref{eq:1006}) into the new (\ref{eq:10024}).
\begin{eqnarray}
y(\varsigma )&=&  Hl\left( 1-\rho ^{-2}, \frac{1}{4}\left( h\rho ^{-2}- \alpha (\alpha +1)\right); \frac{1}{2}(\alpha +1), -\frac{\alpha }{2}, \frac{1}{2}, \frac{1}{2}; \varsigma \right)\nonumber\\
&=& \sum_{i_0=0}^{\infty } \frac{\left(\frac{\alpha }{4} +\frac{1}{4}\right)_{i_0} \left(-\frac{\alpha }{4} \right)_{i_0}}{(1 )_{i_0}\left(\frac{3}{4} \right)_{i_0}} z^{i_0}  + \left\{\sum_{i_0=0}^{\infty }\frac{ i_0^2 + Q}{\left(i_0+ \frac{1}{2} \right)\left(i_0 + \frac{1}{4} \right)}\right. \left. \frac{\left(\frac{\alpha }{4}+\frac{1}{4} \right)_{i_0} \left(-\frac{\alpha }{4} \right)_{i_0}}{(1 )_{i_0}\left(\frac{3}{4} \right)_{i_0}} \sum_{i_1=i_0}^{\infty } \frac{\left(\frac{\alpha }{4}+\frac{3}{4}\right)_{i_1}\left(-\frac{\alpha }{4}+\frac{1}{2} \right)_{i_1}\left(\frac{3}{2} \right)_{i_0}\left( \frac{5}{4} \right)_{i_0}}{\left(\frac{\alpha }{4}+\frac{3}{4}\right)_{i_0}\left(-\frac{\alpha }{4}+\frac{1}{2} \right)_{i_0}\left(\frac{3}{2} \right)_{i_1}\left( \frac{5}{4} \right)_{i_1}} z^{i_1} \right\} \eta \nonumber\\
&&+ \sum_{n=2}^{\infty } \left\{ \sum_{i_0=0}^{\infty } \frac{ i_0^2 + Q}{\left( i_0+\frac{1}{2} \right)\left(i_0 + \frac{1}{4} \right)}
 \frac{\left(\frac{\alpha }{4}+\frac{1}{4} \right)_{i_0} \left(-\frac{\alpha }{4} \right)_{i_0}}{(1 )_{i_0}\left(\frac{3}{4} \right)_{i_0}}\right.\nonumber\\
&&\times \prod _{k=1}^{n-1} \left\{ \sum_{i_k=i_{k-1}}^{\infty } \frac{\left( i_k+\frac{k}{2} \right)^2 + Q}{\left( i_k+ \frac{k}{2}+\frac{1}{2} \right)\left(i_k +\frac{k}{2}+\frac{1}{4} \right)} \right.  \left.\frac{\left(\frac{k}{2}+\frac{\alpha }{4}+\frac{1}{4} \right)_{i_k}\left(\frac{k}{2}-\frac{\alpha }{4} \right)_{i_k}\left( 1+ \frac{k}{2} \right)_{i_{k-1}}\left(\frac{3}{4}+\frac{k}{2} \right)_{i_{k-1}}}{\left(\frac{k}{2}+\frac{\alpha }{4}+\frac{1}{4} \right)_{i_{k-1}}\left(\frac{k}{2}-\frac{\alpha }{4} \right)_{i_{k-1}}\left( 1+\frac{k}{2}\right)_{i_k}\left(\frac{3}{4}+ \frac{k}{2} \right)_{i_k}}\right\} \nonumber\\
&&\times \left. \sum_{i_n= i_{n-1}}^{\infty } \frac{\left(\frac{n}{2}+\frac{\alpha }{4}+\frac{1}{4} \right)_{i_n}\left(\frac{n}{2}-\frac{\alpha }{4} \right)_{i_n}\left( 1+ \frac{n}{2} \right)_{i_{n-1}}\left(\frac{3}{4}+\frac{n}{2} \right)_{i_{n-1}}}{\left(\frac{n}{2}+\frac{\alpha }{4}+\frac{1}{4} \right)_{i_{n-1}}\left(\frac{n}{2}-\frac{\alpha }{4} \right)_{i_{n-1}}\left( 1+\frac{n}{2} \right)_{i_n}\left(\frac{3}{4}+ \frac{n}{2} \right)_{i_n}} z^{i_n} \right\} \eta ^n  \label{eq:10030}
\end{eqnarray}
where
\begin{equation}
Q = \frac{h\rho ^{-2}-\alpha (\alpha +1)}{16(2-\rho ^{-2})} 
\nonumber  
\end{equation}
On (\ref{eq:10029}) and (\ref{eq:10030}),
\begin{equation}
\begin{cases}
\varsigma= 1-\xi \cr
\xi = sn^2(z,\rho ) \cr
\eta =\frac{2-\rho ^{-2}}{1-\rho ^{-2}}\varsigma \cr
z=\frac{-1}{1-\rho ^{-2}}\varsigma ^2 
\end{cases}\nonumber  
\end{equation}
\subsection{ ${\displaystyle (1-x)^{1-\delta } Hl(1-a,-q+(\delta -1)\gamma a+(\alpha -\delta +1)(\beta -\delta +1); \alpha-\delta +1,\beta-\delta +1}$\\${\displaystyle, 2-\delta, \gamma; 1-x)}$}
\subsubsection{Polynomial of type 1}
Replace coefficients $a$, $q$, $\alpha $, $\beta $, $\gamma $, $\delta$ and $x$ by $1-a$, $-q+(\delta -1)\gamma a+(\alpha -\delta +1)(\beta -\delta +1)$, $\alpha-\delta +1 $, $\beta-\delta +1 $, $2-\delta$, $\gamma $ and $1-x$ into (\ref{eq:10023}). Multiply $(1-x)^{1-\delta }$ and (\ref{eq:10023}) together. Put (\ref{eq:1006}) into the new (\ref{eq:10023}) with replacing $\alpha $ by $-2(2\alpha _j+j+1)$ where $j,\alpha _j \in \mathbb{N}_{0}$; apply $\alpha =-2(2\alpha _0+1)$ into sub-power series $y_0(\varsigma)$, apply $-2(2\alpha _0+1)$ into the first summation and $-2(2\alpha _1+2)$ into second summation of sub-power series $y_1(\varsigma)$, apply $-2(2\alpha _0+1)$ into the first summation, $-2(2\alpha _1+2)$ into the second summation and $-2(2\alpha _2+3)$ into the third summation of sub-power series $y_2(\varsigma)$, etc in the new (\ref{eq:10023}).
\begin{eqnarray}
&&\varsigma ^{\frac{1}{2}}y(\varsigma )\nonumber\\
&=& \varsigma ^{\frac{1}{2}} Hl\left( 1-\rho ^{-2}, -\frac{1}{4} (1-h)\rho ^{-2}-4\left( \alpha _j +\frac{j}{2}\right)\left( \alpha _j +\frac{j}{2}+\frac{3}{4}\right) ; -2\alpha _j-j, -2\alpha _j-j, \frac{3}{2}, \frac{1}{2}; \varsigma  \right)\nonumber\\
&=& \varsigma ^{\frac{1}{2}} \left\{ \sum_{i_0=0}^{\alpha _0} \frac{(-\alpha _0)_{i_0} \left( \alpha _0+\frac{3}{4} \right)_{i_0}}{(1)_{i_0}\left(\frac{5}{4} \right)_{i_0}} z^{i_0} \right. \nonumber\\
&&+ \left\{\sum_{i_0=0}^{\alpha _0}\frac{ i_0 \left( i_0+ \frac{1}{2} \right)+ Q_0}{\left(i_0+ \frac{1}{2} \right)\left( i_0 + \frac{3}{4}\right)} \frac{(-\alpha _0)_{i_0} \left(\alpha _0+\frac{3}{4}\right)_{i_0}}{(1 )_{i_0}\left(\frac{5}{4} \right)_{i_0}} \right. \left.\sum_{i_1=i_0}^{\alpha _1} \frac{(-\alpha _1)_{i_1}\left(\alpha _1+\frac{7}{4} \right)_{i_1}\left(\frac{3}{2} \right)_{i_0}\left( \frac{7}{4} \right)_{i_0}}{(-\alpha _1)_{i_0}\left(\alpha _1+\frac{7}{4} \right)_{i_0}\left(\frac{3}{2} \right)_{i_1}\left( \frac{7}{4} \right)_{i_1}} z^{i_1} \right\} \eta \nonumber\\
&&+ \sum_{n=2}^{\infty } \left\{ \sum_{i_0=0}^{\alpha _0} \frac{ i_0 \left( i_0+ \frac{1}{2} \right)+ Q_0}{\left(i_0+ \frac{1}{2} \right)\left( i_0 + \frac{3}{4} \right)}  \frac{(-\alpha _0)_{i_0} \left(\alpha _0+ \frac{3}{4} \right)_{i_0}}{(1 )_{i_0}\left( \frac{5}{4} \right)_{i_0}}\right.\nonumber\\
&&\times \prod _{k=1}^{n-1} \left\{ \sum_{i_k=i_{k-1}}^{\alpha _k} \frac{\left( i_k+\frac{k}{2} \right) \left( i_k+\frac{k}{2}+\frac{1}{2}\right)+ Q_k}{\left(i_k+ \frac{k}{2}+\frac{1}{2} \right)\left(i_k +\frac{k}{2}+\frac{3}{4} \right)}  \frac{(-\alpha _k)_{i_k}\left(\alpha _k+k+\frac{3}{4} \right)_{i_k}\left(1+ \frac{k}{2} \right)_{i_{k-1}}\left(\frac{5}{4}+\frac{k}{2} \right)_{i_{k-1}}}{(-\alpha _k)_{i_{k-1}}\left(\alpha _k+ k +\frac{3}{4} \right)_{i_{k-1}}\left(1+\frac{k}{2} \right)_{i_k}\left(\frac{5}{4}+ \frac{k}{2} \right)_{i_k}}\right\} \nonumber\\
&&\times \left. \left.\sum_{i_n= i_{n-1}}^{\alpha _n} \frac{(-\alpha _n)_{i_n}\left(\alpha _n+n+\frac{3}{4} \right)_{i_n}\left(1+ \frac{n}{2} \right)_{i_{n-1}}\left( \frac{5}{4} +\frac{n}{2}\right)_{i_{n-1}}}{(-\alpha _n)_{i_{n-1}}\left(\alpha _n+n+\frac{3}{4} \right)_{i_{n-1}}\left( 1+\frac{n}{2} \right)_{i_n}\left( \frac{5}{4} +\frac{n}{2} \right)_{i_n}} z^{i_n} \right\} \eta ^n\right\} \label{eq:10031}
\end{eqnarray}
where
\begin{equation}
\begin{cases} 
\alpha = 2\left( 2\alpha _j +j+\frac{1}{2} \right) \;\mbox{or}\; -2\left( 2\alpha _j +j+1\right) \cr
Q_0 = \frac{-\frac{1}{4}(1-h)\rho ^{-2}+4 \alpha _0^2 }{4(2-\rho ^{-2})} \cr
Q_k =  \frac{-\frac{1}{4}(1-h)\rho ^{-2}+4\left( \alpha _k +\frac{k}{2}\right)^2 }{4(2-\rho ^{-2})} 
\end{cases}\nonumber  
\end{equation}
For the minimum value of Lame equation for a polynomial which makes $B_n$ term terminated about $\varsigma =0 $, put $\alpha _0=\alpha _1=\alpha _2=\cdots=0$ in (\ref{eq:10031}).
\begin{eqnarray}
y(\varsigma ) &=&  Hl\left( 1-\rho ^{-2}, -\frac{1}{4} (1-h)\rho ^{-2}- j\left( j+\frac{3}{2}\right) ; -j, -j, \frac{3}{2}, \frac{1}{2}; \varsigma  \right) \nonumber\\
&=& \sum_{n=0}^{\infty } \frac{\prod_{k=0}^{n-1}\left( k(k+1)+\frac{\frac{1}{4}(h-1)\rho ^{-2}+k^2}{2-\rho ^{-2}}\right)}{ n! \left( \frac{3}{2}\right)_n}  \eta ^n  \nonumber\\
&=&  \; _2F_1\left(  \frac{\rho ^2-\frac{1}{2}}{3\rho ^2-1}- A(\rho ,h), \frac{\rho ^2-\frac{1}{2}}{3\rho ^2-1}+ A(\rho ,h), \frac{3}{2}; \frac{(3\rho ^2-1)\eta }{2\rho ^2-1} \right) \label{qq:1}
\end{eqnarray}  
where 
\begin{equation}
A(\rho ,h)= \frac{(\rho ^2-\frac{1}{2})\sqrt{4\rho ^4-(1+3h)\rho ^2 +h}}{|2\rho ^2-1| (3\rho ^2-1)} \nonumber
\end{equation}
(\ref{qq:1}) tells us that Lame polynomials in which makes $B_n$ term terminated, for fixed values of $\alpha $, require $\left| \frac{(3\rho ^2-1)\eta }{2\rho ^2-1}\right|=\left|\frac{3\rho ^2 -1}{ \rho ^2-1} (1-sn^2(z,\rho ))\right| < 1$ for the convergence of the radius.

For the special case, if $\frac{(3\rho ^2-1)\eta }{2\rho ^2-1}= \frac{3\rho ^2 -1}{ \rho ^2-1} (1-sn^2(z,\rho )) =1$ in (\ref{qq:1}),
\begin{eqnarray}
y(\varsigma ) &=&  Hl\left( 1-\rho ^{-2}, -\frac{1}{4} (1-h)\rho ^{-2}- j\left( j+\frac{3}{2}\right) ; -j, -j, \frac{3}{2}, \frac{1}{2}; \varsigma =\frac{\rho ^2-1}{3\rho ^2-1} \right) \nonumber\\
&=&  \; _2F_1\left(  \frac{\rho ^2-\frac{1}{2}}{3\rho ^2-1}- A(\rho ,h), \frac{\rho ^2-\frac{1}{2}}{3\rho ^2-1}+ A(\rho ,h), \frac{3}{2}; 1 \right) \nonumber\\
&=& \frac{\Gamma \left( \frac{3}{2}\right) \Gamma \left( \frac{3}{2}-\frac{2\rho ^2-1}{3\rho ^2-1}\right)}{\Gamma \left( \frac{3}{2}-\frac{ \rho ^2-\frac{1}{2}}{3\rho ^2-1}- \frac{ \sqrt{4\rho ^4-(1+3h)\rho ^2 +h}}{2(3\rho ^2-1)} \right) \Gamma \left( \frac{3}{2}-\frac{ \rho ^2-\frac{1}{2}}{3\rho ^2-1}+\frac{ \sqrt{4\rho ^4-(1+3h)\rho ^2 +h}}{2(3\rho ^2-1)} \right)}\hspace{1cm}\label{qq:2}
\end{eqnarray} 
If $0<\rho <1$ and $ z, sn(z,\rho )\in \mathbb{R}$ are satisfied for the solution in series of Lame equation,  $\mathbb{R}\left(\frac{3}{2}-\frac{2\rho ^2-1}{3\rho ^2-1} \right)>0$ should be required for the analytic solution of (\ref{qq:2}). 
According to these boundary conditions, we obtain the range of a parameter $\rho $ in (\ref{qq:2}) such as $\frac{1}{\sqrt{3}}<\rho <1$. 
\subsubsection{Infinite series}
Replace coefficients $a$, $q$, $\alpha $, $\beta $, $\gamma $, $\delta$ and $x$ by $1-a$, $-q+(\delta -1)\gamma a+(\alpha -\delta +1)(\beta -\delta +1)$, $\alpha-\delta +1 $, $\beta-\delta +1 $, $2-\delta$, $\gamma $ and $1-x$ into (\ref{eq:10024}). Multiply $(1-x)^{1-\delta }$ and (\ref{eq:10024}) together. Put (\ref{eq:1006}) into the new (\ref{eq:10024}).
\begin{eqnarray}
&&\varsigma ^{\frac{1}{2}}y(\varsigma )\nonumber\\
&=& \varsigma ^{\frac{1}{2}} Hl\left( 1-\rho ^{-2}, -\frac{1}{4}\left( (1-h)\rho ^{-2}+(\alpha -1)(\alpha +2)\right); \frac{\alpha }{2}+1, -\frac{\alpha }{2}+\frac{1}{2}, \frac{3}{2}, \frac{1}{2}; \varsigma \right)\nonumber\\
&=& \varsigma ^{\frac{1}{2}} \left\{ \sum_{i_0=0}^{\infty } \frac{\left(\frac{\alpha }{4} +\frac{1}{2}\right)_{i_0} \left(-\frac{\alpha }{4} +\frac{1}{4}\right)_{i_0}}{(1 )_{i_0}\left(\frac{5}{4} \right)_{i_0}} z^{i_0}\right. \nonumber\\
&&+ \left\{\sum_{i_0=0}^{\infty }\frac{ i_0 \left( i_0+ \frac{1}{2} \right)+ Q}{\left( i_0+ \frac{1}{2} \right)\left(i_0 + \frac{3 }{4} \right)}\right. \left. \frac{\left(\frac{\alpha }{4}+\frac{1}{2} \right)_{i_0} \left(-\frac{\alpha }{4}+\frac{1}{4} \right)_{i_0}}{(1 )_{i_0}\left(\frac{5}{4} \right)_{i_0}} \sum_{i_1=i_0}^{\infty } \frac{\left(\frac{\alpha }{4}+1\right)_{i_1}\left(-\frac{\alpha }{4}+\frac{3}{4} \right)_{i_1}\left(\frac{3}{2} \right)_{i_0}\left( \frac{7}{4} \right)_{i_0}}{\left(\frac{\alpha }{4}+1\right)_{i_0}\left(-\frac{\alpha }{4}+\frac{3}{4} \right)_{i_0}\left(\frac{3}{2} \right)_{i_1}\left( \frac{7}{4} \right)_{i_1}} z^{i_1} \right\} \eta \nonumber\\
&&+ \sum_{n=2}^{\infty } \left\{ \sum_{i_0=0}^{\infty } \frac{ i_0 \left( i_0+ \frac{1}{2} \right)+ Q}{\left(i_0+ \frac{1}{2} \right)\left(i_0 + \frac{3}{4} \right)}
 \frac{\left(\frac{\alpha }{4}+\frac{1}{2} \right)_{i_0} \left(-\frac{\alpha }{4}+\frac{1}{4} \right)_{i_0}}{(1 )_{i_0}\left(\frac{5}{4} \right)_{i_0}}\right.\nonumber\\
&&\times \prod _{k=1}^{n-1} \left\{ \sum_{i_k=i_{k-1}}^{\infty } \frac{\left( i_k+\frac{k}{2} \right) \left( i_k+ \frac{k}{2}+\frac{1}{2} \right)+ Q}{\left( i_k+ \frac{k}{2}+\frac{1}{2} \right)\left( i_k +\frac{k}{2}+\frac{3}{4} \right)} \right.  \left.\frac{\left(\frac{k}{2}+\frac{\alpha }{4}+\frac{1}{2} \right)_{i_k}\left(\frac{k}{2}-\frac{\alpha }{4}+\frac{1}{4}  \right)_{i_k}\left( 1+ \frac{k}{2} \right)_{i_{k-1}}\left(\frac{5}{4}+\frac{k}{2} \right)_{i_{k-1}}}{\left(\frac{k}{2}+\frac{\alpha }{4}+\frac{1}{2} \right)_{i_{k-1}}\left(\frac{k}{2}-\frac{\alpha }{4}+\frac{1}{4} \right)_{i_{k-1}}\left(1+\frac{k}{2}\right)_{i_k}\left(\frac{5}{4}+ \frac{k}{2} \right)_{i_k}}\right\} \nonumber\\
&&\times \left. \left. \sum_{i_n= i_{n-1}}^{\infty } \frac{\left(\frac{n}{2}+\frac{\alpha }{4}+\frac{1}{2} \right)_{i_n}\left(\frac{n}{2}-\frac{\alpha }{4}+\frac{1}{4}\right)_{i_n}\left( 1+ \frac{n}{2} \right)_{i_{n-1}}\left(\frac{5}{4}+\frac{n}{2} \right)_{i_{n-1}}}{\left(\frac{n}{2}+\frac{\alpha }{4}+\frac{1}{2} \right)_{i_{n-1}}\left(\frac{n}{2}-\frac{\alpha }{4}+\frac{1}{4} \right)_{i_{n-1}}\left( 1+\frac{n}{2} \right)_{i_n}\left(\frac{5}{4}+ \frac{n}{2} \right)_{i_n}} z^{i_n} \right\} \eta ^n \right\} \label{eq:10032}
\end{eqnarray}
where
\begin{equation}
Q= -\frac{ (1-h)\rho ^{-2}+(\alpha -1)(\alpha +2)}{16(2-\rho ^{-2})} 
\nonumber  
\end{equation}
On (\ref{eq:10031}) and (\ref{eq:10032}),
\begin{equation}
\begin{cases} 
\varsigma= 1-\xi \cr
\xi = sn^2(z,\rho ) \cr
\eta =\frac{2-\rho ^{-2}}{1-\rho ^{-2}}\varsigma \cr
z=\frac{-1}{1-\rho ^{-2}} \varsigma^2 
\end{cases}\nonumber  
\end{equation}
\subsection{ ${\displaystyle x^{-\alpha } Hl\left(\frac{1}{a},\frac{q+\alpha [(\alpha -\gamma -\delta +1)a-\beta +\delta ]}{a}; \alpha , \alpha -\gamma +1, \alpha -\beta +1,\delta ;\frac{1}{x}\right)}$}
\subsubsection{Infinite series}
Replace coefficients $a$, $q$, $\beta $, $\gamma $ and $x$ by $\frac{1}{a}$, $\frac{q+\alpha [(\alpha -\gamma -\delta +1)a-\beta +\delta ]}{a}$, $\alpha-\gamma +1 $, $\alpha -\beta +1 $ and $\frac{1}{x}$ into (\ref{eq:10024}). Multiply $x^{-\alpha }$ and (\ref{eq:10024}) together. Put (\ref{eq:1006}) into the new (\ref{eq:10024}).
\begin{eqnarray}
&&\varsigma ^{\frac{1}{2}(\alpha +1)} y(\varsigma )\nonumber\\
&=& \varsigma ^{\frac{1}{2}(\alpha +1)} Hl\left(\rho ^2,-\frac{1}{4}\left( h-(1+\rho ^2)(\alpha +1)^2\right); \frac{1}{2}(\alpha +1), \frac{1}{2}(\alpha +2),\alpha +\frac{3}{2}, \frac{1}{2}; \varsigma \right) \nonumber\\
&=& \varsigma ^{\frac{1}{2}(\alpha +1)} \left\{ \sum_{i_0=0}^{\infty } \frac{\left(\frac{\alpha }{4} +\frac{1}{4}\right)_{i_0} \left(-\frac{\alpha }{4} +\frac{1}{2}\right)_{i_0}}{(1 )_{i_0}\left(\frac{\alpha }{2} +\frac{5}{4} \right)_{i_0}} z^{i_0}\right. \nonumber\\
&&+ \left\{ \sum_{i_0=0}^{\infty }\frac{ i_0 \left( i_0+ \frac{\alpha }{2} + \frac{1}{2}\right)+ Q}{\left( i_0+ \frac{1}{2} \right)\left( i_0+ \frac{\alpha }{2} + \frac{3}{4} \right)}  \frac{\left(\frac{\alpha }{4}+\frac{1}{4} \right)_{i_0} \left( \frac{\alpha }{4}+\frac{1}{2} \right)_{i_0}}{(1)_{i_0}\left(\frac{\alpha }{2}+\frac{5}{4} \right)_{i_0}} \sum_{i_1=i_0}^{\infty } \frac{\left(\frac{\alpha }{4}+\frac{3}{4}\right)_{i_1}\left( \frac{\alpha }{4}+1\right)_{i_1}\left(\frac{3}{2} \right)_{i_0}\left( \frac{\alpha }{2}+\frac{7}{4} \right)_{i_0}}{\left(\frac{\alpha }{4}+\frac{3}{4}\right)_{i_0}\left( \frac{\alpha }{4}+1\right)_{i_0}\left(\frac{3}{2} \right)_{i_1}\left( \frac{\alpha }{2}+\frac{7}{4} \right)_{i_1}} z^{i_1} \right\} \eta \nonumber\\
&&+ \sum_{n=2}^{\infty } \left\{ \sum_{i_0=0}^{\infty } \frac{ i_0 \left( i_0+ \frac{\alpha }{2}+ \frac{1}{2} \right)+ Q}{\left(i_0+ \frac{1}{2} \right)\left(i_0 + \frac{\alpha }{2}+ \frac{3}{4} \right)}
 \frac{\left(\frac{\alpha }{4}+\frac{1}{4} \right)_{i_0} \left( \frac{\alpha }{4}+\frac{1}{2} \right)_{i_0}}{(1 )_{i_0}\left(\frac{\alpha }{2}+\frac{5}{4}\right)_{i_0}}\right.\nonumber\\
&&\times \prod _{k=1}^{n-1} \left\{ \sum_{i_k=i_{k-1}}^{\infty } \frac{\left( i_k+\frac{k}{2} \right) \left( i_k+ \frac{k}{2}+\frac{\alpha }{2}+\frac{1}{2} \right)+ Q}{\left( i_k+ \frac{k}{2}+\frac{1}{2} \right)\left( i_k +\frac{k}{2}+ \frac{\alpha }{2}+\frac{3}{4} \right)} \right.  \left.\frac{\left(\frac{k}{2}+\frac{\alpha }{4}+\frac{1}{4} \right)_{i_k}\left(\frac{k}{2}+\frac{\alpha }{4}+\frac{1}{2} \right)_{i_k}\left( 1+ \frac{k}{2} \right)_{i_{k-1}}\left(\frac{k}{2}+\frac{\alpha }{2}+\frac{5}{4} \right)_{i_{k-1}}}{\left(\frac{k}{2}+\frac{\alpha }{4}+\frac{1}{4} \right)_{i_{k-1}}\left(\frac{k}{2}+\frac{\alpha }{4}+\frac{1}{2} \right)_{i_{k-1}}\left(1+\frac{k}{2}\right)_{i_k}\left(\frac{k}{2}+\frac{\alpha }{2}+\frac{5}{4} \right)_{i_k}}\right\} \nonumber\\
&&\times \left. \left. \sum_{i_n= i_{n-1}}^{\infty } \frac{\left(\frac{n}{2}+\frac{\alpha }{4}+\frac{1}{4} \right)_{i_n}\left(\frac{n}{2}+\frac{\alpha }{4}+\frac{1}{2} \right)_{i_n}\left( 1+ \frac{n}{2} \right)_{i_{n-1}}\left(\frac{n}{2}+\frac{\alpha }{2}+\frac{5}{4} \right)_{i_{n-1}}}{\left(\frac{n}{2}+\frac{\alpha }{4}+\frac{1}{4} \right)_{i_{n-1}}\left(\frac{n}{2}+\frac{\alpha }{4}+\frac{1}{2} \right)_{i_{n-1}}\left(1+\frac{n}{2}\right)_{i_n}\left(\frac{n}{2}+\frac{\alpha }{2}+\frac{5}{4}\right)_{i_n}} z^{i_n} \right\} \eta ^n \right\} \label{eq:10033}
\end{eqnarray}
where
\begin{equation}
\begin{cases} 
\varsigma =\xi ^{-1} \cr
\xi = sn^2(z,\rho ) \cr
\eta =  (1+\rho ^{-2}) \varsigma \cr
z = -\rho ^{-2}\varsigma ^2 \cr
Q= -\frac{1}{16}\left( h(1+\rho ^2)^{-1}-(\alpha +1)^2\right)
\end{cases}\nonumber  
\end{equation}
\subsection{ ${\displaystyle \left(1-\frac{x}{a} \right)^{-\beta } Hl\left(1-a, -q+\gamma \beta; -\alpha +\gamma +\delta, \beta, \gamma, \delta; \frac{(1-a)x}{x-a} \right)}$}
\subsubsection{Infinite series}
Replace coefficients $a$, $q$, $\alpha $ and $x$ by $1-a$, $-q+\gamma \beta $, $-\alpha+\gamma +\delta $ and $\frac{(1-a)x}{x-a}$ into (\ref{eq:10024}). Multiply $\left(1-\frac{x}{a} \right)^{-\beta }$ and (\ref{eq:10024}) together. Put (\ref{eq:1006}) into the new (\ref{eq:10024}).
\begin{eqnarray}
&&(1-\rho ^2 \xi)^{\frac{\alpha }{2}} y(\varsigma )\nonumber\\
&=& (1-\rho ^2 \xi)^{\frac{\alpha }{2}} Hl\left(1-\rho ^{-2}, \frac{1}{4}\left( h\rho ^{-2} - \alpha \right); -\frac{\alpha }{2}+\frac{1}{2}, -\frac{\alpha }{2}, \frac{1}{2}, \frac{1}{2}; \varsigma \right) \nonumber\\
&=& (1-\rho ^2 \xi)^{\frac{\alpha }{2}} \left\{ \sum_{i_0=0}^{\infty } \frac{\left(-\frac{\alpha }{4} +\frac{1}{4}\right)_{i_0} \left(-\frac{\alpha }{4} \right)_{i_0}}{(1 )_{i_0}\left( \frac{3}{4} \right)_{i_0}} z^{i_0}\right. \nonumber\\
&&+ \left\{ \sum_{i_0=0}^{\infty }\frac{ i_0 \left( i_0+ \Gamma _0\right)+ Q}{\left( i_0+ \frac{1}{2} \right)\left( i_0 + \frac{1}{4} \right)}  \frac{\left(-\frac{\alpha }{4}+\frac{1}{4} \right)_{i_0} \left( -\frac{\alpha }{4} \right)_{i_0}}{(1)_{i_0}\left( \frac{3}{4} \right)_{i_0}} \sum_{i_1=i_0}^{\infty } \frac{\left(-\frac{\alpha }{4}+\frac{3}{4}\right)_{i_1}\left( -\frac{\alpha }{4}+\frac{1}{2}\right)_{i_1}\left(\frac{3}{2} \right)_{i_0}\left( \frac{5}{4} \right)_{i_0}}{\left(-\frac{\alpha }{4}+\frac{3}{4}\right)_{i_0}\left( -\frac{\alpha }{4}+\frac{1}{2}\right)_{i_0}\left(\frac{3}{2} \right)_{i_1}\left( \frac{5}{4} \right)_{i_1}} z^{i_1} \right\} \eta \nonumber\\
&&+ \sum_{n=2}^{\infty } \left\{ \sum_{i_0=0}^{\infty } \frac{ i_0 \left( i_0+ \Gamma _0 \right)+ Q}{\left(i_0+ \frac{1}{2} \right)\left(i_0 + \frac{1}{4} \right)}
 \frac{\left(-\frac{\alpha }{4}+\frac{1}{4} \right)_{i_0} \left( -\frac{\alpha }{4} \right)_{i_0}}{(1 )_{i_0}\left( \frac{3}{4}\right)_{i_0}}\right.\nonumber\\
&&\times \prod _{k=1}^{n-1} \left\{ \sum_{i_k=i_{k-1}}^{\infty } \frac{\left( i_k+\frac{k}{2} \right) \left( i_k+ \Gamma _k \right)+ Q}{\left( i_k+ \frac{k}{2}+\frac{1}{2} \right)\left( i_k +\frac{k}{2} +\frac{1}{4} \right)} \right.  \left.\frac{\left(\frac{k}{2}-\frac{\alpha }{4}+\frac{1}{4} \right)_{i_k}\left(\frac{k}{2}-\frac{\alpha }{4} \right)_{i_k}\left( 1+ \frac{k}{2} \right)_{i_{k-1}}\left( \frac{3}{4} +\frac{k}{2} \right)_{i_{k-1}}}{\left(\frac{k}{2}-\frac{\alpha }{4}+\frac{1}{4} \right)_{i_{k-1}}\left(\frac{k}{2}-\frac{\alpha }{4} \right)_{i_{k-1}}\left( 1+\frac{k}{2}\right)_{i_k}\left( \frac{3}{4} +\frac{k}{2} \right)_{i_k}}\right\} \nonumber\\
&&\times \left. \left. \sum_{i_n= i_{n-1}}^{\infty } \frac{\left(\frac{n}{2}-\frac{\alpha }{4}+\frac{1}{4} \right)_{i_n}\left(\frac{n}{2}-\frac{\alpha }{4} \right)_{i_n}\left( 1+ \frac{n}{2} \right)_{i_{n-1}}\left( \frac{3}{4} +\frac{n}{2} \right)_{i_{n-1}}}{\left(\frac{n}{2}-\frac{\alpha }{4}+\frac{1}{4} \right)_{i_{n-1}}\left(\frac{n}{2}-\frac{\alpha }{4} \right)_{i_{n-1}}\left(1+\frac{n}{2}\right)_{i_n}\left( \frac{3}{4} +\frac{n}{2}\right)_{i_n}} z^{i_n} \right\} \eta ^n \right\} \label{eq:10034}
\end{eqnarray}
where
\begin{equation}
\begin{cases} 
\varsigma =\frac{(1-\rho ^{-2})\xi }{\xi -\rho ^{-2}} \cr
\xi = sn^2(z,\rho ) \cr
\eta =  \frac{2-\rho ^{-2}}{1-\rho ^{-2}} \varsigma \cr
z = \frac{-1}{1-\rho ^{-2}}\varsigma ^2 \cr
\Gamma _0 = -\frac{\alpha }{2(2-\rho ^{-2})}\cr
\Gamma _k = \frac{k}{2} -\frac{\alpha }{2(2-\rho ^{-2})}\cr
Q=  \frac{h\rho ^{-2}-\alpha }{16(2-\rho ^{-2})} 
\end{cases}\nonumber  
\end{equation}
\subsection{ ${\displaystyle (1-x)^{1-\delta }\left(1-\frac{x}{a} \right)^{-\beta+\delta -1} Hl\Bigg(1-a, -q+\gamma [(\delta -1)a+\beta -\delta +1]; -\alpha +\gamma +1}$\\ ${\displaystyle, \beta -\delta+1, \gamma, 2-\delta; \frac{(1-a)x}{x-a} \Bigg)}$ }
\subsubsection{Infinite series}
Replace coefficients $a$, $q$, $\alpha $, $\beta $, $\delta $ and $x$ by $1-a$, $-q+\gamma [(\delta -1)a+\beta -\delta +1]$, $-\alpha +\gamma +1$, $\beta -\delta+1$, $2-\delta $ and $\frac{(1-a)x}{x-a}$ into (\ref{eq:10024}). Multiply $(1-x)^{1-\delta }\left(1-\frac{x}{a} \right)^{-\beta+\delta -1}$ and (\ref{eq:10024}) together. Put (\ref{eq:1006}) into the new (\ref{eq:10024}).
\begin{eqnarray}
&&(1-\xi )^{\frac{1}{2}}(1-\rho ^2 \xi)^{\frac{1}{2}(\alpha -1)} y(\varsigma )\nonumber\\
&=& (1-\xi )^{\frac{1}{2}}(1-\rho ^2 \xi)^{\frac{1}{2}(\alpha -1)} Hl\left( 1-\rho ^{-2}, \frac{1}{4}\left( (h-1)\rho ^{-2} +1- \alpha \right); -\frac{\alpha }{2}+1, -\frac{\alpha }{2}+\frac{1}{2}, \frac{1}{2}, \frac{3}{2}; \varsigma \right) \nonumber\\
&=& (1-\xi )^{\frac{1}{2}}(1-\rho ^2 \xi)^{\frac{1}{2}(\alpha -1)} \left\{ \sum_{i_0=0}^{\infty } \frac{\left(-\frac{\alpha }{4} +\frac{1}{2}\right)_{i_0} \left(-\frac{\alpha }{4} +\frac{1}{4}\right)_{i_0}}{(1 )_{i_0}\left( \frac{3}{4} \right)_{i_0}} z^{i_0}\right. \nonumber\\
&&+ \left\{ \sum_{i_0=0}^{\infty }\frac{ i_0 \left( i_0+ \Gamma _0\right)+ Q}{\left( i_0+ \frac{1}{2} \right)\left( i_0 + \frac{1}{4} \right)}  \frac{\left(-\frac{\alpha }{4}+\frac{1}{2} \right)_{i_0} \left( -\frac{\alpha }{4}+ \frac{1}{4} \right)_{i_0}}{(1)_{i_0}\left( \frac{3}{4} \right)_{i_0}} \sum_{i_1=i_0}^{\infty } \frac{\left(-\frac{\alpha }{4}+1\right)_{i_1}\left( -\frac{\alpha }{4}+\frac{3}{4}\right)_{i_1}\left(\frac{3}{2} \right)_{i_0}\left( \frac{5}{4} \right)_{i_0}}{\left(-\frac{\alpha }{4}+1\right)_{i_0}\left( -\frac{\alpha }{4}+\frac{3}{4}\right)_{i_0}\left(\frac{3}{2} \right)_{i_1}\left( \frac{5}{4} \right)_{i_1}} z^{i_1} \right\} \eta \nonumber\\
&&+ \sum_{n=2}^{\infty } \left\{ \sum_{i_0=0}^{\infty } \frac{ i_0 \left( i_0+ \Gamma _0 \right)+ Q}{\left(i_0+ \frac{1}{2} \right)\left(i_0 + \frac{1}{4} \right)}
 \frac{\left(-\frac{\alpha }{4}+\frac{1}{2} \right)_{i_0} \left( -\frac{\alpha }{4}+\frac{1}{4} \right)_{i_0}}{(1)_{i_0}\left( \frac{3}{4}\right)_{i_0}}\right.\nonumber\\
&&\times \prod _{k=1}^{n-1} \left\{ \sum_{i_k=i_{k-1}}^{\infty } \frac{\left( i_k+\frac{k}{2} \right) \left( i_k+ \Gamma _k \right)+ Q}{\left( i_k+ \frac{k}{2}+\frac{1}{2} \right)\left( i_k +\frac{k}{2} +\frac{1}{4} \right)} \right.  \left.\frac{\left(\frac{k}{2}-\frac{\alpha }{4}+\frac{1}{2} \right)_{i_k}\left(\frac{k}{2}-\frac{\alpha }{4} +\frac{1}{4}\right)_{i_k}\left( 1+ \frac{k}{2} \right)_{i_{k-1}}\left( \frac{3}{4} +\frac{k}{2} \right)_{i_{k-1}}}{\left(\frac{k}{2}-\frac{\alpha }{4}+\frac{1}{2} \right)_{i_{k-1}}\left(\frac{k}{2}-\frac{\alpha }{4} +\frac{1}{4}\right)_{i_{k-1}}\left( 1+\frac{k}{2}\right)_{i_k}\left( \frac{3}{4} +\frac{k}{2} \right)_{i_k}}\right\} \nonumber\\
&&\times \left. \left. \sum_{i_n= i_{n-1}}^{\infty } \frac{\left(\frac{n}{2}-\frac{\alpha }{4}+\frac{1}{2} \right)_{i_n}\left(\frac{n}{2}-\frac{\alpha }{4} +\frac{1}{4}\right)_{i_n}\left( 1+ \frac{n}{2} \right)_{i_{n-1}}\left( \frac{3}{4} +\frac{n}{2} \right)_{i_{n-1}}}{\left(\frac{n}{2}-\frac{\alpha }{4}+\frac{1}{2} \right)_{i_{n-1}}\left(\frac{n}{2}-\frac{\alpha }{4} +\frac{1}{4}\right)_{i_{n-1}}\left( 1+\frac{n}{2}\right)_{i_n}\left( \frac{3}{4} +\frac{n}{2}\right)_{i_n}} z^{i_n} \right\} \eta ^n \right\} \label{eq:10035}
\end{eqnarray}
where
\begin{equation}
\begin{cases} 
\varsigma =\frac{(1-\rho ^{-2})\xi }{\xi -\rho ^{-2}} \cr
\xi = sn^2(z,\rho ) \cr
\eta =  \frac{2-\rho ^{-2}}{1-\rho ^{-2}} \varsigma \cr
z = \frac{-1}{1-\rho ^{-2}}\varsigma ^2 \cr
\Gamma _0 = -\frac{\alpha -1+\rho ^{-2}}{2(2-\rho ^{-2})}\cr
\Gamma _k = \frac{k}{2} -\frac{\alpha -1+\rho ^{-2}}{2(2-\rho ^{-2})}\cr
Q=  \frac{ (h-1)\rho ^{-2}+1-\alpha}{16(2-\rho ^{-2})} 
\end{cases}\nonumber  
\end{equation}
\subsection{ ${\displaystyle x^{-\alpha } Hl\left(\frac{a-1}{a}, \frac{-q+\alpha (\delta a+\beta -\delta )}{a}; \alpha, \alpha -\gamma +1, \delta , \alpha -\beta +1; \frac{x-1}{x} \right)}$}
\subsubsection{Infinite series}
Replace coefficients $a$, $q$, $\beta $, $\gamma $, $\delta $ and $x$ by $\frac{a-1}{a}$, $\frac{-q+\alpha (\delta a+\beta -\delta )}{a}$, $\alpha -\gamma +1$, $\delta $, $\alpha -\beta +1$ and $\frac{x-1}{x}$ into (\ref{eq:10024}). Multiply $x^{-\alpha }$ and (\ref{eq:10024}) together.
Put (\ref{eq:1006}) into the new (\ref{eq:10024}).
\begin{eqnarray}
&&\xi ^{-\frac{1}{2}(\alpha +1)} y(\varsigma )\nonumber\\
&=& \xi ^{-\frac{1}{2}(\alpha +1)} Hl\left( 1-\rho ^2, \frac{1}{4}\left[ h+ (\alpha +1)\left( 1-(\alpha +1)\rho ^2\right) \right];  \frac{1}{2}(\alpha +1), \frac{1}{2}(\alpha +2), \frac{1}{2}, \alpha +\frac{3}{2}; \varsigma \right) \nonumber\\
&=& \xi ^{-\frac{1}{2}(\alpha +1)} \left\{ \sum_{i_0=0}^{\infty } \frac{\left( \frac{\alpha }{4} +\frac{1}{4}\right)_{i_0} \left( \frac{\alpha }{4} +\frac{1}{2}\right)_{i_0}}{(1)_{i_0}\left( \frac{3}{4} \right)_{i_0}} z^{i_0}\right. \nonumber\\
&&+ \left\{ \sum_{i_0=0}^{\infty }\frac{ i_0 \left( i_0+ \Gamma _0\right)+ Q}{\left( i_0+ \frac{1}{2} \right)\left( i_0 + \frac{1}{4} \right)}  \frac{\left( \frac{\alpha }{4}+\frac{1}{4} \right)_{i_0} \left( \frac{\alpha }{4}+ \frac{1}{2} \right)_{i_0}}{(1)_{i_0}\left( \frac{3}{4} \right)_{i_0}} \sum_{i_1=i_0}^{\infty } \frac{\left( \frac{\alpha }{4}+\frac{3}{4}\right)_{i_1}\left( \frac{\alpha }{4}+1\right)_{i_1}\left(\frac{3}{2} \right)_{i_0}\left( \frac{5}{4} \right)_{i_0}}{\left( \frac{\alpha }{4}+\frac{3}{4}\right)_{i_0}\left( \frac{\alpha }{4}+1\right)_{i_0}\left(\frac{3}{2} \right)_{i_1}\left( \frac{5}{4} \right)_{i_1}} z^{i_1} \right\} \eta \nonumber\\
&&+ \sum_{n=2}^{\infty } \left\{ \sum_{i_0=0}^{\infty } \frac{ i_0 \left( i_0+ \Gamma _0 \right)+ Q}{\left( i_0+ \frac{1}{2} \right)\left( i_0 + \frac{1}{4} \right)}
 \frac{\left( \frac{\alpha }{4}+\frac{1}{4} \right)_{i_0} \left( \frac{\alpha }{4}+\frac{1}{2} \right)_{i_0}}{(1)_{i_0}\left( \frac{3}{4}\right)_{i_0}}\right.\nonumber\\
&&\times \prod _{k=1}^{n-1} \left\{ \sum_{i_k=i_{k-1}}^{\infty } \frac{\left( i_k+\frac{k}{2} \right) \left( i_k+ \Gamma _k \right)+ Q}{\left( i_k+ \frac{k}{2}+\frac{1}{2} \right)\left( i_k +\frac{k}{2} +\frac{1}{4} \right)} \right.  \left.\frac{\left(\frac{k}{2}+\frac{\alpha }{4}+\frac{1}{4} \right)_{i_k}\left(\frac{k}{2}+\frac{\alpha }{4} +\frac{1}{2}\right)_{i_k}\left( 1+ \frac{k}{2} \right)_{i_{k-1}}\left( \frac{3}{4} +\frac{k}{2} \right)_{i_{k-1}}}{\left(\frac{k}{2}+\frac{\alpha }{4}+\frac{1}{4} \right)_{i_{k-1}}\left(\frac{k}{2}+\frac{\alpha }{4} +\frac{1}{2}\right)_{i_{k-1}}\left( 1+\frac{k}{2}\right)_{i_k}\left( \frac{3}{4} +\frac{k}{2} \right)_{i_k}}\right\} \nonumber\\
&&\times \left. \left. \sum_{i_n= i_{n-1}}^{\infty } \frac{\left(\frac{n}{2}+\frac{\alpha }{4}+\frac{1}{4} \right)_{i_n}\left(\frac{n}{2}+\frac{\alpha }{4} +\frac{1}{2}\right)_{i_n}\left( 1+ \frac{n}{2} \right)_{i_{n-1}}\left( \frac{3}{4} +\frac{n}{2} \right)_{i_{n-1}}}{\left(\frac{n}{2}+\frac{\alpha }{4}+\frac{1}{4} \right)_{i_{n-1}}\left(\frac{n}{2}+\frac{\alpha }{4} +\frac{1}{2}\right)_{i_{n-1}}\left( 1+\frac{n}{2}\right)_{i_n}\left( \frac{3}{4} +\frac{n}{2}\right)_{i_n}} z^{i_n} \right\} \eta ^n \right\} \label{eq:10036}
\end{eqnarray}
where
\begin{equation}
\begin{cases} 
\varsigma =\frac{ \xi -1}{\xi } \cr
\xi = sn^2(z,\rho ) \cr
\eta =  \frac{2-\rho ^2}{1-\rho ^2} \varsigma \cr
z = \frac{-1}{1-\rho ^2}\varsigma ^2 \cr
\Gamma _0 =  \frac{ (1-\rho ^2)(\alpha +1)}{2(2-\rho ^2)}\cr
\Gamma _k = \frac{k}{2} +\frac{ (1-\rho ^2)(\alpha +1)}{2(2-\rho ^2)}\cr
Q=  \frac{h+(\alpha +1)(1-(\alpha +1)\rho ^2)}{16(2-\rho ^2)} 
\end{cases}\nonumber  
\end{equation}
\subsection{ ${\displaystyle \left(\frac{x-a}{1-a} \right)^{-\alpha } Hl\left(a, q-(\beta -\delta )\alpha ; \alpha , -\beta+\gamma +\delta , \delta , \gamma; \frac{a(x-1)}{x-a} \right)}$}
\subsubsection{Infinite series}
Replace coefficients $q$, $\beta $, $\gamma $, $\delta $ and $x$  by $q-(\beta -\delta )\alpha $, $-\beta+\gamma +\delta $, $\delta $,  $\gamma $ and $\frac{a(x-1)}{x-a}$ into (\ref{eq:10024}). Multiply $\left(\frac{x-a}{1-a} \right)^{-\alpha }$ and (\ref{eq:10024}) together. Put (\ref{eq:1006}) into the new (\ref{eq:10024}).
\begin{eqnarray}
&&\left(\frac{\xi-\rho ^{-2}}{1-\rho ^{-2}} \right)^{-\frac{1}{2}(\alpha +1)} y(\varsigma )\nonumber\\
&=& \left(\frac{\xi-\rho ^{-2}}{1-\rho ^{-2}} \right)^{-\frac{1}{2}(\alpha +1)} Hl\left( \rho ^{-2}, -\frac{1}{4}\left( h\rho ^{-2}- (\alpha +1)^2\right);  \frac{1}{2}(\alpha +1), -\frac{1}{2}(\alpha -2), \frac{1}{2}, \frac{1}{2}; \varsigma \right) \nonumber\\
&=& \left(\frac{\xi-\rho ^{-2}}{1-\rho ^{-2}} \right)^{-\frac{1}{2}(\alpha +1)} \left\{ \sum_{i_0=0}^{\infty } \frac{\left( \frac{\alpha }{4} +\frac{1}{4}\right)_{i_0} \left( \frac{\alpha }{4} +\frac{1}{2}\right)_{i_0}}{(1)_{i_0}\left( \frac{3}{4} \right)_{i_0}} z^{i_0}\right. \nonumber\\
&&+ \left\{ \sum_{i_0=0}^{\infty }\frac{ i_0 \left( i_0+ \Gamma _0\right)+ Q}{\left( i_0+ \frac{1}{2} \right)\left( i_0 + \frac{1}{4} \right)}  \frac{\left( \frac{\alpha }{4}+\frac{1}{4} \right)_{i_0} \left( \frac{\alpha }{4}+ \frac{1}{2} \right)_{i_0}}{(1)_{i_0}\left( \frac{3}{4} \right)_{i_0}} \sum_{i_1=i_0}^{\infty } \frac{\left( \frac{\alpha }{4}+\frac{3}{4}\right)_{i_1}\left( \frac{\alpha }{4}+1\right)_{i_1}\left(\frac{3}{2} \right)_{i_0}\left( \frac{5}{4} \right)_{i_0}}{\left( \frac{\alpha }{4}+\frac{3}{4}\right)_{i_0}\left( \frac{\alpha }{4}+1\right)_{i_0}\left(\frac{3}{2} \right)_{i_1}\left( \frac{5}{4} \right)_{i_1}} z^{i_1} \right\} \eta \nonumber\\
&&+ \sum_{n=2}^{\infty } \left\{ \sum_{i_0=0}^{\infty } \frac{ i_0 \left( i_0+ \Gamma _0 \right)+ Q}{\left( i_0+ \frac{1}{2} \right)\left( i_0 + \frac{1}{4} \right)}
 \frac{\left( \frac{\alpha }{4}+\frac{1}{4} \right)_{i_0} \left( \frac{\alpha }{4}+\frac{1}{2} \right)_{i_0}}{(1)_{i_0}\left( \frac{3}{4}\right)_{i_0}}\right.\nonumber\\
&&\times \prod _{k=1}^{n-1} \left\{ \sum_{i_k=i_{k-1}}^{\infty } \frac{\left( i_k+\frac{k}{2} \right) \left( i_k+ \Gamma _k \right)+ Q}{\left( i_k+ \frac{k}{2}+\frac{1}{2} \right)\left( i_k +\frac{k}{2} +\frac{1}{4} \right)} \right.  \left.\frac{\left(\frac{k}{2}+\frac{\alpha }{4}+\frac{1}{4} \right)_{i_k}\left(\frac{k}{2}+\frac{\alpha }{4} +\frac{1}{2}\right)_{i_k}\left( 1+ \frac{k}{2} \right)_{i_{k-1}}\left( \frac{3}{4} +\frac{k}{2} \right)_{i_{k-1}}}{\left(\frac{k}{2}+\frac{\alpha }{4}+\frac{1}{4} \right)_{i_{k-1}}\left(\frac{k}{2}+\frac{\alpha }{4} +\frac{1}{2}\right)_{i_{k-1}}\left( 1+\frac{k}{2}\right)_{i_k}\left( \frac{3}{4} +\frac{k}{2} \right)_{i_k}}\right\} \nonumber\\
&&\times \left. \left. \sum_{i_n= i_{n-1}}^{\infty } \frac{\left(\frac{n}{2}+\frac{\alpha }{4}+\frac{1}{4} \right)_{i_n}\left(\frac{n}{2}+\frac{\alpha }{4} +\frac{1}{2}\right)_{i_n}\left( 1+ \frac{n}{2} \right)_{i_{n-1}}\left( \frac{3}{4} +\frac{n}{2} \right)_{i_{n-1}}}{\left(\frac{n}{2}+\frac{\alpha }{4}+\frac{1}{4} \right)_{i_{n-1}}\left(\frac{n}{2}+\frac{\alpha }{4} +\frac{1}{2}\right)_{i_{n-1}}\left( 1+\frac{n}{2}\right)_{i_n}\left( \frac{3}{4} +\frac{n}{2}\right)_{i_n}} z^{i_n} \right\} \eta ^n \right\} \label{eq:10037}
\end{eqnarray}
where
\begin{equation}
\begin{cases}
\varsigma =\frac{\xi -1}{\rho ^2(\xi-\rho ^{-2})} \cr
\xi = sn^2(z,\rho ) \cr
\eta =  (1+\rho ^2) \varsigma \cr
z = -\rho ^2 \varsigma ^2 \cr
\Gamma _0 =  \frac{\alpha +1}{2(1+\rho ^{-2})} \cr
\Gamma _k = \frac{k}{2} +\frac{\alpha +1}{2(1+\rho ^{-2})} \cr
Q=  \frac{-h\rho ^{-2}+(\alpha +1)^2}{16(1+\rho ^{-2})} 
\end{cases}\nonumber  
\end{equation}
\section{Asymptotic behavior}
In Ref.\cite{Chou2012d}, an asymptotic representation of Heun function about $x=0$ for an infinite series is given by
\begin{equation}
\lim_{n\gg 1} Hl\left( a, q; \alpha, \beta, \gamma, \delta; x\right) = \frac{1}{1-\left(-\frac{1}{a}x^2 +\frac{1+a}{a}x\right)}  \label{eq:10038}
\end{equation}
(\ref{eq:10038}) is geometric series. The condition of convergence of (\ref{eq:10038}) is
\begin{equation}
\left|-\frac{1}{a}x^2 \right| +\left| \frac{1+a}{a}x \right|<1 \label{eq:10039}
\end{equation}
\subsection{ ${\displaystyle (1-x)^{1-\delta } Hl(a, q - (\delta  - 1)\gamma a; \alpha - \delta  + 1, \beta - \delta + 1, \gamma ,2 - \delta ; x)}$ }
Replace coefficients $q$, $\alpha$, $\beta$ and $\delta$ by $q - (\delta - 1)\gamma a $, $\alpha - \delta  + 1 $, $\beta - \delta + 1$ and $2 - \delta$ into (\ref{eq:10038}) and (\ref{eq:10039}). Put (\ref{eq:1006}) into the new (\ref{eq:10038}) and (\ref{eq:10039}).

For an infinite series,
\begin{eqnarray}
&&\lim_{n\gg 1} Hl\left(\rho ^{-2}, -\frac{1}{4}(h-1)\rho ^{-2}; \frac{\alpha }{2}+1, -\frac{\alpha }{2}+\frac{1}{2}, \frac{1}{2},\frac{3}{2};  sn^2(z,\rho ) \right) \nonumber\\
&&= \frac{1}{1+ \rho ^2 sn^4(z,\rho )-(1+\rho ^2) sn^2(z,\rho ) } \label{eq:10044}
\end{eqnarray}
The condition of convergence of (\ref{eq:10044}) is
\begin{equation}
\left| \rho ^2 sn^4(z,\rho ) \right| +\left| (1+\rho ^2) sn^2(z,\rho ) \right|<1  \label{eq:10045}
\end{equation}
For the case of $ z, sn(z,\rho )\in \mathbb{R}$ where $ 0< \rho < 1 $, the boundary condition of $sn^2(z,\rho )$ in (\ref{eq:10045}) is given by
\begin{equation}
0 \leq sn^2(z,\rho )< \frac{-(1+\rho ^2)+\sqrt{\rho ^4+6\rho ^2+1}}{2 \rho ^2}\nonumber
\end{equation}
 In the case of $\rho \approx 0$ assuming $\rho $ is approximately close to $0$, (\ref{eq:10044}) turns to be
\begin{eqnarray}
&&\lim_{\substack{n\gg 1\\ \rho \approx 0}} Hl\left(\rho ^{-2}, -\frac{1}{4}(h-1)\rho ^{-2}; \frac{\alpha }{2}+1, -\frac{\alpha }{2}+\frac{1}{2}, \frac{1}{2},\frac{3}{2}; \xi  \right) \approx  \frac{1}{1 - \sin^2 z}\nonumber
\end{eqnarray}
where  $\xi \approx \sin^2 z$ and $\left| \sin^2 z \right|<1$. If $ z \in \mathbb{R}$, its radius of convergence is $ 0\leq\sin^2 z <1$.
\subsection{ ${\displaystyle x^{1-\gamma } (1-x)^{1-\delta } Hl(a, q-(\gamma +\delta -2)a-(\gamma -1)(\alpha +\beta -\gamma -\delta +1); \alpha - \gamma -\delta +2}$ \\${\displaystyle, \beta - \gamma -\delta +2, 2-\gamma, 2 - \delta ; x)}$}
Replace coefficients $q$, $\alpha$, $\beta$, $\gamma $ and $\delta$ by $q-(\gamma +\delta -2)a-(\gamma -1)(\alpha +\beta -\gamma -\delta +1)$, $\alpha - \gamma -\delta +2$, $\beta - \gamma -\delta +2, 2-\gamma$ and $2 - \delta$ into (\ref{eq:10038}) and (\ref{eq:10039}). Put (\ref{eq:1006}) into the new  (\ref{eq:10038}) and (\ref{eq:10039}).

For an infinite series,
\begin{eqnarray}
&&\lim_{n\gg 1} Hl\left(\rho ^{-2}, -\frac{1}{4}\left( (h-4)\rho ^{-2}-1\right); \frac{\alpha }{2}+\frac{3}{2}, -\frac{\alpha }{2} +1, \frac{3}{2},\frac{3}{2}; sn^2(z,\rho ) \right) \nonumber\\
&&= \frac{1}{1+ \rho ^2 sn^4(z,\rho )-(1+\rho ^2) sn^2(z,\rho )} \label{eq:10048}
\end{eqnarray}
The condition of convergence of (\ref{eq:10048}) is
\begin{equation}
\left| \rho ^2 sn^4(z,\rho ) \right| +\left| (1+\rho ^2) sn^2(z,\rho ) \right|<1 \label{eq:10049}
\end{equation}
For the case of $ z, sn(z,\rho )\in \mathbb{R}$ where $ 0< \rho < 1 $, the boundary condition of $sn^2(z,\rho )$ in (\ref{eq:10049}) is given by
\begin{equation}
0 \leq sn^2(z,\rho )< \frac{-(1+\rho ^2)+\sqrt{\rho ^4+6\rho ^2+1}}{2 \rho ^2}\nonumber
\end{equation}
For $\rho \approx 0$, (\ref{eq:10048}) turns to be
\begin{equation}
\lim_{\substack{n\gg 1\\ \rho \approx 0}} Hl\left(\rho ^{-2}, -\frac{1}{4}\left( (h-4)\rho ^{-2}-1\right); \frac{\alpha }{2}+\frac{3}{2}, -\frac{\alpha }{2} +1, \frac{3}{2},\frac{3}{2}; \xi \right)\approx  \frac{1}{1 - \sin^2 z} \nonumber
\end{equation}
where  $\xi \approx \sin^2 z$ and $\left| \sin^2 z \right|<1$. If $ z \in \mathbb{R}$, its radius of convergence is $ 0\leq\sin^2 z <1$.
\subsection{ ${\displaystyle  Hl(1-a,-q+\alpha \beta; \alpha,\beta, \delta, \gamma; 1-x)}$} 
Replace coefficients $a$, $q$, $\gamma $, $\delta$ and $x$ by $1-a$, $-q +\alpha \beta $, $\delta $, $\gamma $ and $1-x$ into (\ref{eq:10038}) and (\ref{eq:10039}). Put (\ref{eq:1006}) into the new (\ref{eq:10038}) and (\ref{eq:10039}).

For an infinite series,
\begin{eqnarray}
&&\lim_{n\gg 1}Hl\left( 1-\rho ^{-2}, \frac{1}{4}\left( h\rho ^{-2}- \alpha (\alpha +1)\right); \frac{1}{2}(\alpha +1), -\frac{\alpha }{2}, \frac{1}{2}, \frac{1}{2}; 1-sn^2(z,\rho ) \right)\nonumber\\
&& = \frac{1}{1+\frac{(1-sn^2(z,\rho ))^2}{1-\rho ^{-2}}-\frac{2-\rho ^{-2}}{1-\rho ^{-2}}(1-sn^2(z,\rho )) }  \label{eq:10052}
\end{eqnarray}
 The condition of convergence of (\ref{eq:10052}) is
\begin{equation}
\left| \frac{(1-sn^2(z,\rho ))^2}{1-\rho ^{-2}}\right| +\left|\frac{2-\rho ^{-2}}{1-\rho ^{-2}}(1-sn^2(z,\rho )) \right|<1 \label{eq:10053}
\end{equation}
For the case of $ z, sn(z,\rho )\in \mathbb{R}$ where $ 0< \rho < 1 $, the boundary condition of $sn^2(z,\rho )$ in (\ref{eq:10053}) is given by
\begin{table}[h]
\begin{center}
\tabcolsep 5.8pt
\begin{tabular}{l*{6}{c}|r}
Range of the coefficient $\rho$ & Range of the independent variable $sn^2(z,\rho )$ \\
\hline 
As $\frac{1}{\sqrt{2}}<\rho<1$ & $2-\rho^{-2}<sn^2(z,\rho )\leq 1$ \\ 
As $0<\rho<\frac{1}{\sqrt{2}}$  & $0<sn^2(z,\rho )\leq 1$  \\ 
\end{tabular}
\end{center}
\caption{The radius of convergence for $sn^2(z,\rho )$}
\end{table}

For $\rho =1/\sqrt{2} $, (\ref{eq:10052}) turns to be   
\begin{equation}
\lim_{ n\gg 1} Hl\left( 1-\rho ^{-2}=-1, \frac{1}{4}\left( h\rho ^{-2}- \alpha (\alpha +1)\right); \frac{1}{2}(\alpha +1), -\frac{\alpha }{2}, \frac{1}{2}, \frac{1}{2}; 1-sn^2(z,1/\sqrt{2}) \right) = \frac{1}{ 1-\left( 1-sn^2(z,1/\sqrt{2})\right)^2 } \nonumber
\end{equation}
where $\left| 1-sn^2(z,1/\sqrt{2}) \right|<1$. If $ z,sn^2(z,1/\sqrt{2}) \in \mathbb{R}$, its radius of convergence is $0<sn^2(z,1/\sqrt{2})\leq 1$.

For $\rho \approx 0$, (\ref{eq:10052}) turns to be  
\begin{equation}
 \lim_{\substack{n\gg 1\\\rho \approx 0}}  Hl\left( 1-\rho ^{-2}, \frac{1}{4}\left( h\rho ^{-2}- \alpha (\alpha +1)\right); \frac{1}{2}(\alpha +1), -\frac{\alpha }{2}, \frac{1}{2}, \frac{1}{2}; 1-\xi \right)  \approx \frac{1}{ \sin^2 z }  \nonumber
\end{equation}
where  $\xi \approx \sin^2 z$ and $\left| 1-\sin^2 z \right|<1$. If $ z \in \mathbb{R}$, its radius of convergence is $ 0<\sin^2 z \leq 1$.
\subsection{ ${\displaystyle (1-x)^{1-\delta } Hl(1-a,-q+(\delta -1)\gamma a+(\alpha -\delta +1)(\beta -\delta +1); \alpha-\delta +1,\beta-\delta +1}$\\${\displaystyle, 2-\delta, \gamma; 1-x)}$}
Replace coefficients $a$, $q$, $\alpha $, $\beta $, $\gamma $, $\delta$ and $x$ by $1-a$, $-q+(\delta -1)\gamma a+(\alpha -\delta +1)(\beta -\delta +1)$, $\alpha-\delta +1 $, $\beta-\delta +1 $, $2-\delta$, $\gamma $ and $1-x$ into (\ref{eq:10038}) and (\ref{eq:10039}). Put (\ref{eq:1006}) into the new (\ref{eq:10038}) and (\ref{eq:10039}).

For an infinite series,
\begin{eqnarray}
&&\lim_{n\gg 1} Hl\left( 1-\rho ^{-2}, -\frac{1}{4}\left( (1-h)\rho ^{-2}+(\alpha -1)(\alpha +2)\right); \frac{\alpha }{2}+1, -\frac{\alpha }{2}+\frac{1}{2}, \frac{3}{2}, \frac{1}{2}; 1-sn^2(z,\rho ) \right)\nonumber\\
&& = \frac{1}{1+\frac{(1-sn^2(z,\rho ))^2}{1-\rho ^{-2}}-\frac{2-\rho ^{-2}}{1-\rho ^{-2}}(1-sn^2(z,\rho )) }  \label{eq:10058}
\end{eqnarray}
 The condition of convergence of (\ref{eq:10058}) is
\begin{equation}
\left| \frac{(1-sn^2(z,\rho ))^2}{1-\rho ^{-2}}\right| +\left| \frac{2-\rho ^{-2}}{1-\rho ^{-2}}(1-sn^2(z,\rho )) \right|<1 \label{eq:10059}
\end{equation}
For the case of $ z, sn(z,\rho )\in \mathbb{R}$ where $ 0< \rho < 1 $, the boundary condition of $sn^2(z,\rho )$ in (\ref{eq:10059}) is given by
\begin{table}[h]
\begin{center}
\tabcolsep 5.8pt
\begin{tabular}{l*{6}{c}|r}
Range of the coefficient $\rho$ & Range of the independent variable $sn^2(z,\rho )$ \\
\hline 
As $\frac{1}{\sqrt{2}}<\rho<1$ & $2-\rho^{-2}<sn^2(z,\rho )\leq 1$ \\ 
As $0<\rho<\frac{1}{\sqrt{2}}$  & $0<sn^2(z,\rho )\leq 1$  \\ 
\end{tabular}
\end{center}
\caption{The radius of convergence for $sn^2(z,\rho )$}
\end{table}

For $\rho =1/\sqrt{2}$, (\ref{eq:10058}) turns to be   
\begin{equation}
\lim_{ n\gg 1} Hl\left( 1-\rho ^{-2}=-1, \frac{1}{4}\left( h\rho ^{-2}- \alpha (\alpha +1)\right); \frac{1}{2}(\alpha +1), -\frac{\alpha }{2}, \frac{1}{2}, \frac{1}{2}; 1-sn^2(z,1/\sqrt{2}) \right) = \frac{1}{ 1-\left( 1-sn^2(z,1/\sqrt{2})\right)^2 } \nonumber
\end{equation}
where $\left| 1-sn^2(z,1/\sqrt{2}) \right|<1$. If $ z,sn^2(z,1/\sqrt{2}) \in \mathbb{R}$, its radius of convergence is $0<sn^2(z,1/\sqrt{2})\leq 1$.

For $\rho \approx 0$, (\ref{eq:10058}) turns to be  
\begin{equation}
 \lim_{\substack{n\gg 1\\\rho \approx 0}}  Hl\left( 1-\rho ^{-2}, \frac{1}{4}\left( h\rho ^{-2}- \alpha (\alpha +1)\right); \frac{1}{2}(\alpha +1), -\frac{\alpha }{2}, \frac{1}{2}, \frac{1}{2}; 1-\xi \right)  \approx \frac{1}{ \sin^2 z }  \nonumber
\end{equation}
where  $\xi \approx \sin^2 z$ and $\left| 1-\sin^2 z \right|<1$. If $ z \in \mathbb{R}$, its radius of convergence is $ 0<\sin^2 z \leq 1$.
\subsection{ ${\displaystyle x^{-\alpha } Hl\left(\frac{1}{a},\frac{q+\alpha [(\alpha -\gamma -\delta +1)a-\beta +\delta ]}{a}; \alpha , \alpha -\gamma +1, \alpha -\beta +1,\delta ;\frac{1}{x}\right)}$}
Replace coefficients $a$, $q$, $\beta $, $\gamma $ and $x$ by $\frac{1}{a}$, $\frac{q+\alpha [(\alpha -\gamma -\delta +1)a-\beta +\delta ]}{a}$, $\alpha-\gamma +1 $, $\alpha -\beta +1 $ and $\frac{1}{x}$ into  (\ref{eq:10038}) and (\ref{eq:10039}). Put (\ref{eq:1006}) into the new  (\ref{eq:10038}) and (\ref{eq:10039}).

For an infinite series,
\begin{eqnarray}
&&\lim_{n\gg 1} Hl\left(\rho ^2,-\frac{1}{4}\left( h-(1+\rho ^2)(\alpha +1)^2\right); \frac{1}{2}(\alpha +1), \frac{1}{2}(\alpha +2),\alpha +\frac{3}{2}, \frac{1}{2}; sn^{-2}(z,\rho )\right)\nonumber\\
&&= \frac{1}{1+\rho ^{-2}sn^{-4}(z,\rho )-(1+\rho ^{-2})sn^{-2}(z,\rho )}  \label{eq:10064}
\end{eqnarray}
The condition of convergence of (\ref{eq:10064}) is
\begin{equation}
\left| \rho ^{-2}sn^{-4}(z,\rho )\right| +\left| (1+\rho ^{-2})sn^{-2}(z,\rho ) \right|<1 \label{eq:10065}
\end{equation}
For the case of $ z, sn(z,\rho )\in \mathbb{R}$ where $ 0< \rho < 1 $, the boundary condition of $sn^2(z,\rho )$ in (\ref{eq:10065}) is given by
\begin{equation}
0\leq sn^2(z,\rho )\leq 1 \nonumber
\end{equation}
\subsection{ ${\displaystyle \left(1-\frac{x}{a} \right)^{-\beta } Hl\left(1-a, -q+\gamma \beta; -\alpha +\gamma +\delta, \beta, \gamma, \delta; \frac{(1-a)x}{x-a} \right)}$}
Replace coefficients $a$, $q$, $\alpha $ and $x$ by $1-a$, $-q+\gamma \beta $, $-\alpha+\gamma +\delta $ and $\frac{(1-a)x}{x-a}$ into (\ref{eq:10038}) and (\ref{eq:10039}).  Put (\ref{eq:1006}) into the new (\ref{eq:10038}) and (\ref{eq:10039}).

For an infinite series,
\begin{eqnarray}
&&\lim_{n\gg 1} Hl\left(1-\rho ^{-2}, \frac{1}{4}\left( h\rho ^{-2} - \alpha \right); -\frac{\alpha }{2}+\frac{1}{2}, -\frac{\alpha }{2}, \frac{1}{2}, \frac{1}{2}; \frac{(1-\rho ^{-2})sn^2(z,\rho ) }{sn^2(z,\rho ) -\rho ^{-2}}\right) \nonumber\\
&&= \frac{1}{1+\frac{(1-\rho ^{-2})sn^4(z,\rho )}{(sn^2(z,\rho )-\rho ^{-2})^2}- \frac{(2-\rho ^{-2})sn^2(z,\rho )}{(sn^2(z,\rho )-\rho ^{-2})} }  \label{eq:10066}
\end{eqnarray}
The condition of convergence of (\ref{eq:10066}) is
\begin{equation}
\left| \frac{(1-\rho ^{-2})sn^4(z,\rho )}{(sn^2(z,\rho )-\rho ^{-2})^2}\right|+ \left|\frac{(2-\rho ^{-2})sn^2(z,\rho )}{(sn^2(z,\rho )-\rho ^{-2})} \right|<1 \label{eq:10067}
\end{equation}
For the case of $ z, sn(z,\rho )\in \mathbb{R}$ where $ 0< \rho < 1 $, the boundary condition of $sn^2(z,\rho )$ in (\ref{eq:10067}) is given by
\begin{table}[h]
\begin{center}
\tabcolsep 5.8pt
\begin{tabular}{l*{6}{c}|r}
Range of the coefficient $\rho$ & Range of the independent variable $sn^2(z,\rho )$ \\
\hline 
As $\frac{1}{\sqrt{2}}<\rho<1$ & $0\leq  sn^2(z,\rho )< \frac{1}{2 \rho ^2}$ \\ 
As $0<\rho<\frac{1}{\sqrt{2}}$  & $0\leq sn^2(z,\rho )<1$  \\ 
\end{tabular}
\end{center}
\caption{The radius of convergence for $sn^2(z,\rho )$}
\end{table}

For $\rho =1/\sqrt{2}$, (\ref{eq:10066}) turns to be   
\begin{equation}
\lim_{ n\gg 1} Hl\left(1-\rho ^{-2}=-1, \frac{1}{4}\left( h\rho ^{-2} - \alpha \right); -\frac{\alpha }{2}+\frac{1}{2}, -\frac{\alpha }{2}, \frac{1}{2}, \frac{1}{2}; \frac{(1-\rho ^{-2})sn^2(z,1/\sqrt{2})}{sn^2(z,1/\sqrt{2}) -\rho ^{-2}}\right) = \frac{1}{1-\frac{sn^4(z,1/\sqrt{2})}{(sn^2(z,1/\sqrt{2})-2)^2}  } \nonumber
\end{equation}
where $\left|\frac{sn^4(z,1/\sqrt{2})}{(sn^2(z,1/\sqrt{2})-2)^2}\right|<1$. If $ z, sn(z,1/\sqrt{2} ) \in \mathbb{R}$, its radius of convergence is $0\leq sn^2(z,1/\sqrt{2})< 1$.

For $\rho \approx 0$, (\ref{eq:10066}) turns to be  
\begin{equation}
\lim_{\substack{n\gg 1\\ \rho \approx 0}} Hl\left(1-\rho ^{-2}, \frac{1}{4}\left( h\rho ^{-2} - \alpha \right); -\frac{\alpha }{2}+\frac{1}{2}, -\frac{\alpha }{2}, \frac{1}{2}, \frac{1}{2}; \frac{(1-\rho ^{-2})\xi }{\xi -\rho ^{-2}}\right) \approx  \frac{1}{1- \sin^2 z }  \nonumber
\end{equation}
where $\xi\approx \sin^2 z$ and $\left|\sin^2 z\right|<1$. If $ z \in \mathbb{R}$, its radius of convergence is $0\leq  \sin^2 z<1$.

For $\rho \approx 1$, (\ref{eq:10066}) turns to be  
\begin{equation}
\lim_{\substack{n\gg 1\\ \rho \approx 1}} Hl\left(1-\rho ^{-2}, \frac{1}{4}\left( h\rho ^{-2} - \alpha \right); -\frac{\alpha }{2}+\frac{1}{2}, -\frac{\alpha }{2}, \frac{1}{2}, \frac{1}{2}; \frac{(1-\rho ^{-2})\xi }{\xi -\rho ^{-2}}\right)\approx  \frac{1}{1- \frac{\tanh^2 z}{\tanh^2 z-1} }  \nonumber
\end{equation}
where $\xi \approx \tanh^2 z$ and $ \left| \frac{\tanh^2 z}{\tanh^2 z-1}\right|< 1$. More precisely, for $ z \in \mathbb{R}$, the range of an independent $z$ is derived by $z<\tanh^{-1}(1/\sqrt{2})$.
\subsection{ ${\displaystyle (1-x)^{1-\delta }\left(1-\frac{x}{a} \right)^{-\beta+\delta -1} Hl\Bigg(1-a, -q+\gamma [(\delta -1)a+\beta -\delta +1]; -\alpha +\gamma +1}$\\ ${\displaystyle, \beta -\delta+1, \gamma, 2-\delta; \frac{(1-a)x}{x-a} \Bigg)}$ }
Replace coefficients $a$, $q$, $\alpha $, $\beta $, $\delta $ and $x$ by $1-a$, $-q+\gamma [(\delta -1)a+\beta -\delta +1]$, $-\alpha +\gamma +1$, $\beta -\delta+1$, $2-\delta $ and $\frac{(1-a)x}{x-a}$ into (\ref{eq:10038}) and (\ref{eq:10039}). Put (\ref{eq:1006}) into the new (\ref{eq:10038}) and (\ref{eq:10039}).

For an infinite series,
\begin{eqnarray}
&&\lim_{n\gg 1} Hl\left( 1-\rho ^{-2}, \frac{1}{4}\left( (h-1)\rho ^{-2} +1- \alpha \right); -\frac{\alpha }{2}+1, -\frac{\alpha }{2}+\frac{1}{2}, \frac{1}{2}, \frac{3}{2}; \frac{(1-\rho ^{-2})sn^2(z,\rho ) }{sn^2(z,\rho ) -\rho ^{-2}}\right) \nonumber\\
&&= \frac{1}{1+\frac{(1-\rho ^{-2})sn^4(z,\rho )}{(sn^2(z,\rho )-\rho ^{-2})^2}- \frac{(2-\rho ^{-2})sn^2(z,\rho )}{(sn^2(z,\rho )-\rho ^{-2})} }  \label{eq:10072}
\end{eqnarray}
The condition of convergence of (\ref{eq:10072}) is
\begin{equation}
\left| \frac{(1-\rho ^{-2})sn^4(z,\rho )}{(sn^2(z,\rho )-\rho ^{-2})^2}\right| + \left| \frac{(2-\rho ^{-2})sn^2(z,\rho )}{sn^2(z,\rho )-\rho ^{-2}} \right|<1 \label{eq:10073}
\end{equation}
For the case of $ z, sn(z,\rho )\in \mathbb{R}$ where $ 0< \rho < 1 $, the boundary condition of $sn^2(z,\rho )$ in (\ref{eq:10073}) is given by
\begin{table}[h]
\begin{center}
\tabcolsep 5.8pt
\begin{tabular}{l*{6}{c}|r}
Range of the coefficient $\rho$ & Range of the independent variable $sn^2(z,\rho )$ \\
\hline 
As $\frac{1}{\sqrt{2}}<\rho<1$ & $0\leq  sn^2(z,\rho )< \frac{1}{2 \rho ^2}$ \\ 
As $0<\rho<\frac{1}{\sqrt{2}}$  & $0\leq sn^2(z,\rho )<1$  \\ 
\end{tabular}
\end{center}
\caption{The radius of convergence for $sn^2(z,\rho )$}
\end{table}

For $\rho =1/\sqrt{2}$, (\ref{eq:10072}) turns to be   
\begin{eqnarray}
&&\lim_{ n\gg 1} Hl\left( 1-\rho ^{-2}=-1, \frac{1}{4}\left( (h-1)\rho ^{-2} +1- \alpha \right); -\frac{\alpha }{2}+1, -\frac{\alpha }{2}+\frac{1}{2}, \frac{1}{2}, \frac{3}{2}; \frac{(1-\rho ^{-2})sn^2(z,1/\sqrt{2}) }{sn^2(z,1/\sqrt{2}) -\rho ^{-2}}\right) \nonumber\\
&&= \frac{1}{1-\frac{sn^4(z,1/\sqrt{2})}{(sn^2(z,1/\sqrt{2})-2)^2}  } \nonumber
\end{eqnarray}
where $\left|\frac{sn^4(z,1/\sqrt{2})}{(sn^2(z,1/\sqrt{2})-2)^2}\right|<1$. If $z, sn(z,1/\sqrt{2} ) \in \mathbb{R}$, its radius of convergence is $0\leq sn^2(z,1/\sqrt{2})< 1$.

For $\rho \approx 0$, (\ref{eq:10072}) turns to be  
\begin{equation}
\lim_{\substack{n\gg 1\\ \rho \approx 0}} Hl\left( 1-\rho ^{-2}, \frac{1}{4}\left( (h-1)\rho ^{-2} +1- \alpha \right); -\frac{\alpha }{2}+1, -\frac{\alpha }{2}+\frac{1}{2}, \frac{1}{2}, \frac{3}{2}; \frac{(1-\rho ^{-2})\xi }{\xi -\rho ^{-2}}\right) \approx  \frac{1}{1- \sin^2 z }  \nonumber
\end{equation}
where $\xi\approx \sin^2 z$ and $\left|\sin^2 z\right|<1$. If $ z \in \mathbb{R}$, its radius of convergence is $0\leq  \sin^2 z<1$.

For $\rho \approx 1$, (\ref{eq:10072}) turns to be  
\begin{equation}
\lim_{\substack{n\gg 1\\ \rho \approx 1}} Hl\left( 1-\rho ^{-2}, \frac{1}{4}\left( (h-1)\rho ^{-2} +1- \alpha \right); -\frac{\alpha }{2}+1, -\frac{\alpha }{2}+\frac{1}{2}, \frac{1}{2}, \frac{3}{2}; \frac{(1-\rho ^{-2})\xi }{\xi -\rho ^{-2}}\right) \approx  \frac{1}{1- \frac{\tanh^2 z}{\tanh^2 z-1} }  \nonumber
\end{equation}
where $\xi \approx \tanh^2 z$ and $ \left| \frac{\tanh^2 z}{\tanh^2 z-1}\right|< 1$. More precisely, for $ z \in \mathbb{R}$, the range of an independent $z$ is derived by $z<\tanh^{-1}(1/\sqrt{2})$.
\subsection{ ${\displaystyle x^{-\alpha } Hl\left(\frac{a-1}{a}, \frac{-q+\alpha (\delta a+\beta -\delta )}{a}; \alpha, \alpha -\gamma +1, \delta , \alpha -\beta +1; \frac{x-1}{x} \right)}$}
Replace coefficients $a$, $q$, $\beta $, $\gamma $, $\delta $ and $x$ by $\frac{a-1}{a}$, $\frac{-q+\alpha (\delta a+\beta -\delta )}{a}$, $\alpha -\gamma +1$, $\delta $, $\alpha -\beta +1$ and $\frac{x-1}{x}$ into (\ref{eq:10038}) and (\ref{eq:10039}). Put (\ref{eq:1006}) into the new (\ref{eq:10038}) and (\ref{eq:10039}).

For an infinite series,
\begin{eqnarray}
&&\lim_{n\gg 1} Hl\left( 1-\rho ^2, \frac{1}{4}\left[ h+ (\alpha +1)\left( 1-(\alpha +1)\rho ^2\right) \right];  \frac{1}{2}(\alpha +1), \frac{1}{2}(\alpha +2), \frac{1}{2}, \alpha +\frac{3}{2}; \frac{sn^2(z,\rho ) -1}{sn^2(z,\rho )}\right)  \nonumber\\
&&= \frac{1}{1+ \frac{1}{1-\rho ^2}\left( \frac{sn^2(z,\rho )-1}{sn^2(z,\rho )}\right)^2 -\frac{2-\rho ^2}{1-\rho ^2} \left( \frac{sn^2(z,\rho )-1}{sn^2(z,\rho )}\right) }  \label{eq:10078}
\end{eqnarray}
 The condition of convergence of (\ref{eq:10078}) is
\begin{equation}
\left|  \frac{1}{1-\rho ^2}\left( \frac{sn^2(z,\rho )-1}{sn^2(z,\rho )}\right)^2\right| +\left|\frac{2-\rho ^2}{1-\rho ^2} \left( \frac{sn^2(z,\rho )-1}{sn^2(z,\rho )}\right) \right|<1 \label{eq:10079}
\end{equation}
For the case of $ z, sn(z,\rho )\in \mathbb{R}$ where $ 0< \rho < 1 $, the boundary condition of $sn^2(z,\rho )$ in (\ref{eq:10079}) is given by
\begin{equation}
\frac{\rho ^2-\sqrt{\rho ^4-8\rho ^2+8}}{4(\rho ^2-1)}< sn(z,\rho )\leq 1 \nonumber
\end{equation}
\subsection{ ${\displaystyle \left(\frac{x-a}{1-a} \right)^{-\alpha } Hl\left(a, q-(\beta -\delta )\alpha ; \alpha , -\beta+\gamma +\delta , \delta , \gamma; \frac{a(x-1)}{x-a} \right)}$}
Replace coefficients $q$, $\beta $, $\gamma $, $\delta $ and $x$  by $q-(\beta -\delta )\alpha $, $-\beta+\gamma +\delta $, $\delta $,  $\gamma $ and $\frac{a(x-1)}{x-a}$ into (\ref{eq:10038}) and (\ref{eq:10039}). Put (\ref{eq:1006}) into the new (\ref{eq:10038}) and (\ref{eq:10039}).

For an infinite series,
\begin{eqnarray}
&&\lim_{n\gg 1} Hl\left( \rho ^{-2}, -\frac{1}{4}\left( h\rho ^{-2}- (\alpha +1)^2\right);  \frac{1}{2}(\alpha +1), -\frac{1}{2}(\alpha -2), \frac{1}{2}, \frac{1}{2}; \frac{sn^2(z,\rho ) -1}{\rho ^2(sn^2(z,\rho )-\rho ^{-2})}\right) \nonumber\\
&&= \frac{1}{1+\frac{(sn^2(z,\rho )-1)^2}{\rho ^2(sn^2(z,\rho )-\rho ^{-2})^2 } -\frac{(1+\rho ^2) (sn^2(z,\rho )-1)}{\rho ^2(sn^2(z,\rho )-\rho ^{-2})} }  \label{eq:10080}
\end{eqnarray}
 The condition of convergence of (\ref{eq:10080}) is
\begin{equation}
\left| \frac{(sn^2(z,\rho )-1)^2}{\rho ^2(sn^2(z,\rho )-\rho ^{-2})^2 } \right| +\left|\frac{(1+\rho ^2) (sn^2(z,\rho )-1)}{\rho ^2(sn^2(z,\rho )-\rho ^{-2})} \right|<1 \label{eq:10081}
\end{equation}
For the case of $ z, sn(z,\rho )\in \mathbb{R}$ where $ 0< \rho < 1 $, the boundary condition of $sn^2(z,\rho )$ in (\ref{eq:10081}) is given by
\begin{equation}
\frac{(1+\rho ^2)^2-(1-\rho ^2)\sqrt{\rho ^4+6\rho ^2+1}}{4\rho ^2}< sn(z,\rho )\leq 1 \nonumber
\end{equation}
For $\rho \approx 0$, (\ref{eq:10080}) is approximately equal to
\begin{equation}
  \lim_{\substack{n\gg 1\\ \rho \approx 0}} Hl\left( \rho ^{-2}, -\frac{1}{4}\left( h\rho ^{-2}- (\alpha +1)^2\right);  \frac{1}{2}(\alpha +1), -\frac{1}{2}(\alpha -2), \frac{1}{2}, \frac{1}{2}; \frac{\xi -1}{\rho ^2(\xi-\rho ^{-2})}\right) \approx  \frac{1}{\sin^2 z}  \nonumber
\end{equation}
where $\xi\approx \sin^2 z$ and $\left|1-\sin^2 z\right|<1$. If $ z \in \mathbb{R}$, its radius of convergence is $0< \sin^2 z\leq 1$.
\section{Integral representation}
In Ref.\cite{Chou2012d}, the integral representation of Heun equation about $x=0$ of the first kind for polynomial of type 1 as $\alpha = -2\alpha_j -j $ where $j, \alpha _j =0,1,2,\cdots$ is given by
\begin{eqnarray}
 y(x)&=& HF_{\alpha _j, \beta }\left( \alpha _j =-\frac{1}{2}(\alpha +j)\big|_{j\in \mathbb{N}_{0}}; \eta = \frac{(1+a)}{a} x ; z= -\frac{1}{a} x^2 \right) \nonumber\\
&=& _2F_1 \left(-\alpha _0, \frac{\beta }{2};\frac{1}{2}+\frac{\gamma }{2}; z \right) + \sum_{n=1}^{\infty } \Bigg\{\prod _{k=0}^{n-1} \Bigg\{ \int_{0}^{1} dt_{n-k}\;t_{n-k}^{\frac{1}{2}(n-k-2)} \int_{0}^{1} du_{n-k}\;u_{n-k}^{\frac{1}{2}(n-k-3+\gamma )} \nonumber\\
&&\times \frac{1}{2\pi i}  \oint dv_{n-k} \frac{1}{v_{n-k}} \left( 1-\frac{1}{v_{n-k}}\right)^{\alpha _{n-k}} \left( 1- \overleftrightarrow {w}_{n-k+1,n}v_{n-k}(1-t_{n-k})(1-u_{n-k})\right)^{-\frac{1}{2}(n-k+\beta )}\nonumber\\
&&\times  \left( \overleftrightarrow {w}_{n-k,n}^{-\frac{1}{2}(n-k-1)}\left(  \overleftrightarrow {w}_{n-k,n} \partial _{ \overleftrightarrow {w}_{n-k,n}}\right) \overleftrightarrow {w}_{n-k,n}^{\frac{1}{2}(n-k-1)}\left( \overleftrightarrow {w}_{n-k,n} \partial _{ \overleftrightarrow {w}_{n-k,n}} + \Omega _{n-k-1}^{(S)}\right) +Q\right) \Bigg\}\nonumber\\
&&\times  _2F_1 \left(-\alpha _0, \frac{\beta }{2};\frac{1}{2}+\frac{\gamma }{2}; \overleftrightarrow {w}_{1,n} \right) \Bigg\} \eta ^n \label{eq:10084}
\end{eqnarray}
where
\begin{equation}
\begin{cases} z = -\frac{1}{a}x^2 \cr
\eta = \frac{(1+a)}{a} x \cr
\alpha _i\leq \alpha _j \;\;\mbox{only}\;\mbox{if}\;i\leq j\;\;\mbox{where}\;i,j= 0,1,2,\cdots
\end{cases}\nonumber 
\end{equation}
and
\begin{equation}
\begin{cases} 
\Omega _{n-k-1}^{(S)} = \frac{1}{2(1+a)}(-2\alpha _{n-k-1}+\beta -\delta +a(\delta +\gamma +n-k-2)) \cr
Q= \frac{q}{4(1+a)}
\end{cases}\nonumber 
\end{equation}
In Ref.\cite{Chou2012d}, the integral representation of Heun equation about $x=0$ of the first kind for infinite series is given by
\begin{eqnarray}
 y(x)&=& HF_{\alpha , \beta }\left( \eta = \frac{(1+a)}{a} x ; z= -\frac{1}{a} x^2 \right) \nonumber\\
&=& _2F_1 \left( \frac{\alpha }{2}, \frac{\beta }{2};\frac{1}{2}+\frac{\gamma }{2}; z \right) + \sum_{n=1}^{\infty } \Bigg\{\prod _{k=0}^{n-1} \Bigg\{ \int_{0}^{1} dt_{n-k}\;t_{n-k}^{\frac{1}{2}(n-k-2)} \int_{0}^{1} du_{n-k}\;u_{n-k}^{\frac{1}{2}(n-k-3+\gamma )} \nonumber\\
&&\times\frac{1}{2\pi i}  \oint dv_{n-k} \frac{1}{v_{n-k}} \left( 1-\frac{1}{v_{n-k}}\right)^{-\frac{1}{2}(n-k+\alpha )}  \left( 1- \overleftrightarrow {w}_{n-k+1,n}v_{n-k}(1-t_{n-k})(1-u_{n-k})\right)^{-\frac{1}{2}(n-k+\beta )}\nonumber\\
&&\times  \left( \overleftrightarrow {w}_{n-k,n}^{-\frac{1}{2}(n-k-1)}\left(  \overleftrightarrow {w}_{n-k,n} \partial _{ \overleftrightarrow {w}_{n-k,n}}\right) \overleftrightarrow {w}_{n-k,n}^{\frac{1}{2}(n-k-1)}\left( \overleftrightarrow {w}_{n-k,n} \partial _{ \overleftrightarrow {w}_{n-k,n}}  + \Omega _{n-k-1}^{(I)} \right) +Q \right) \Bigg\}\nonumber\\
&&\times _2F_1 \left( \frac{\alpha }{2}, \frac{\beta }{2};\frac{1}{2}+\frac{\gamma }{2}; \overleftrightarrow {w}_{1,n} \right) \Bigg\} \eta ^n \label{eq:10085}
\end{eqnarray}
where
\begin{equation}
\begin{cases} 
\Omega _{n-k-1}^{(I)} = \frac{1}{2(1+a)}(\alpha +\beta -\delta +n-k-1 +a(\delta +\gamma +n-k-2)) \cr
Q= \frac{q}{4(1+a)}
\end{cases}\nonumber 
\end{equation}
On (\ref{eq:10084}) and (\ref{eq:10085}),
\begin{equation}\overleftrightarrow {w}_{i,j}=
\begin{cases} \displaystyle {\frac{v_i}{(v_i-1)}\; \frac{\overleftrightarrow w_{i+1,j} t_i u_i}{1- \overleftrightarrow w_{i+1,j} v_i (1-t_i)(1-u_i)}}\;\;\mbox{where}\; i\leq j\cr
z \;\;\mbox{only}\;\mbox{if}\; i>j
\end{cases}\nonumber 
\end{equation}
\subsection{ ${\displaystyle (1-x)^{1-\delta } Hl(a, q - (\delta  - 1)\gamma a; \alpha - \delta  + 1, \beta - \delta + 1, \gamma ,2 - \delta ; x)}$ }
\subsubsection{Polynomial of type 1}
Replace coefficients $q$, $\alpha$, $\beta$ and $\delta$ by $q - (\delta - 1)\gamma a $, $\alpha - \delta  + 1 $, $\beta - \delta + 1$ and $2 - \delta$ into (\ref{eq:10084}). Multiply $(1-x)^{1-\delta }$ and (\ref{eq:10084}) together. Put (\ref{eq:1006}) into the new (\ref{eq:10084}) with replacing $\alpha $ by $-2(2\alpha _j+j+1)$ where $j,\alpha _j \in \mathbb{N}_{0}$; apply $\alpha =-2(2\alpha _0+1)$ into sub-integral $y_0(\xi)$, apply $\alpha =-2(2\alpha _0+1)$ into the first summation and $\alpha =-2(2\alpha _1+2)$ into second summation of sub-integral $y_1(\xi)$, apply $\alpha =-2(2\alpha _0+1)$ into the first summation, $\alpha =-2(2\alpha _1+2)$ into the second summation and $\alpha =-2(2\alpha _2+3)$ into the third summation of sub-integral $y_2(\xi)$, etc in the new (\ref{eq:10084}).
\begin{eqnarray}
&& (1-\xi )^{\frac{1}{2}} y(\xi )\nonumber\\
&=& (1-\xi )^{\frac{1}{2}} Hl\left(\rho ^{-2}, -\frac{1}{4}(h-1)\rho ^{-2}; -2\alpha _j-j, -2\alpha _j-j, \frac{1}{2},\frac{3}{2}; \xi \right)\nonumber\\
&=& (1-\xi )^{\frac{1}{2}} \left\{ _2F_1 \left(-\alpha _0, \alpha _0+\frac{3}{4}; \frac{3}{4}; z \right) \right. + \sum_{n=1}^{\infty } \Bigg\{\prod _{k=0}^{n-1} \Bigg\{ \int_{0}^{1} dt_{n-k}\;t_{n-k}^{\frac{1}{2}(n-k-2)} \int_{0}^{1} du_{n-k}\;u_{n-k}^{\frac{1}{2}(n-k-\frac{5}{2})} \nonumber\\
&&\times \frac{1}{2\pi i}  \oint dv_{n-k} \frac{1}{v_{n-k}} \left( 1-\frac{1}{v_{n-k}}\right)^{\alpha _{n-k}} \left( 1- \overleftrightarrow {w}_{n-k+1,n}v_{n-k}(1-t_{n-k})(1-u_{n-k})\right)^{\alpha _{n-k}}\nonumber\\
&&\times  \left( \overleftrightarrow {w}_{n-k,n}^{-\frac{1}{2}(n-k-1)}\left(  \overleftrightarrow {w}_{n-k,n} \partial _{ \overleftrightarrow {w}_{n-k,n}}\right) \overleftrightarrow {w}_{n-k,n}^{\frac{1}{2}(n-k-1)}\left( \overleftrightarrow {w}_{n-k,n} \partial _{ \overleftrightarrow {w}_{n-k,n}} + \Omega _{n-k-1} \right) +Q\right) \Bigg\}\nonumber\\
&&\times \left. _2F_1 \left(-\alpha _0, \alpha _0+\frac{3}{4};\frac{3}{4}; \overleftrightarrow {w}_{1,n} \right) \Bigg\} \eta ^n\right\} \label{eq:10086}
\end{eqnarray}
where
\begin{equation} 
\alpha = 2\left( 2\alpha _j +j+\frac{1}{2} \right) \;\mbox{or}\; -2\left( 2\alpha _j +j+1\right) \nonumber  
\end{equation}
\subsubsection{Infinite series}
Replace coefficients $q$, $\alpha$, $\beta$ and $\delta$ by $q - (\delta - 1)\gamma a $, $\alpha - \delta  + 1 $, $\beta - \delta + 1$ and $2 - \delta$ into (\ref{eq:10085}). Multiply $(1-x)^{1-\delta }$ and (\ref{eq:10085}) together. Put (\ref{eq:1006}) into the new (\ref{eq:10085}).
\begin{eqnarray}
&& (1-\xi )^{\frac{1}{2}} y(\xi )\nonumber\\
&=& (1-\xi )^{\frac{1}{2}} Hl\left(\rho ^{-2}, -\frac{1}{4}(h-1)\rho ^{-2}; \frac{\alpha }{2}+1, -\frac{\alpha }{2}+\frac{1}{2}, \frac{1}{2},\frac{3}{2}; \xi \right)\nonumber\\
&=& (1-\xi )^{\frac{1}{2}} \left\{ _2F_1 \left( \frac{\alpha }{4}+\frac{1}{2}, -\frac{\alpha }{4}+\frac{1}{4}; \frac{3}{4}; z \right)\right. + \sum_{n=1}^{\infty } \Bigg\{\prod _{k=0}^{n-1} \Bigg\{ \int_{0}^{1} dt_{n-k}\;t_{n-k}^{\frac{1}{2}(n-k-2)} \int_{0}^{1} du_{n-k}\;u_{n-k}^{\frac{1}{2}(n-k-\frac{5}{2})} \nonumber\\
&&\times\frac{1}{2\pi i}  \oint dv_{n-k} \frac{1}{v_{n-k}} \left( 1-\frac{1}{v_{n-k}}\right)^{-\frac{1}{2}(n-k+1+\frac{\alpha }{2} )}  \left( 1- \overleftrightarrow {w}_{n-k+1,n}v_{n-k}(1-t_{n-k})(1-u_{n-k})\right)^{-\frac{1}{2}(n-k+\frac{1}{2}-\frac{\alpha }{2})}\nonumber\\
&&\times  \left( \overleftrightarrow {w}_{n-k,n}^{-\frac{1}{2}(n-k-1)}\left(  \overleftrightarrow {w}_{n-k,n} \partial _{ \overleftrightarrow {w}_{n-k,n}}\right) \overleftrightarrow {w}_{n-k,n}^{\frac{1}{2}(n-k-1)}\left( \overleftrightarrow {w}_{n-k,n} \partial _{ \overleftrightarrow {w}_{n-k,n}}  + \Omega _{n-k-1} \right) +Q \right) \Bigg\}\nonumber\\
&&\times \left. _2F_1 \left( \frac{\alpha }{4}+\frac{1}{2}, -\frac{\alpha }{4}+\frac{1}{4}; \frac{3}{4}; \overleftrightarrow {w}_{1,n} \right) \Bigg\} \eta ^n \right\} \label{eq:10087}
\end{eqnarray}
On (\ref{eq:10086}) and (\ref{eq:10087}),
\begin{equation}
\begin{cases} 
\eta =(1+\rho ^2)\xi \cr
\xi = sn^2(z,\rho ) \cr
z=-\rho ^2\xi^2 \cr
\Omega _{n-k-1} = \frac{1}{2}\left( n-k-\frac{1}{1+\rho ^{-2}}\right) \cr
Q= \frac{1-h}{16(1+\rho ^2)}
\end{cases}\nonumber 
\end{equation}
\subsection{ ${\displaystyle x^{1-\gamma } (1-x)^{1-\delta } Hl(a, q-(\gamma +\delta -2)a-(\gamma -1)(\alpha +\beta -\gamma -\delta +1); \alpha - \gamma -\delta +2}$ \\${\displaystyle, \beta - \gamma -\delta +2, 2-\gamma, 2 - \delta ; x)}$}
\subsubsection{Polynomial of type 1}
Replace coefficients $q$, $\alpha$, $\beta$, $\gamma $ and $\delta$ by $q-(\gamma +\delta -2)a-(\gamma -1)(\alpha +\beta -\gamma -\delta +1)$, $\alpha - \gamma -\delta +2$, $\beta - \gamma -\delta +2, 2-\gamma$ and $2 - \delta$ into (\ref{eq:10084}). Multiply $x^{1-\gamma } (1-x)^{1-\delta }$ and (\ref{eq:10084}) together. Put (\ref{eq:1006}) into the new (\ref{eq:10084}) with replacing $\alpha $ by $-2(2\alpha _j+j+3/2)$ where $j,\alpha _j \in \mathbb{N}_{0}$; apply $\alpha =-2(2\alpha _0+3/2)$ into sub-integral $y_0(\xi)$, apply $-2(2\alpha _0+3/2)$ into the first summation and $-2(2\alpha _1 +5/2)$ into second summation of sub-integral $y_1(\xi)$, apply $-2(2\alpha _0+3/2)$ into the first summation, $-2(2\alpha _1+ 5/2)$ into the second summation and $-2(2\alpha _2 +7/2)$ into the third summation of sub-integral $y_2(\xi)$, etc in the new (\ref{eq:10084}).
\begin{eqnarray}
&&\xi ^{\frac{1}{2}} (1-\xi )^{\frac{1}{2}} y(\xi )\nonumber\\
&=&\xi ^{\frac{1}{2}} (1-\xi )^{\frac{1}{2}} Hl\left(\rho ^{-2}, -\frac{1}{4}\left( (h-4)\rho ^{-2}-1\right); -2\alpha _j-j, -2\alpha _j-j, \frac{3}{2},\frac{3}{2}; \xi \right)\nonumber\\
&=& \xi ^{\frac{1}{2}} (1-\xi )^{\frac{1}{2}} \left\{ _2F_1 \left(-\alpha _0, \alpha _0+\frac{5}{4};\frac{5}{4}; z \right) \right. + \sum_{n=1}^{\infty } \Bigg\{\prod _{k=0}^{n-1} \Bigg\{ \int_{0}^{1} dt_{n-k}\;t_{n-k}^{\frac{1}{2}(n-k-2)} \int_{0}^{1} du_{n-k}\;u_{n-k}^{\frac{1}{2}(n-k-\frac{3}{2})} \nonumber\\
&&\times \frac{1}{2\pi i}  \oint dv_{n-k} \frac{1}{v_{n-k}} \left( 1-\frac{1}{v_{n-k}}\right)^{\alpha _{n-k}} \left( 1- \overleftrightarrow {w}_{n-k+1,n}v_{n-k}(1-t_{n-k})(1-u_{n-k})\right)^{\alpha _{n-k}}\nonumber\\
&&\times  \left( \overleftrightarrow {w}_{n-k,n}^{-\frac{1}{2}(n-k-1)}\left(  \overleftrightarrow {w}_{n-k,n} \partial _{ \overleftrightarrow {w}_{n-k,n}}\right) \overleftrightarrow {w}_{n-k,n}^{\frac{1}{2}(n-k-1)}\left( \overleftrightarrow {w}_{n-k,n} \partial _{ \overleftrightarrow {w}_{n-k,n}} + \Omega _{n-k-1} \right) +Q\right) \Bigg\}\nonumber\\
&&\times \left. _2F_1 \left( -\alpha _0, \alpha _0+\frac{5}{4};\frac{5}{4}; \overleftrightarrow {w}_{1,n} \right) \Bigg\} \eta ^n \right\}\label{eq:10088}
\end{eqnarray}
where
\begin{equation}
\alpha = 2\left( 2\alpha _j +j+1 \right) \;\mbox{or}\; -2\left( 2\alpha _j +j+\frac{3}{2}\right) \nonumber  
\end{equation}
\subsubsection{Infinite series}
Replace coefficients $q$, $\alpha$, $\beta$, $\gamma $ and $\delta$ by $q-(\gamma +\delta -2)a-(\gamma -1)(\alpha +\beta -\gamma -\delta +1)$, $\alpha - \gamma -\delta +2$, $\beta - \gamma -\delta +2, 2-\gamma$ and $2 - \delta$ into (\ref{eq:10085}). Multiply $x^{1-\gamma } (1-x)^{1-\delta }$ and (\ref{eq:10085}) together. Put (\ref{eq:1006}) into the new (\ref{eq:10085}).
\begin{eqnarray}
&&\xi ^{\frac{1}{2}} (1-\xi )^{\frac{1}{2}} y(\xi )\nonumber\\
&=&\xi ^{\frac{1}{2}} (1-\xi )^{\frac{1}{2}} Hl\left(\rho ^{-2}, -\frac{1}{4}\left( (h-4)\rho ^{-2}-1\right); \frac{\alpha }{2}+\frac{3}{2}, -\frac{\alpha }{2} +1, \frac{3}{2},\frac{3}{2}; \xi \right)\nonumber\\
&=& \xi ^{\frac{1}{2}} (1-\xi )^{\frac{1}{2}} \left\{ _2F_1 \left( \frac{\alpha }{4}+\frac{3}{4}, -\frac{\alpha }{4}+\frac{1}{2}; \frac{5}{4}; z \right) \right. + \sum_{n=1}^{\infty } \Bigg\{\prod _{k=0}^{n-1} \Bigg\{ \int_{0}^{1} dt_{n-k}\;t_{n-k}^{\frac{1}{2}(n-k-2)} \int_{0}^{1} du_{n-k}\;u_{n-k}^{\frac{1}{2}(n-k-\frac{3}{2} )} \nonumber\\
&&\times\frac{1}{2\pi i}  \oint dv_{n-k} \frac{1}{v_{n-k}} \left( 1-\frac{1}{v_{n-k}}\right)^{-\frac{1}{2}(n-k+\frac{3}{2}+\frac{\alpha }{2} )}  \left( 1- \overleftrightarrow {w}_{n-k+1,n}v_{n-k}(1-t_{n-k})(1-u_{n-k})\right)^{-\frac{1}{2}(n-k+1-\frac{\alpha }{2})}\nonumber\\
&&\times  \left( \overleftrightarrow {w}_{n-k,n}^{-\frac{1}{2}(n-k-1)}\left(  \overleftrightarrow {w}_{n-k,n} \partial _{ \overleftrightarrow {w}_{n-k,n}}\right) \overleftrightarrow {w}_{n-k,n}^{\frac{1}{2}(n-k-1)}\left( \overleftrightarrow {w}_{n-k,n} \partial _{ \overleftrightarrow {w}_{n-k,n}}  + \Omega _{n-k-1} \right) +Q \right) \Bigg\}\nonumber\\
&&\times \left. _2F_1 \left( \frac{\alpha }{4}+\frac{3}{4}, -\frac{\alpha }{4}+\frac{1}{2}; \frac{5}{4}; \overleftrightarrow {w}_{1,n} \right) \Bigg\} \eta ^n \right\} \label{eq:10089}
\end{eqnarray}
On (\ref{eq:10088}) and (\ref{eq:10089}), 
\begin{equation}
\begin{cases} 
\eta =(1+\rho ^2)\xi \cr
\xi = sn^2(z,\rho ) \cr
z=-\rho ^2\xi^2 \cr
\Omega _{n-k-1} = \frac{1}{2}\left( n-k+\frac{1}{1+\rho ^2}\right) \cr
Q= \frac{4-h+\rho ^2}{16(1+\rho ^2)}
\end{cases}\nonumber 
\end{equation}
\subsection{ ${\displaystyle  Hl(1-a,-q+\alpha \beta; \alpha,\beta, \delta, \gamma; 1-x)}$} 
\subsubsection{Polynomial of type 1}
Replace coefficients $a$, $q$, $\gamma $, $\delta$ and $x$ by $1-a$, $-q +\alpha \beta $, $\delta $, $\gamma $ and $1-x$ into (\ref{eq:10084}). Put (\ref{eq:1006}) into the new (\ref{eq:10084}) with replacing $\alpha $ by $-2(2\alpha _j+j+1/2)$ where $j,\alpha _j \in \mathbb{N}_{0}$; apply $\alpha =-2(2\alpha _0+1/2)$ into sub-integral $y_0(\varsigma)$, apply $-2(2\alpha _0+1/2)$ into the first summation and $-2(2\alpha _1 +3/2)$ into second summation of sub-integral $y_1(\varsigma)$, apply $-2(2\alpha _0+1/2)$ into the first summation, $-2(2\alpha _1+ 3/2)$ into the second summation and $-2(2\alpha _2 +5/2)$ into the third summation of sub-integral $y_2(\varsigma)$, etc in the new (\ref{eq:10084}).
\begin{eqnarray}
y(\varsigma )&=&  Hl\left( 1-\rho ^{-2}, \frac{1}{4}h\rho ^{-2} -4\left( \alpha _j+\frac{j}{2}\right)\left( \alpha _j+\frac{j}{2}+\frac{1}{4}\right) ; -2\alpha _j-j, -2\alpha _j-j, \frac{1}{2}, \frac{1}{2}; \varsigma \right)\nonumber\\
&=&  _2F_1 \left( -\alpha _0, \alpha _0+\frac{1}{4};\frac{3}{4}; z \right) + \sum_{n=1}^{\infty } \Bigg\{\prod _{k=0}^{n-1} \Bigg\{ \int_{0}^{1} dt_{n-k}\;t_{n-k}^{\frac{1}{2}(n-k-2)} \int_{0}^{1} du_{n-k}\;u_{n-k}^{\frac{1}{2}(n-k-\frac{5}{2})} \nonumber\\
&&\times \frac{1}{2\pi i}  \oint dv_{n-k} \frac{1}{v_{n-k}} \left( 1-\frac{1}{v_{n-k}}\right)^{\alpha _{n-k}} \left( 1- \overleftrightarrow {w}_{n-k+1,n}v_{n-k}(1-t_{n-k})(1-u_{n-k})\right)^{\alpha _{n-k}}\nonumber\\
&&\times  \left( \overleftrightarrow {w}_{n-k,n}^{-\frac{1}{2}(n-k-1)}\left(  \overleftrightarrow {w}_{n-k,n} \partial _{ \overleftrightarrow {w}_{n-k,n}}\right) \overleftrightarrow {w}_{n-k,n}^{\frac{1}{2}(n-k-1)}\left( \overleftrightarrow {w}_{n-k,n} \partial _{ \overleftrightarrow {w}_{n-k,n}} + \frac{1}{2}(n-k-1)\right) +Q_{n-k-1} \right) \Bigg\}\nonumber\\
&&\times  _2F_1 \left( -\alpha _0, \alpha _0+\frac{1}{4};\frac{3}{4}; \overleftrightarrow {w}_{1,n} \right) \Bigg\} \eta ^n \label{eq:10090}
\end{eqnarray}
where
\begin{equation} 
\begin{cases} 
\alpha = 2\left( 2\alpha _j +j \right) \;\mbox{or}\; -2\left( 2\alpha _j +j+\frac{1}{2}\right) \cr
Q_{n-k-1} = \frac{1}{16(2-\rho ^{-2})}\left( h\rho ^{-2}+4 (2\alpha _{n-k-1}+n-k-1)^2 \right) 
\end{cases} \nonumber 
\end{equation}
\subsubsection{Infinite series}
Replace coefficients $a$, $q$, $\gamma $, $\delta$ and $x$ by $1-a$, $-q +\alpha \beta $, $\delta $, $\gamma $ and $1-x$ into (\ref{eq:10085}). Put (\ref{eq:1006}) into the new (\ref{eq:10085}).
\begin{eqnarray}
y(\varsigma )&=&  Hl\left( 1-\rho ^{-2}, \frac{1}{4}\left( h\rho ^{-2}- \alpha (\alpha +1)\right); \frac{1}{2}(\alpha +1), -\frac{\alpha }{2}, \frac{1}{2}, \frac{1}{2}; \varsigma \right)\nonumber\\
&=& _2F_1 \left( \frac{\alpha }{4}+\frac{1}{4}, -\frac{\alpha }{4}; \frac{3}{4}; z \right) + \sum_{n=1}^{\infty } \Bigg\{\prod _{k=0}^{n-1} \Bigg\{ \int_{0}^{1} dt_{n-k}\;t_{n-k}^{\frac{1}{2}(n-k-2)} \int_{0}^{1} du_{n-k}\;u_{n-k}^{\frac{1}{2}(n-k-\frac{5}{2})} \nonumber\\
&&\times\frac{1}{2\pi i}  \oint dv_{n-k} \frac{1}{v_{n-k}} \left( 1-\frac{1}{v_{n-k}}\right)^{-\frac{1}{2}(n-k+\frac{1}{2}+\frac{\alpha }{2})}  \left( 1- \overleftrightarrow {w}_{n-k+1,n}v_{n-k}(1-t_{n-k})(1-u_{n-k})\right)^{-\frac{1}{2}(n-k-\frac{\alpha }{2} )}\nonumber\\
&&\times  \left( \overleftrightarrow {w}_{n-k,n}^{-\frac{1}{2}(n-k-1)}\left(  \overleftrightarrow {w}_{n-k,n} \partial _{ \overleftrightarrow {w}_{n-k,n}}\right) \overleftrightarrow {w}_{n-k,n}^{\frac{1}{2}(n-k-1)}\left( \overleftrightarrow {w}_{n-k,n} \partial _{ \overleftrightarrow {w}_{n-k,n}}  + \frac{1}{2}(n-k-1) \right) +Q \right) \Bigg\}\nonumber\\
&&\times _2F_1 \left( \frac{\alpha }{4}+\frac{1}{4}, -\frac{\alpha }{4}; \frac{3}{4}; \overleftrightarrow {w}_{1,n} \right) \Bigg\} \eta ^n \label{eq:10091}
\end{eqnarray}
where
\begin{equation}
Q = \frac{1}{16(2-\rho ^{-2})}\left( h\rho ^{-2}-\alpha (\alpha +1)\right)
\nonumber  
\end{equation}
On (\ref{eq:10090}) and (\ref{eq:10091}),
\begin{equation}
\begin{cases}
\varsigma= 1-\xi \cr
\xi = sn^2(z,\rho ) \cr
\eta =\frac{2-\rho ^{-2}}{1-\rho ^{-2}}\varsigma \cr
z=\frac{-1}{1-\rho ^{-2}}\varsigma ^2 
\end{cases}\nonumber  
\end{equation}
\subsection{ ${\displaystyle (1-x)^{1-\delta } Hl(1-a,-q+(\delta -1)\gamma a+(\alpha -\delta +1)(\beta -\delta +1); \alpha-\delta +1,\beta-\delta +1}$\\${\displaystyle, 2-\delta, \gamma; 1-x)}$}
\subsubsection{Polynomial of type 1}
Replace coefficients $a$, $q$, $\alpha $, $\beta $, $\gamma $, $\delta$ and $x$ by $1-a$, $-q+(\delta -1)\gamma a+(\alpha -\delta +1)(\beta -\delta +1)$, $\alpha-\delta +1 $, $\beta-\delta +1 $, $2-\delta$, $\gamma $ and $1-x$ into (\ref{eq:10084}). Multiply $(1-x)^{1-\delta }$ and (\ref{eq:10084}) together. Put (\ref{eq:1006}) into the new (\ref{eq:10084}) with replacing $\alpha $ by $-2(2\alpha _j+j+1)$ where $j,\alpha _j \in \mathbb{N}_{0}$; apply $\alpha =-2(2\alpha _0+1)$ into sub-integral $y_0(\varsigma)$, apply $-2(2\alpha _0+1)$ into the first summation and $-2(2\alpha _1+2)$ into second summation of sub-integral $y_1(\varsigma)$, apply $-2(2\alpha _0+1)$ into the first summation, $-2(2\alpha _1+2)$ into the second summation and $-2(2\alpha _2+3)$ into the third summation of sub-integral $y_2(\varsigma)$, etc in the new (\ref{eq:10084}).
\begin{eqnarray}
&&\varsigma ^{\frac{1}{2}}y(\varsigma )\nonumber\\
&=& \varsigma ^{\frac{1}{2}} Hl\left( 1-\rho ^{-2}, -\frac{1}{4} (1-h)\rho ^{-2}-4\left( \alpha _j +\frac{j}{2}\right)\left( \alpha _j +\frac{j}{2}+\frac{3}{4}\right); -2\alpha _j-j, -2\alpha _j-j, \frac{3}{2}, \frac{1}{2}; \varsigma \right)\nonumber\\
&=& \varsigma ^{\frac{1}{2}} \left\{  _2F_1 \left( -\alpha _0, \alpha _0+\frac{3}{4};\frac{5}{4}; z \right)\right. + \sum_{n=1}^{\infty } \Bigg\{\prod _{k=0}^{n-1} \Bigg\{ \int_{0}^{1} dt_{n-k}\;t_{n-k}^{\frac{1}{2}(n-k-2)} \int_{0}^{1} du_{n-k}\;u_{n-k}^{\frac{1}{2}(n-k- \frac{3}{2})} \nonumber\\
&&\times \frac{1}{2\pi i}  \oint dv_{n-k} \frac{1}{v_{n-k}} \left( 1-\frac{1}{v_{n-k}}\right)^{\alpha _{n-k}} \left( 1- \overleftrightarrow {w}_{n-k+1,n}v_{n-k}(1-t_{n-k})(1-u_{n-k})\right)^{\alpha _{n-k}}\nonumber\\
&&\times  \left( \overleftrightarrow {w}_{n-k,n}^{-\frac{1}{2}(n-k-1)}\left(  \overleftrightarrow {w}_{n-k,n} \partial _{ \overleftrightarrow {w}_{n-k,n}}\right) \overleftrightarrow {w}_{n-k,n}^{\frac{1}{2}(n-k-1)}\left( \overleftrightarrow {w}_{n-k,n} \partial _{ \overleftrightarrow {w}_{n-k,n}} + \frac{1}{2}(n-k)\right) +Q_{n-k-1} \right) \Bigg\}\nonumber\\
&&\times \left. _2F_1 \left( -\alpha _0, \alpha _0+\frac{3}{4};\frac{5}{4}; \overleftrightarrow {w}_{1,n} \right) \Bigg\} \eta ^n \right\} \label{eq:10092}
\end{eqnarray}
where
\begin{equation}
\begin{cases} 
\alpha = 2\left( 2\alpha _j +j +\frac{1}{2}\right) \;\mbox{or}\; -2\left( 2\alpha _j +j+1\right) \cr
Q_{n-k-1} = \frac{-1}{4(2-\rho ^{-2})}\left( \frac{1}{4}(1-h)\rho ^{-2} -4 \left(  \alpha _{n-k-1}+\frac{1}{2}(n-k-1) \right)^2 \right) 
\end{cases} \nonumber 
\end{equation}
\subsubsection{Infinite series}
Replace coefficients $a$, $q$, $\alpha $, $\beta $, $\gamma $, $\delta$ and $x$ by $1-a$, $-q+(\delta -1)\gamma a+(\alpha -\delta +1)(\beta -\delta +1)$, $\alpha-\delta +1 $, $\beta-\delta +1 $, $2-\delta$, $\gamma $ and $1-x$ into (\ref{eq:10085}). Multiply $(1-x)^{1-\delta }$ and (\ref{eq:10085}) together. Put (\ref{eq:1006}) into the new (\ref{eq:10085}).
\begin{eqnarray}
&&\varsigma ^{\frac{1}{2}}y(\varsigma )\nonumber\\
&=& \varsigma ^{\frac{1}{2}} Hl\left( 1-\rho ^{-2}, -\frac{1}{4}\left( (1-h)\rho ^{-2}+(\alpha -1)(\alpha +2)\right); \frac{\alpha }{2}+1, -\frac{\alpha }{2}+\frac{1}{2}, \frac{3}{2}, \frac{1}{2}; \varsigma \right)\nonumber\\
&=& \varsigma ^{\frac{1}{2}} \left\{ _2F_1 \left( \frac{\alpha }{4}+\frac{1}{2}, -\frac{\alpha }{4}+\frac{1}{4}; \frac{5}{4}; z \right) \right. + \sum_{n=1}^{\infty } \Bigg\{\prod _{k=0}^{n-1} \Bigg\{ \int_{0}^{1} dt_{n-k}\;t_{n-k}^{\frac{1}{2}(n-k-2)} \int_{0}^{1} du_{n-k}\;u_{n-k}^{\frac{1}{2}(n-k-\frac{3}{2} )} \nonumber\\
&&\times\frac{1}{2\pi i}  \oint dv_{n-k} \frac{1}{v_{n-k}} \left( 1-\frac{1}{v_{n-k}}\right)^{-\frac{1}{2}(n-k+1+\frac{\alpha}{2} )}  \left( 1- \overleftrightarrow {w}_{n-k+1,n}v_{n-k}(1-t_{n-k})(1-u_{n-k})\right)^{-\frac{1}{2}(n-k+\frac{1}{2}-\frac{\alpha}{2})}\nonumber\\
&&\times  \left( \overleftrightarrow {w}_{n-k,n}^{-\frac{1}{2}(n-k-1)}\left(  \overleftrightarrow {w}_{n-k,n} \partial _{ \overleftrightarrow {w}_{n-k,n}}\right) \overleftrightarrow {w}_{n-k,n}^{\frac{1}{2}(n-k-1)}\left( \overleftrightarrow {w}_{n-k,n} \partial _{ \overleftrightarrow {w}_{n-k,n}}  + \frac{1}{2}(n-k) \right) +Q \right) \Bigg\}\nonumber\\
&&\times \left. _2F_1 \left( \frac{\alpha }{4}+\frac{1}{2}, -\frac{\alpha }{4}+\frac{1}{4}; \frac{5}{4}; \overleftrightarrow {w}_{1,n} \right) \Bigg\} \eta ^n \right\}\label{eq:10093}
\end{eqnarray}
where
\begin{equation}
Q= -\frac{1}{16(2-\rho ^{-2})} \left( (1-h)\rho ^{-2} +(\alpha -1)(\alpha +2)\right)
\nonumber  
\end{equation}
On (\ref{eq:10092}) and (\ref{eq:10093}),
\begin{equation}
\begin{cases} 
\varsigma= 1-\xi \cr
\xi = sn^2(z,\rho ) \cr
\eta =\frac{2-\rho ^{-2}}{1-\rho ^{-2}}\varsigma \cr
z=\frac{-1}{1-\rho ^{-2}} \varsigma^2 
\end{cases}\nonumber  
\end{equation}
\subsection{ ${\displaystyle x^{-\alpha } Hl\left(\frac{1}{a},\frac{q+\alpha [(\alpha -\gamma -\delta +1)a-\beta +\delta ]}{a}; \alpha , \alpha -\gamma +1, \alpha -\beta +1,\delta ;\frac{1}{x}\right)}$}
\subsubsection{Infinite series}
Replace coefficients $a$, $q$, $\beta $, $\gamma $ and $x$ by $\frac{1}{a}$, $\frac{q+\alpha [(\alpha -\gamma -\delta +1)a-\beta +\delta ]}{a}$, $\alpha-\gamma +1 $, $\alpha -\beta +1 $ and $\frac{1}{x}$ into (\ref{eq:10085}). Multiply $x^{-\alpha }$ and (\ref{eq:10085}) together. Put (\ref{eq:1006}) into the new (\ref{eq:10085}).
\begin{eqnarray}
&&\varsigma ^{\frac{1}{2}(\alpha +1)} y(\varsigma )\nonumber\\
&=& \varsigma ^{\frac{1}{2}(\alpha +1)} Hl\left(\rho ^2,-\frac{1}{4}\left( h-(1+\rho ^2)(\alpha +1)^2\right); \frac{1}{2}(\alpha +1), \frac{1}{2}(\alpha +2),\alpha +\frac{3}{2}, \frac{1}{2}; \varsigma \right) \nonumber\\
&=& \varsigma ^{\frac{1}{2}(\alpha +1)} \left\{ _2F_1 \left( \frac{\alpha }{4}+\frac{1}{4}, \frac{\alpha }{4}+\frac{1}{2}; \frac{\alpha }{2}+\frac{5}{4}; z \right) \right. + \sum_{n=1}^{\infty } \Bigg\{\prod _{k=0}^{n-1} \Bigg\{ \int_{0}^{1} dt_{n-k}\;t_{n-k}^{\frac{1}{2}(n-k-2)} \int_{0}^{1} du_{n-k}\;u_{n-k}^{\frac{1}{2}(n-k-\frac{3}{2}+\alpha )} \nonumber\\
&&\times\frac{1}{2\pi i}  \oint dv_{n-k} \frac{1}{v_{n-k}} \left( 1-\frac{1}{v_{n-k}}\right)^{-\frac{1}{2}(n-k+\frac{1}{2}+\frac{\alpha }{2})}  \left( 1- \overleftrightarrow {w}_{n-k+1,n}v_{n-k}(1-t_{n-k})(1-u_{n-k})\right)^{-\frac{1}{2}(n-k+1+\frac{\alpha }{2})}\nonumber\\
&&\times  \left( \overleftrightarrow {w}_{n-k,n}^{-\frac{1}{2}(n-k-1)}\left(  \overleftrightarrow {w}_{n-k,n} \partial _{ \overleftrightarrow {w}_{n-k,n}}\right) \overleftrightarrow {w}_{n-k,n}^{\frac{1}{2}(n-k-1)}\left( \overleftrightarrow {w}_{n-k,n} \partial _{ \overleftrightarrow {w}_{n-k,n}}  +\frac{1}{2}(n-k+\alpha ) \right) +Q \right) \Bigg\}\nonumber\\
&&\times \left. _2F_1 \left(  \frac{\alpha }{4}+\frac{1}{4}, \frac{\alpha }{4}+\frac{1}{2}; \frac{\alpha }{2}+\frac{5}{4}; \overleftrightarrow {w}_{1,n} \right) \Bigg\} \eta ^n \right\} \label{eq:10094}
\end{eqnarray}
where
\begin{equation}
\begin{cases} 
\varsigma =\xi ^{-1} \cr
\xi = sn^2(z,\rho ) \cr
\eta =  (1+\rho ^{-2}) \varsigma \cr
z = -\rho ^{-2}\varsigma ^2 \cr
Q= -\frac{1}{16}\left( h(1+\rho ^2)^{-1}-(\alpha +1)^2\right)
\end{cases}\nonumber  
\end{equation}
\subsection{ ${\displaystyle \left(1-\frac{x}{a} \right)^{-\beta } Hl\left(1-a, -q+\gamma \beta; -\alpha +\gamma +\delta, \beta, \gamma, \delta; \frac{(1-a)x}{x-a} \right)}$}
\subsubsection{Infinite series}
Replace coefficients $a$, $q$, $\alpha $ and $x$ by $1-a$, $-q+\gamma \beta $, $-\alpha+\gamma +\delta $ and $\frac{(1-a)x}{x-a}$ into (\ref{eq:10085}). Multiply $\left(1-\frac{x}{a} \right)^{-\beta }$ and (\ref{eq:10085}) together. Put (\ref{eq:1006}) into the new (\ref{eq:10085}).
\begin{eqnarray}
&&(1-\rho ^2 \xi)^{\frac{\alpha }{2}} y(\varsigma )\nonumber\\
&=& (1-\rho ^2 \xi)^{\frac{\alpha }{2}} Hl\left(1-\rho ^{-2}, \frac{1}{4}\left( h\rho ^{-2} - \alpha \right); -\frac{\alpha }{2}+\frac{1}{2}, -\frac{\alpha }{2}, \frac{1}{2}, \frac{1}{2}; \varsigma \right) \nonumber\\
&=& (1-\rho ^2 \xi)^{\frac{\alpha }{2}} \left\{ _2F_1 \left( -\frac{\alpha }{4}+\frac{1}{4}, -\frac{\alpha }{4}; \frac{3}{4}; z \right)\right. + \sum_{n=1}^{\infty } \Bigg\{\prod _{k=0}^{n-1} \Bigg\{ \int_{0}^{1} dt_{n-k}\;t_{n-k}^{\frac{1}{2}(n-k-2)} \int_{0}^{1} du_{n-k}\;u_{n-k}^{\frac{1}{2}(n-k-\frac{5}{2})} \nonumber\\
&&\times\frac{1}{2\pi i}  \oint dv_{n-k} \frac{1}{v_{n-k}} \left( 1-\frac{1}{v_{n-k}}\right)^{-\frac{1}{2}(n-k+\frac{1}{2}-\frac{\alpha }{2} )}  \left( 1- \overleftrightarrow {w}_{n-k+1,n}v_{n-k}(1-t_{n-k})(1-u_{n-k})\right)^{-\frac{1}{2}(n-k-\frac{\alpha }{2} )}\nonumber\\
&&\times  \left( \overleftrightarrow {w}_{n-k,n}^{-\frac{1}{2}(n-k-1)}\left(  \overleftrightarrow {w}_{n-k,n} \partial _{ \overleftrightarrow {w}_{n-k,n}}\right) \overleftrightarrow {w}_{n-k,n}^{\frac{1}{2}(n-k-1)}\left( \overleftrightarrow {w}_{n-k,n} \partial _{ \overleftrightarrow {w}_{n-k,n}}  + \Omega _{n-k-1} \right) +Q \right) \Bigg\}\nonumber\\
&&\times \left. _2F_1 \left( -\frac{\alpha }{4}+\frac{1}{4}, -\frac{\alpha }{4}; \frac{3}{4}; \overleftrightarrow {w}_{1,n} \right) \Bigg\} \eta ^n \right\} \label{eq:10095}
\end{eqnarray}
where
\begin{equation}
\begin{cases} 
\varsigma =\frac{(1-\rho ^{-2})\xi }{\xi -\rho ^{-2}} \cr
\xi = sn^2(z,\rho ) \cr
\eta =  \frac{2-\rho ^{-2}}{1-\rho ^{-2}} \varsigma \cr
z = \frac{-1}{1-\rho ^{-2}}\varsigma ^2 \cr
\Omega _{n-k-1}  = \frac{-\alpha }{2(2-\rho ^{-2})}+\frac{1}{2}(n-k-1) \cr
Q= \frac{h\rho ^{-2}-\alpha }{16(2-\rho ^{-2})} 
\end{cases}\nonumber  
\end{equation}
\subsection{ ${\displaystyle (1-x)^{1-\delta }\left(1-\frac{x}{a} \right)^{-\beta+\delta -1} Hl\Bigg(1-a, -q+\gamma [(\delta -1)a+\beta -\delta +1]; -\alpha +\gamma +1}$\\ ${\displaystyle, \beta -\delta+1, \gamma, 2-\delta; \frac{(1-a)x}{x-a} \Bigg)}$ }
\subsubsection{Infinite series}
Replace coefficients $a$, $q$, $\alpha $, $\beta $, $\delta $ and $x$ by $1-a$, $-q+\gamma [(\delta -1)a+\beta -\delta +1]$, $-\alpha +\gamma +1$, $\beta -\delta+1$, $2-\delta $ and $\frac{(1-a)x}{x-a}$ into (\ref{eq:10085}). Multiply $(1-x)^{1-\delta }\left(1-\frac{x}{a} \right)^{-\beta+\delta -1}$ and (\ref{eq:10085}) together. Put (\ref{eq:1006}) into the new (\ref{eq:10085}).
\begin{eqnarray}
&&(1-\xi )^{\frac{1}{2}}(1-\rho ^2 \xi)^{\frac{1}{2}(\alpha -1)} y(\varsigma )\nonumber\\
&=& (1-\xi )^{\frac{1}{2}}(1-\rho ^2 \xi)^{\frac{1}{2}(\alpha -1)} Hl\left( 1-\rho ^{-2}, \frac{1}{4}\left( (h-1)\rho ^{-2} +1- \alpha \right); -\frac{\alpha }{2}+1, -\frac{\alpha }{2}+\frac{1}{2}, \frac{1}{2}, \frac{3}{2}; \varsigma \right) \nonumber\\
&=& (1-\xi )^{\frac{1}{2}}(1-\rho ^2 \xi)^{\frac{1}{2}(\alpha -1)} \left\{ _2F_1 \left( -\frac{\alpha }{4}+\frac{1}{2}, -\frac{\alpha }{4}+\frac{1}{4}; \frac{3}{4}; z \right) \right. + \sum_{n=1}^{\infty } \Bigg\{\prod _{k=0}^{n-1} \Bigg\{ \int_{0}^{1} dt_{n-k}\;t_{n-k}^{\frac{1}{2}(n-k-2)} \int_{0}^{1} du_{n-k}\;u_{n-k}^{\frac{1}{2}(n-k-\frac{5}{2} )} \nonumber\\
&&\times\frac{1}{2\pi i}  \oint dv_{n-k} \frac{1}{v_{n-k}} \left( 1-\frac{1}{v_{n-k}}\right)^{-\frac{1}{2}(n-k+1-\frac{\alpha }{2} )}  \left( 1- \overleftrightarrow {w}_{n-k+1,n}v_{n-k}(1-t_{n-k})(1-u_{n-k})\right)^{-\frac{1}{2}(n-k+\frac{1}{2}-\frac{\alpha }{2} )}\nonumber\\
&&\times  \left( \overleftrightarrow {w}_{n-k,n}^{-\frac{1}{2}(n-k-1)}\left(  \overleftrightarrow {w}_{n-k,n} \partial _{ \overleftrightarrow {w}_{n-k,n}}\right) \overleftrightarrow {w}_{n-k,n}^{\frac{1}{2}(n-k-1)}\left( \overleftrightarrow {w}_{n-k,n} \partial _{ \overleftrightarrow {w}_{n-k,n}}  + \Omega _{n-k-1} \right) +Q \right) \Bigg\}\nonumber\\
&&\times \left. _2F_1 \left( -\frac{\alpha }{4}+\frac{1}{2}, -\frac{\alpha }{4}+\frac{1}{4}; \frac{3}{4}; \overleftrightarrow {w}_{1,n} \right) \Bigg\} \eta ^n \right\} \label{eq:10096}
\end{eqnarray}
where
\begin{equation}
\begin{cases}
\varsigma =\frac{(1-\rho ^{-2})\xi }{\xi -\rho ^{-2}} \cr
\xi = sn^2(z,\rho ) \cr
\eta =  \frac{2-\rho ^{-2}}{1-\rho ^{-2}} \varsigma \cr
z = \frac{-1}{1-\rho ^{-2}}\varsigma ^2 \cr
\Omega _{n-k-1} = \frac{1}{2}\left( n-k -\frac{\alpha +1}{2-\rho ^{-2}}\right) \cr
Q= \frac{ (h-1)\rho ^{-2}+1-\alpha}{16(2-\rho ^{-2})} 
\end{cases}\nonumber  
\end{equation}
\subsection{ ${\displaystyle x^{-\alpha } Hl\left(\frac{a-1}{a}, \frac{-q+\alpha (\delta a+\beta -\delta )}{a}; \alpha, \alpha -\gamma +1, \delta , \alpha -\beta +1; \frac{x-1}{x} \right)}$}
\subsubsection{Infinite series}
Replace coefficients $a$, $q$, $\beta $, $\gamma $, $\delta $ and $x$ by $\frac{a-1}{a}$, $\frac{-q+\alpha (\delta a+\beta -\delta )}{a}$, $\alpha -\gamma +1$, $\delta $, $\alpha -\beta +1$ and $\frac{x-1}{x}$ into (\ref{eq:10085}). Multiply $x^{-\alpha }$ and (\ref{eq:10085}) together.
Put (\ref{eq:1006}) into the new (\ref{eq:10085}).
\begin{eqnarray}
&&\xi ^{-\frac{1}{2}(\alpha +1)} y(\varsigma )\nonumber\\
&=& \xi ^{-\frac{1}{2}(\alpha +1)} Hl\left( 1-\rho ^2, \frac{1}{4}\left[ h+ (\alpha +1)\left( 1-(\alpha +1)\rho ^2\right) \right];  \frac{1}{2}(\alpha +1), \frac{1}{2}(\alpha +2), \frac{1}{2}, \alpha +\frac{3}{2}; \varsigma \right) \nonumber\\
&=& \xi ^{-\frac{1}{2}(\alpha +1)} \left\{ _2F_1 \left( \frac{\alpha }{4}+\frac{1}{4}, \frac{\alpha }{4}+\frac{1}{2}; \frac{3}{4}; z \right) \right. + \sum_{n=1}^{\infty } \Bigg\{\prod _{k=0}^{n-1} \Bigg\{ \int_{0}^{1} dt_{n-k}\;t_{n-k}^{\frac{1}{2}(n-k-2)} \int_{0}^{1} du_{n-k}\;u_{n-k}^{\frac{1}{2}(n-k-\frac{5}{2} )} \nonumber\\
&&\times\frac{1}{2\pi i}  \oint dv_{n-k} \frac{1}{v_{n-k}} \left( 1-\frac{1}{v_{n-k}}\right)^{-\frac{1}{2}(n-k+\frac{1}{2}+\frac{\alpha }{2} )}  \left( 1- \overleftrightarrow {w}_{n-k+1,n}v_{n-k}(1-t_{n-k})(1-u_{n-k})\right)^{-\frac{1}{2}(n-k+1+\frac{\alpha }{2})}\nonumber\\
&&\times  \left( \overleftrightarrow {w}_{n-k,n}^{-\frac{1}{2}(n-k-1)}\left(  \overleftrightarrow {w}_{n-k,n} \partial _{ \overleftrightarrow {w}_{n-k,n}}\right) \overleftrightarrow {w}_{n-k,n}^{\frac{1}{2}(n-k-1)}\left( \overleftrightarrow {w}_{n-k,n} \partial _{ \overleftrightarrow {w}_{n-k,n}}  + \Omega _{n-k-1} \right) +Q \right) \Bigg\}\nonumber\\
&&\times \left. _2F_1 \left( \frac{\alpha }{4}+\frac{1}{4}, \frac{\alpha }{4}+\frac{1}{2}; \frac{3}{4}; \overleftrightarrow {w}_{1,n} \right) \Bigg\} \eta ^n \right\} \label{eq:10097}
\end{eqnarray}
where
\begin{equation}
\begin{cases} 
\varsigma =\frac{ \xi -1}{\xi } \cr
\xi = sn^2(z,\rho ) \cr
\eta =  \frac{2-\rho ^2}{1-\rho ^2} \varsigma \cr
z = \frac{-1}{1-\rho ^2}\varsigma ^2 \cr
\Omega _{n-k-1} = \frac{1}{2}\left( n-k +\frac{\alpha (1-\rho ^2)-1}{2-\rho ^2}\right) \cr
Q= \frac{h+(\alpha +1)(1-(\alpha +1)\rho ^2)}{16(2-\rho ^2)} 
\end{cases}\nonumber  
\end{equation}
\subsection{ ${\displaystyle \left(\frac{x-a}{1-a} \right)^{-\alpha } Hl\left(a, q-(\beta -\delta )\alpha ; \alpha , -\beta+\gamma +\delta , \delta , \gamma; \frac{a(x-1)}{x-a} \right)}$}
\subsubsection{Infinite series}
Replace coefficients $q$, $\beta $, $\gamma $, $\delta $ and $x$  by $q-(\beta -\delta )\alpha $, $-\beta+\gamma +\delta $, $\delta $,  $\gamma $ and $\frac{a(x-1)}{x-a}$ into (\ref{eq:10085}). Multiply $\left(\frac{x-a}{1-a} \right)^{-\alpha }$ and (\ref{eq:10085}) together. Put (\ref{eq:1006}) into the new (\ref{eq:10085}).
\begin{eqnarray}
&&\left(\frac{\xi-\rho ^{-2}}{1-\rho ^{-2}} \right)^{-\frac{1}{2}(\alpha +1)} y(\varsigma )\nonumber\\
&=& \left(\frac{\xi-\rho ^{-2}}{1-\rho ^{-2}} \right)^{-\frac{1}{2}(\alpha +1)} Hl\left( \rho ^{-2}, -\frac{1}{4}\left( h\rho ^{-2}- (\alpha +1)^2\right);  \frac{1}{2}(\alpha +1), -\frac{1}{2}(\alpha -2), \frac{1}{2}, \frac{1}{2}; \varsigma \right) \nonumber\\
&=& \left(\frac{\xi-\rho ^{-2}}{1-\rho ^{-2}} \right)^{-\frac{1}{2}(\alpha +1)} \left\{ _2F_1 \left( \frac{\alpha }{4}+\frac{1}{4}, \frac{\alpha }{4}+\frac{1}{2}; \frac{3}{4}; z \right) \right. + \sum_{n=1}^{\infty } \Bigg\{\prod _{k=0}^{n-1} \Bigg\{ \int_{0}^{1} dt_{n-k}\;t_{n-k}^{\frac{1}{2}(n-k-2)} \int_{0}^{1} du_{n-k}\;u_{n-k}^{\frac{1}{2}(n-k-3-\frac{5}{2} )} \nonumber\\
&&\times\frac{1}{2\pi i}  \oint dv_{n-k} \frac{1}{v_{n-k}} \left( 1-\frac{1}{v_{n-k}}\right)^{-\frac{1}{2}(n-k+\frac{1}{2}+\frac{\alpha }{2})}  \left( 1- \overleftrightarrow {w}_{n-k+1,n}v_{n-k}(1-t_{n-k})(1-u_{n-k})\right)^{-\frac{1}{2}(n-k+1+\frac{\alpha }{2} )}\nonumber\\
&&\times  \left( \overleftrightarrow {w}_{n-k,n}^{-\frac{1}{2}(n-k-1)}\left(  \overleftrightarrow {w}_{n-k,n} \partial _{ \overleftrightarrow {w}_{n-k,n}}\right) \overleftrightarrow {w}_{n-k,n}^{\frac{1}{2}(n-k-1)}\left( \overleftrightarrow {w}_{n-k,n} \partial _{ \overleftrightarrow {w}_{n-k,n}}  + \Omega _{n-k-1} \right) +Q \right) \Bigg\}\nonumber\\
&&\times \left. _2F_1 \left( \frac{\alpha }{4}+\frac{1}{4}, \frac{\alpha }{4}+\frac{1}{2}; \frac{3}{4}; \overleftrightarrow {w}_{1,n} \right) \Bigg\} \eta ^n \right\} \label{eq:10098}
\end{eqnarray}
where
\begin{equation}
\begin{cases} 
\varsigma =\frac{\xi -1}{\rho ^2(\xi-\rho ^{-2})} \cr
\xi = sn^2(z,\rho ) \cr
\eta =  (1+\rho ^2) \varsigma \cr
z = -\rho ^2 \varsigma ^2 \cr
\Omega _{n-k-1} = \frac{1}{2}\left( n-k-1 +\frac{ \alpha +1 }{1+\rho ^{-2}}\right) \cr
Q= \frac{-h\rho ^{-2}+ (\alpha +1)^2}{16(1+\rho ^{-2})} 
\end{cases}\nonumber  
\end{equation}

\section*{Acknowledgment}
I thank Bogdan Nicolescu. The discussions I had with him on number theory was of great joy. 
\vspace{3mm}

\bibliographystyle{amsplain}
 
\end{document}